\documentclass[a4paper,11pt]{article}
\pdfoutput=1 % if your are submitting a pdflatex (i.e. if you have
\usepackage{jinstpub} % for details on the use of the package, please
\usepackage{lineno}
\usepackage{siunitx}
\usepackage{subfig}
\usepackage{xspace}
\usepackage{multirow}
\usepackage{soul}
\newcommand{\ohmcm}{\ensuremath{\Omega\cdot\rm{cm}}}
\newcommand{\onemev}{\ensuremath{\rm{n_{eq}}/cm^2}}

\title{Electrical Characterization of \SI[detect-weight]{180}{\nano\meter} ATLASPix2 HV-CMOS Monolithic Prototypes for the High-Luminosity LHC}

\author[a,1]{D M S Sultan,\note{Corresponding author.}}
\author[a]{S. Gonzalez-Sevilla,}
\author[a]{D. Ferrere,}
\author[a]{G. Iacobucci,}
\author[a]{E. Zaffaroni,}
\author[a]{W. Wong,}
\author[a]{M. Kiehn,}
\author[a]{and M. Benoit}

\affiliation[a]{Department of Particle and Nuclear Physics, University of Geneva,\\24, quai Ernest-Ansermet, CH-1211 Gen\`eve 4, Switzerland}
\emailAdd{dms.sultan@unige.ch}

\abstract{
We report on the experimental study made on a successive prototype of High-Voltage Complementary Metal-Oxide-Semiconductor ATLASPix2 sensor for the tracking detector application, developed with 180 nm feature size. These sensors are to qualify mainly the peripheral data processing blocks ({\em e.g.} Command Decoder, Trigger Buffer, etc.). Over a decade, the monolithic pixelated sensors for a foreseen application to the ATLAS High-Luminosity LHC upgrade are being investigated as a viable option and a significant R\&D progress made. It is a smaller version of $24\times 36$ pixelated sensor in comparison to the earlier generation of ATLASPix1 fabricated in both ams AG, Austria, and TSI Semiconductors, USA. While ams produced ATLASPix2 showed breakdown voltage $\sim$50 V in nonirradiated condition as it was seen on its predecessors ATLASpix1, TSI produced prototypes reported breakdown voltage greater than 100 V. The chosen wafer of MCz 20~\ohmcm\ P-type substrate resistivity can deplete a few tenths of $\mu m$, where the process-driven surface damage can have a greater impact on device operating conditions before and after irradiation. In an aim to understand device intrinsic performance at the irradiated case, a dedicated neutron irradiation campaign has been made at Jo\v{z}ef Stefan Institute in Slovenia for different fluences. Characterizations have been performed at different temperatures after irradiation to analyze the leakage current and breakdown voltage before and after irradiation. TSI prototypes showed a breakdown voltage decrease ($\sim$\SI{90}{\volt}) due to impact ionization and enhanced effective doping concentration. Results demonstrated for the neutron-irradiated devices up to the fluence of $2\times 10^{15}~\onemev$ can still safely be operated at a voltage high enough to allow for high efficiency. Accelerated annealing steps were also made on selective irradiated ATLASPix2 samples, equivalent to more than two years of room-temperature annealing (at \SI{20}{\celsius}), and they showed the reassuring expected breakdown voltage increase and damage constant rate $\alpha^*$ (geometry dependent) decrease, driven by the beneficial annealing. 
}
\keywords{Radiation-hard detectors; Solid state detectors; Detector design and construction
technologies and materials; Radiation damage to detector materials (solid state).}

\begin{document}
\maketitle
\flushbottom

\section{Introduction}
\label{sec:introduction}

The ATLAS experiment will perform a major detector upgrade for the High-Luminosity LHC (HL-LHC), where protons will collide at a center of mass energy of $\sqrt{s}=$\SI{14}{\tera\electronvolt} and the instantaneous luminosity will reach $\mathcal{L}=7.5\times 10^{34}\,\rm{cm^{-2}\,s^{-1}}$. The new ATLAS Inner Tracker (ITk) will be an all-silicon tracker composed of pixel detectors in the innermost layers and of silicon micro-strip modules at outer radii. ITk will provide excellent tracking capabilities in a very dense particle environment with up to $\sim 200$ inelastic proton-proton collisions per bunch-crossing~\cite{strip-upgrade-tdr}. The pixel detectors  for ITk will need to withstand an extreme radiation environment, with expected fluences ranging from $\sim 10^{15}~\onemev$ at the outermost pixel layer ($R\sim$ \SI{290}{\milli\meter}) up to $\sim 10^{16}~\onemev$ at the innermost radius \mbox{($R\sim$\SI{39}{\milli\meter})}, including a safety factor of 1.5~\cite{pixel-upgrade-tdr}. High-Voltage Complementary Metal-Oxide-Semiconductor (HV-CMOS) sensors have been considered as a viable option for the outermost pixel layer because this technology offers interesting features such as high-granularity, low-power consumption, low system-level cost and high radiation tolerance. In this context, we have developed the ATLASPix2 chip, a fully monolithic HV-CMOS chip produced in \SI{180}{\nano\meter} at two different foundries, ams AG~\cite{ams} and TSI Semiconductors~\cite{tsi}. Both productions were received at the last quarter of 2018, and some chips were subsequently irradiated with neutrons in march 2019 to fluences up to $2\times 10^{15}$~\onemev. 

In this paper, we report on the electrical characteristics of both non-irradiated and irradiated ATLASPix2 prototype chips. The paper is organized as follows. In section~\ref{sec:atlaspix2} the ATLASPix2 chip is introduced. The small design differences between the ams AG and the TSI samples are described. The neutron irradiation campaign performed at the Jo\v{z}ef Stefan Institute (JSI) TRIGA reactor is explained in section~\ref{sec:irradiation}. Then the experimental setup used for the electrical characterization of the samples is described in section~\ref{sec:setup}. Electrical results obtained before and after neutron irradiation are shown in section~\ref{sec:results}. Results on a systematic accelerated annealing study are also presented. Finally, conclusions are given in section~\ref{sec:conclusions}.

\section{The ATLASPix2 HV-CMOS monolithic chip}
\label{sec:atlaspix2}

\begin{figure}[!htbp]
\centering 
\includegraphics[width=1.0\textwidth]{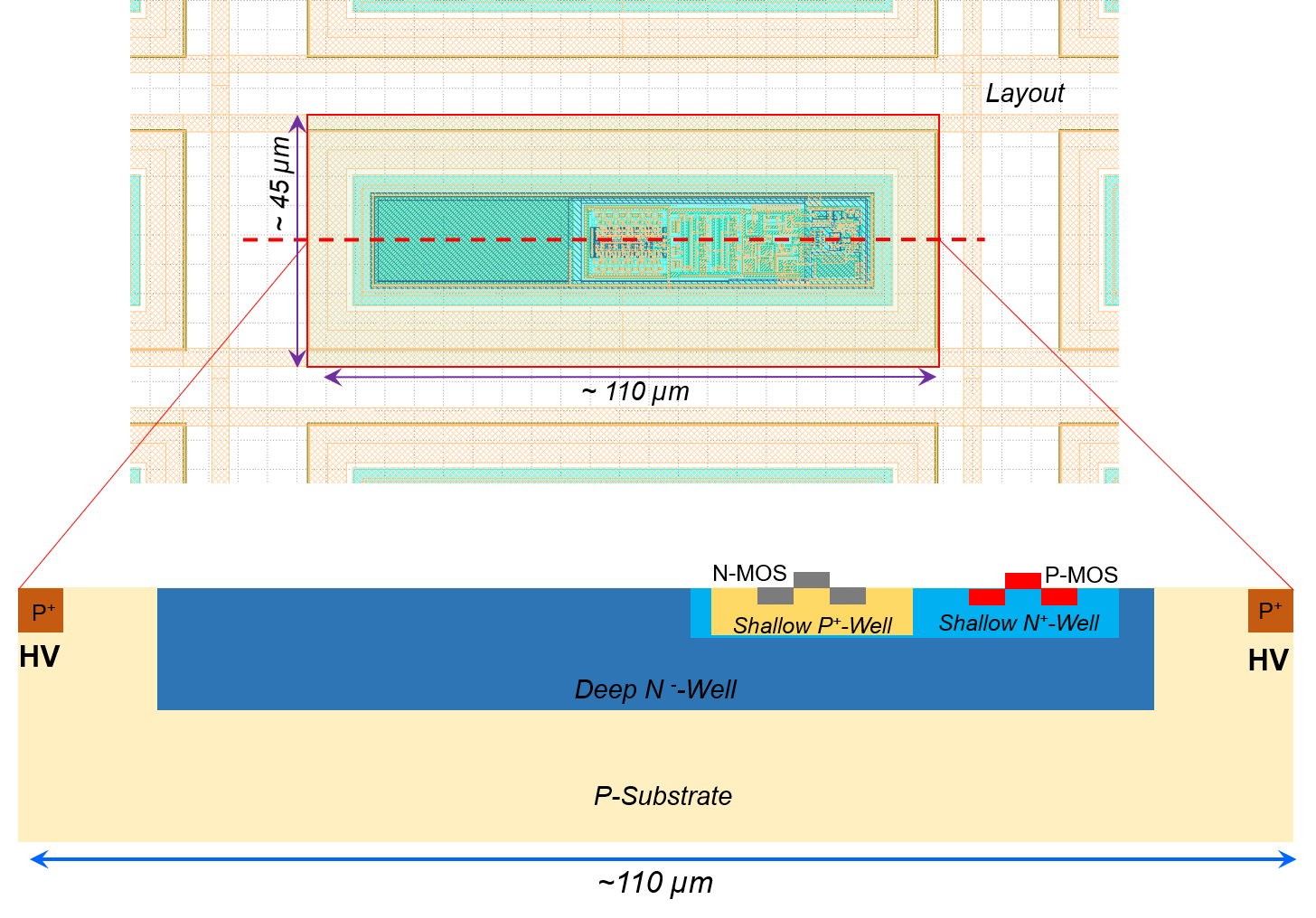}
\caption{\label{f:atlaspix2-layout} Top layout of the ATLASPix2 chip where single pixel-pitch has been denoted with the red rectangular boundary. The dashed red line indicates the lateral view reference where the corresponding cross-section is shown (below).}
\end{figure}

The ATLASPix2 is a fully monolithic \SI{180}{\nano\meter} HV-CMOS chip. Its design is similar to that of its predecessor, the ATLASPix1 chip~\cite{ivan1, sultan1}, with main differences being a smaller pixel matrix of $24\times 36$ pixels,  several improved peripheral blocks (e.g. command decoder) and a single triggered readout scheme. Figure~\ref{f:atlaspix2-layout} shows the top layout and the schematic cross-section of a single ATLASPix2 pixel. The pixel active area is $\sim 110\times\SI{45}{\micro\meter\squared}$. A deep $\rm{N^-}$-well acts both as the charge collection electrode and as the substrate for PMOS transistors. Shallow $\rm{N^+}$ and $\rm{P^+}$-wells, embedded in the deep $\rm{N^-}$-well, hold the analog electronics (charge amplifier and discriminator). A High-Voltage $\rm{P^+}$-implant placed outside the deep $\rm{N^-}$-well is used to deplete the $\rm{P}$-Substrate. The distance between the $\rm{N^-}$-well and the $\rm{P^+}$-implant has been carefully chosen using numerical TCAD simulations to achieve a breakdown greater than \SI{120}{\volt} for non-irradiated devices~\cite{sultan1}. The ATLASPix2 holds a  single $\rm{P^+}$-implant guard-ring around the pixel matrix. The dicing edge to the guard-ring has been kept larger than \SI{50}{\micro\meter}.

\begin{figure}[!htbp]
\centering
\subfloat[]{\label{f:atlaspix2-a}\includegraphics[width=.47\textwidth]{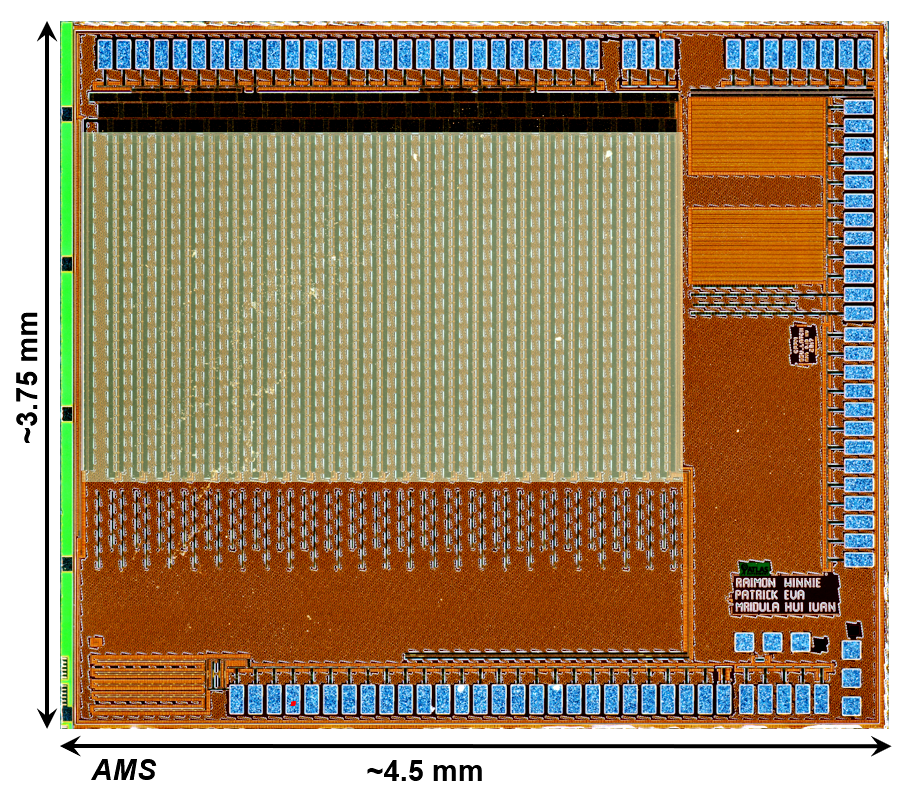}}
\subfloat[]{\label{f:atlaspix2-b}\includegraphics[width=.52\textwidth]{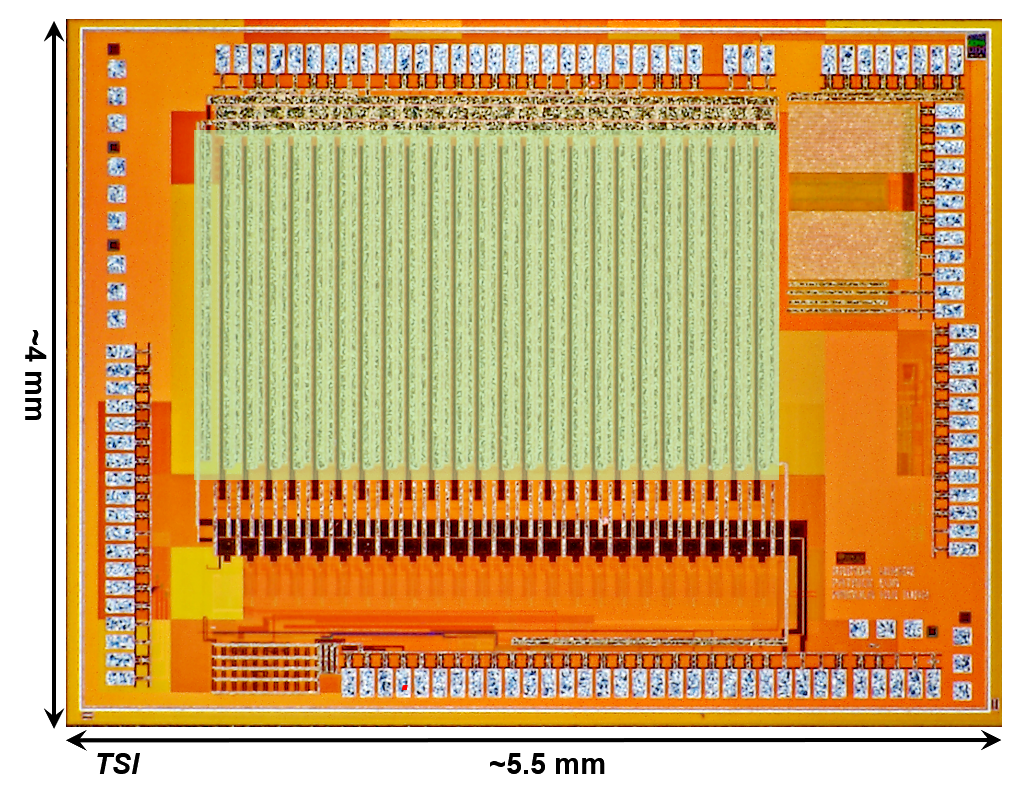}}
\caption{\label{f:atlaspix2} Micrographs of ATLASPix2 prototype chips: (a) fabricated at ams AG, and (b) fabricated at TSI Semiconductors.  In the batch produced by TSI, additional pads were added.}
\end{figure}

ATLASPix2 prototype chips have been fabricated at two different foundries, ams AG~\cite{ams} and TSI Semiconductors~\cite{tsi}. Both batches were produced with P-type silicon substrate from the ingot TOPSIL~\cite{topsil}. %With the aim of exploring competitive alternatives to ams AG, TSI Semiconductors was chosen because offering shorter production-times and compatible design kits. 
In both cases, the wafers were of Magnetic Czochralski (MCz) $\langle 100\rangle$ crystal orientation, \SI{200}{\milli\meter} in diameter and standard substrate resistivity of 10-20~\ohmcm. No higher substrate resistivity has been explored. Figure~\ref{f:atlaspix2} shows the micrograph of the two ATLASPix2 prototype chips.
The diced reticle of the TSI prototype is slightly larger in size ($\sim 4.0\times\SI{5.5}{\milli\meter\squared}$) in comparison to the ams one ($\sim 3.75\times\SI{4.5}{\milli\meter\squared}$). The 180 nm HV-CMOS process from TSI allows for one additional metal-layer in comparison to the ams technology, and therefore a few additional signal-pads for testing purposes have been integrated in the TSI chip. No differences in the electrical characteristics of the chips produced from the two foundries are expected from the inclusion of these additional test pads. After production, the wafers from ams AG (TSI) were thinned down to $220\pm \SI{10}{\micro\meter}$ ($254\pm \SI{10}{\micro\meter}$) by a Dicing Before Grinding (DBG) process.
%\hl{The diced reticle of TSI prototype is slightly larger in size ($\sim 4.0\times\SI{5.5}{\micro\meter\squared}$) in comparison to AMS one ($\sim 3.75\times\SI{4.5}{\micro\meter\squared}$). TSI technology offers one additional metal-layer and so, a few addiotional test signal-pads ('SerialPowerRegulator', 'HitBusOR' etc.) have been integrated in the TSI production. It is worth noting, no potential difference of electrical characteristics of sensor-diodes (of these two batches) have been expected in terms of the layout-design.} After production, the wafers from ams AG (TSI) were thinned down to $220\pm \SI{10}{\micro\meter}$  ($254\pm \SI{10}{\micro\meter}$) by a Dicing Before Grinding (DBG) process. 

\section{Irradiation}
\label{sec:irradiation}

Several ATLASPix2 chips (viz. Table~\ref{t:samples}) have been irradiated at the Jo\v{z}ef Stefan Institute (JSI) Neutron Irradiation facility at the \SI{250}{\kilo\watt} TRIGA Mark II reactor~\cite{ambrozic}. In terms of neutron spectra and total flux values, the JSI TRIGA reactor has become a reference center for neutron irradiation of particle detectors. The pool-type reactor core has an annular configuration with a diameter of \SI{44}{\centi\meter}, 91 in-core positions for fuel elements and numerous irradiation positions. The samples were irradiated unbiased at the full reactor power (\SI{250}{\kilo\watt}) with a uniform neutron flux of $\Phi_\text{eq} = 1.54\times 10^{12}~\onemev$, with an uncertainty of 10\%. To reach the target cumulated fluences of $1\times 10^{15}$, $1.5\times 10^{15}$ and $2\times 10^{15}$~\onemev, irradiation steps spanned few tens of minutes (e.g. $\sim$10.84 minutes for $1\times 10^{15}$~\onemev). As soon as the extreme activation dropped to acceptable levels (typically $\sim$10 mins after irradiation), samples were stored in a fridge to avoid any annealing effect. While the ambient temperature at the reactor pool remained at $\sim$\SI{20}{\celsius}, the temperature of the irradiated samples increased up to $\sim$\SI{40}{\celsius} due to the high radiation backgrounds~\cite{mandic1}.

\begin{table}[htbp]
\renewcommand{\arraystretch}{1.4}
\centering
\caption{\label{t:samples}List of irradiated ATLASPix2 chips.}
\smallskip
\begin{tabular}{|c|c|c|c|}
\hline
{\bf Sensor ID} & {\bf Foundry} & {\bf Fluence} \\ 
&  & {\bf [\onemev]} \\ \hline
AP2-AMS-08 & ams AG & $1\times 10^{15}$ \\ \hline
AP2-AMS-09 & ams AG & $1\times 10^{15}$ \\ \hline
AP2-TSI-05 & TSI  & $1\times 10^{15}$ \\ \hline
AP2-TSI-11 & TSI & $1\times 10^{15}$ \\ \hline
AP2-TSI-06 & TSI & $1.5\times 10^{15}$ \\ \hline
AP2-TSI-24 & TSI & $1.5\times 10^{15}$ \\ \hline
AP2-TSI-08 & TSI & $2\times 10^{15}$ \\ \hline
AP2-TSI-17 & TSI & $2\times 10^{15}$ \\ \hline
\end{tabular}
\end{table}
\section{Electrical characterization}
\label{sec:setup}

%---------------------------------------
\subsection{Experimental setup}

\begin{figure}[!tbp]
\centering
\subfloat[]{\label{f:setup-a}\includegraphics[width=.52\textwidth]{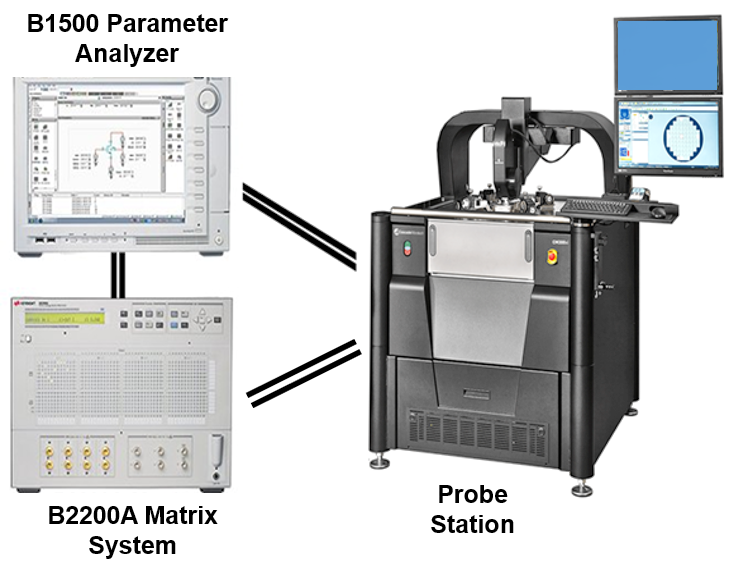}}
\subfloat[]{\label{f:setup-b}\includegraphics[width=.47\textwidth]{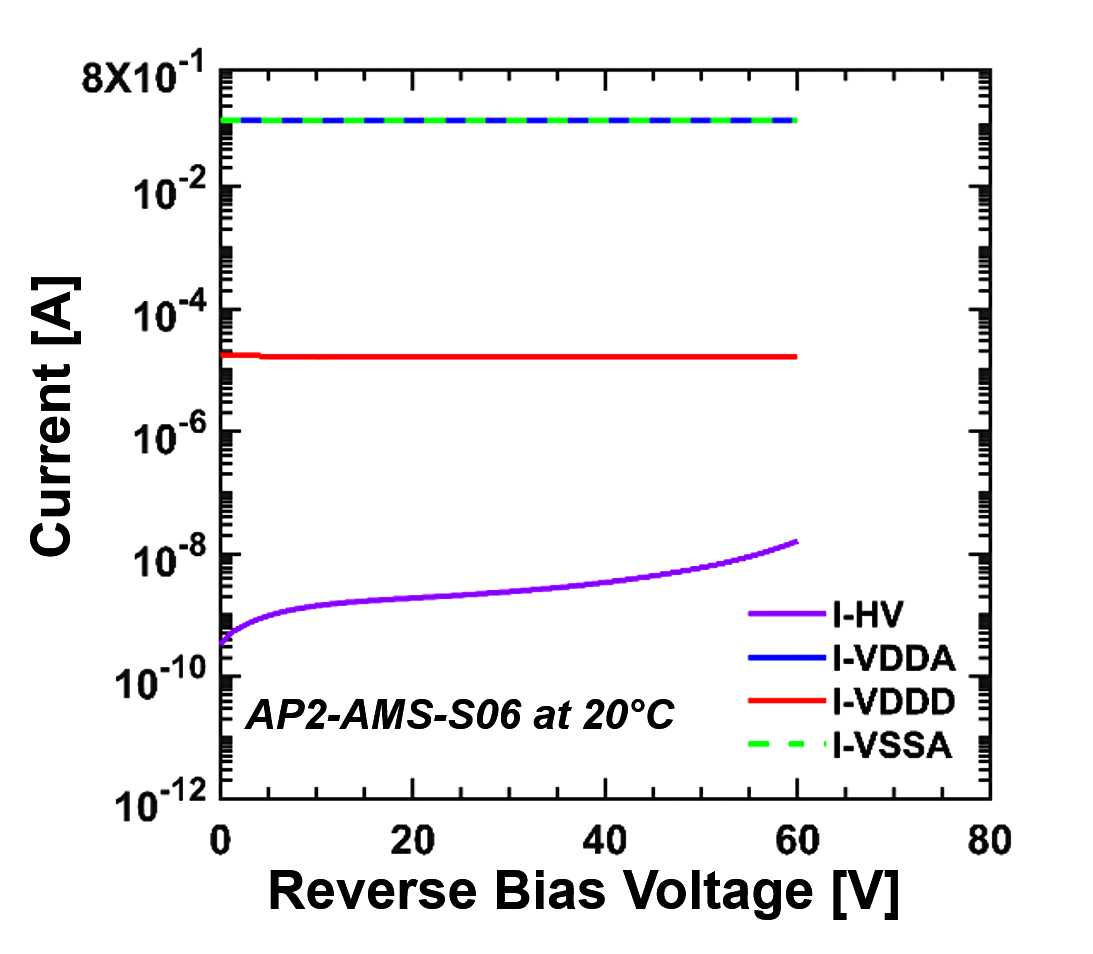}}
\caption{\label{f:probe} (a) Probe-system used for the eletrical characterization. (b) Example of I-V curves for a qualified ATLASPix2 chip.}
\end{figure}

The electrical characterization of the different samples has been performed in the clean-room of the Department of Particle and Nuclear Physics (DPNC) of the University of Geneva with a Cascade Microtech CM300 semi-automatic probe station~\cite{formfactor}. This system provides high attenuation ($>120$ dB) for precise low leakage current measurements and allows temperature probing in the range between \SI{-60}{\celsius} and \SI{200}{\celsius}. The chuck used to hold the dies is a Nickel-coated Aluminum plate with a planarity of $\pm$\SI{5}{\micro\meter}. Figure~\ref{f:setup-a} shows the main components of the setup. In addition to the probe-station itself, a parametric analyzer (model Keysight B1500) and a logical switching matrix (model Keysight B2200A) are used. The Keysight B1500 can provide four Source Measure Unit (SMU) channels with an accuracy of $\pm$\SI{20}{\femto\ampere} each. For the I-V measurements, a Keysight B1510A High Power Unit provides a bias voltage accuracy of $\pm(0.018\% + 30~\rm{mV})$ on the applied voltage range ($\pm$\SI{200}{\volt}). The accuracy of the current measurement depends on both the range of the current and of the applied voltage, but the uncertainity of measured data is $\pm 1\%$ only for a nonirradiated device. For the results presented in this paper, the reverse bias voltage was typically swept between 0 and 120 V. 

The Capacitance-Voltage (C-V) measurements have been performed using a Multi-Frequency Capacitance Measurement Unit (MFCMU). Since the MFCMU has a DC bias range limited to $\pm$\SI{25}{\volt}, a separate SMU was used for higher-range voltage sweeps with additional decoupling capacitors. All C-V data have been acquired at \SI{1}{\kilo\hertz} with a $\pm$\SI{100}{\milli\volt} input AC-signal. The total uncertainty of the C-V measurements includes the measuring instrument uncertainty, the discrete component's tolerances (used in the decoupling box) and the fitting errors to the C-V curves. This total uncertainty is estimated to be  $\pm 20\%$ of the reported values.

During the I-V measurements, all power lines of the readout electronics have been kept at their nominal values: \SI{1.8}{\volt} for the digital (VDDD) and analog (VDDA) voltage supply, and \SI{1}{\volt} for the analog front-end ground (VSSA). %A sensor is said to be qualified if the current of all power-lines remained unaffected before the avalanche breakdown, see figure~\ref{f:setup-b}. 
Figure~\ref{f:setup-b} shows all different measured currents for a qualified ATLASPix2 prototype. 

%-------------------------------------------
\subsection{Definition of breakdown voltage}
%-------------------------------------------

%From the measured leakage from the applied reverse bias (I-V), the breakdown voltage $V_\text{bd}$ of the device is then calculated through an adimensional function as of equation below:

From the I-V characteristics, the breakdown voltage $V_\text{bd}$ is determined as the maximum bias voltage for which $K(I,V) < V_\text{bd}$, with the adimensional function $K(I,V)$ being defined as
\begin{equation}
K(I,V) = \frac{\Delta I}{\Delta V}\cdot\frac{V}{I}
\end{equation}
where $I$ is the sensor reverse current, $V$ is the applied bias voltage and $\Delta I/\Delta V$ is the slope of the measured I–V curve. This definition is independent of the depletion voltage and especially suitable for cases, where the leakage current shows a smooth and continuous rise due to  defects existing in the original wafer or introduced in the silicon lattice after irradiation~\cite{bacchetta}. A $K$-value less than 1 corresponds to the ohmic state of the sensor before breakdown. The case $K(I,V)\gg 1$ represents a real avalanche characteristics. In this paper, a $K$-value equal to 4 has been chosen for the breakdown~\cite{betta}. The error associated to the resulting breakdown voltage ($V_\text{bd}$) is determined from the accuracy of the applied voltage and of the current sensing. Given the very high accuracy of current measurements in the range of interest, uncertainties in the breakdown voltage were mainly driven by the step chosen for the voltage ramp, which goes up to 2 V. The worst-case value ($\pm$\SI{2}{\volt}) has been taken as an upper limit of $V_\text{bd}$ for all measurements.

%The breakdown voltage $V_\text{bd}$ is evaluated at the maximum value of bias $V$ for the limit of the sensitivity of the current $K_\text{bd}$. A $K$-value less than 1 corresponds to the ohmic state of the sensor before breakdown. A $K$-value greater than 1 represents real avalanche characteristics. We have chosen a $K$-value equal to 4 for the breakdown~\cite{betta}. Keeping account the instrumental accuracy for an applied bias, Vbd would have uncertainty of $\pm$2 V.

%Capacitance-Voltage (C-V) measurements have been performed using Multi-Frequency Capacitance Measurement Unit (MFCMU). Since the MFCMU has DC bias range limited to ±25 V, a separate high power SMU was used for voltage sweep with the integration of a capacitive coupling box. All C-V data acquired for 1 kHz frequency with a ±100 mV small AC signal (Root Mean Square value). The similar P-substrate used in AMS HVCMOS of 350 nm process showed depletion depth ~10-20 µm for substrate resistivity of 20 Ω·cm~\cite{ettore}. The total uncertainty of the C-V measurements is estimated to te $\pm 20\%$ of the reported values.       

\section{Results}
\label{sec:results}

%------------------------------------------
\subsection{Results before irradiation}
%------------------------------------------
\begin{figure}[htbp]
\centering
\subfloat[]{\label{f:fig4a}\includegraphics[width=0.48\textwidth]{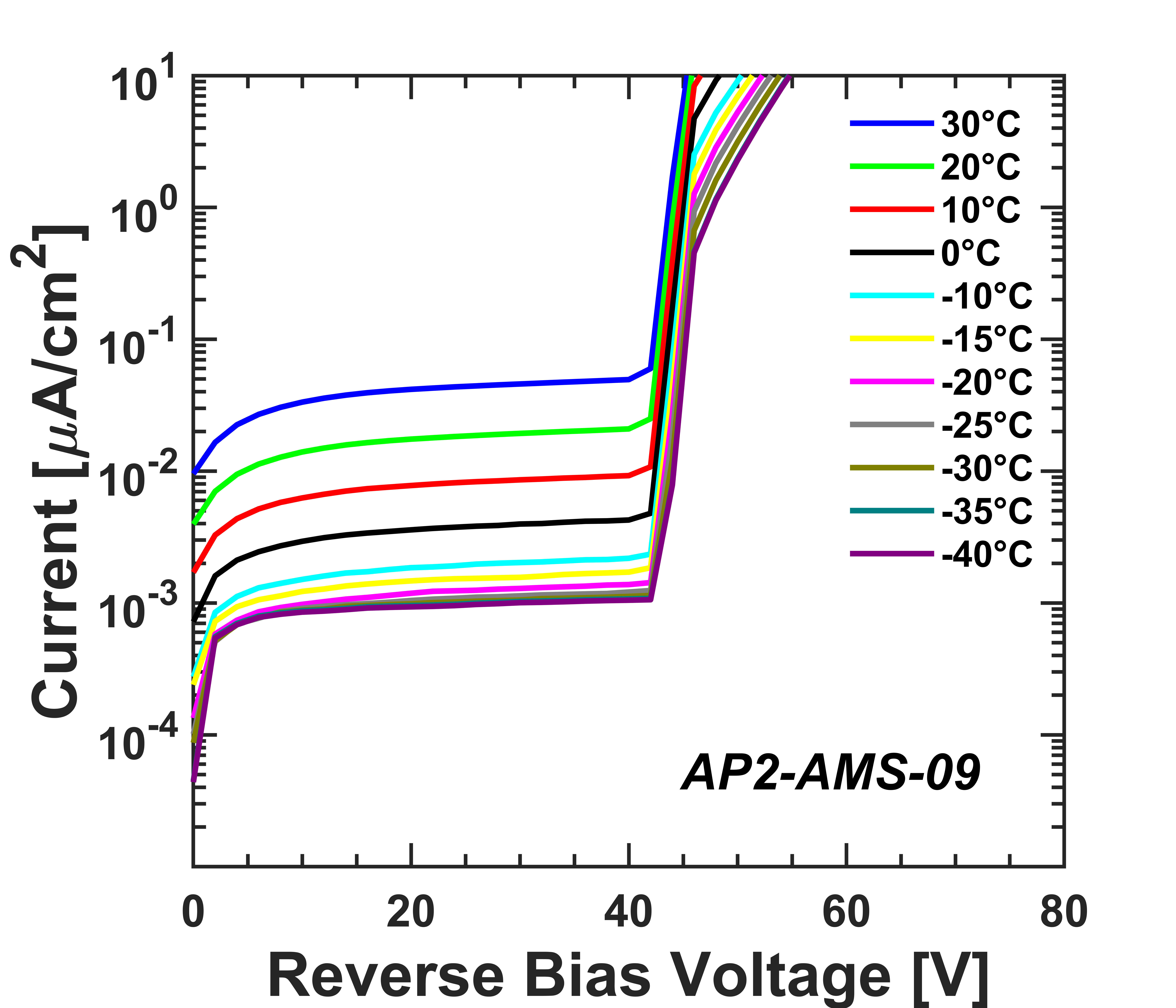}}
\subfloat[]{\label{f:fig4b}\includegraphics[width=0.48\textwidth]{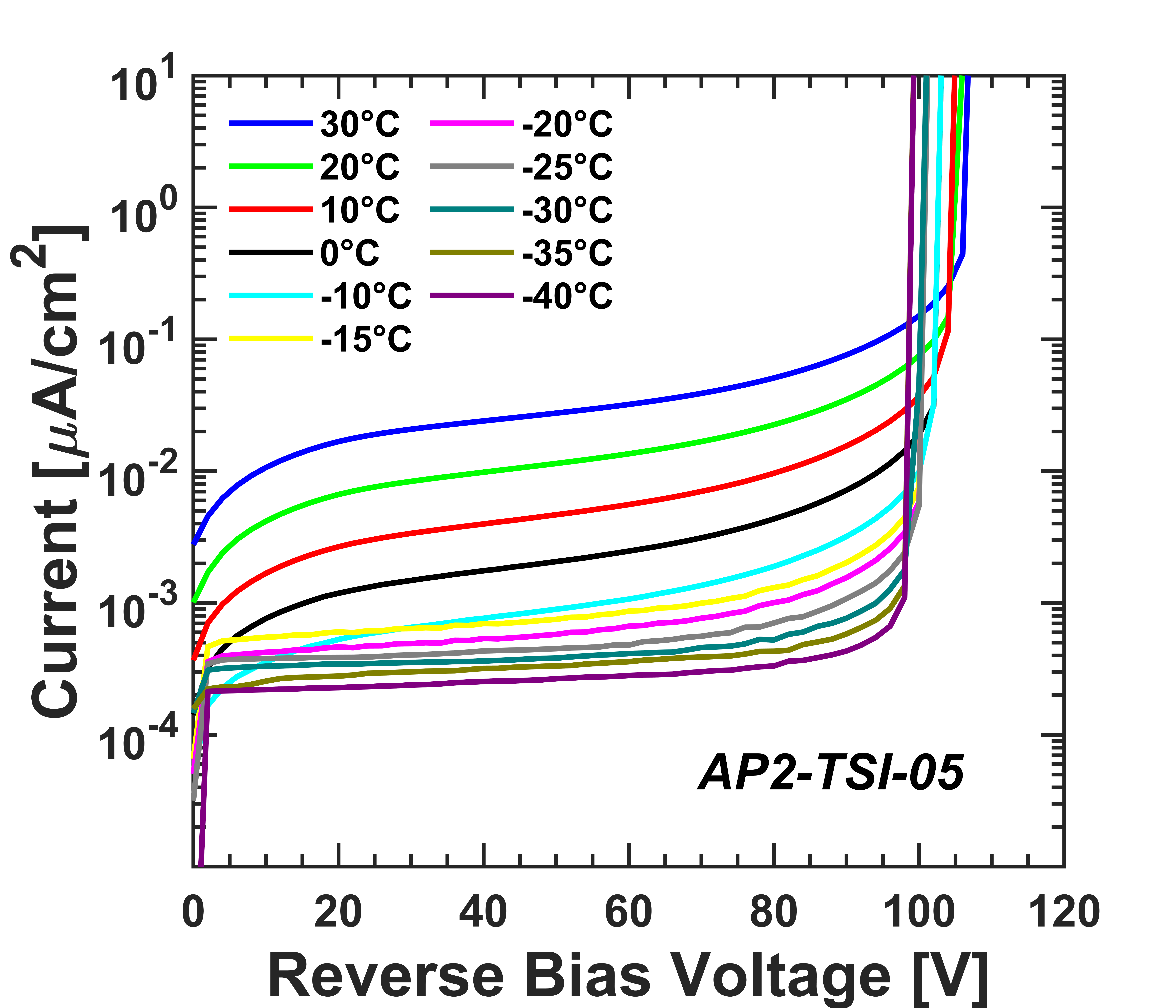}}\\ 
\subfloat[]{\label{f:fig4c}\includegraphics[width=0.48\textwidth]{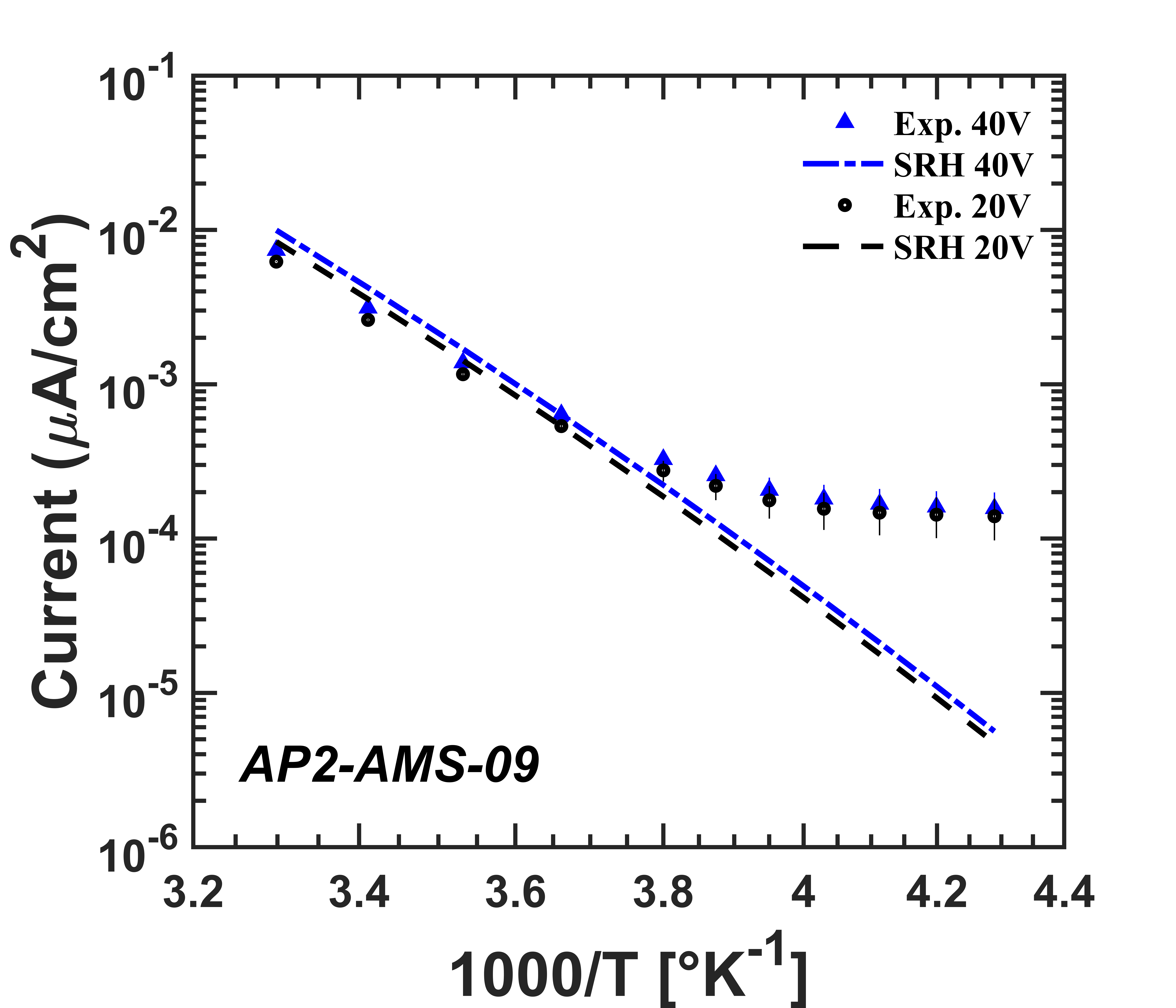}}
\subfloat[]{\label{f:fig4d}\includegraphics[width=0.48\textwidth]{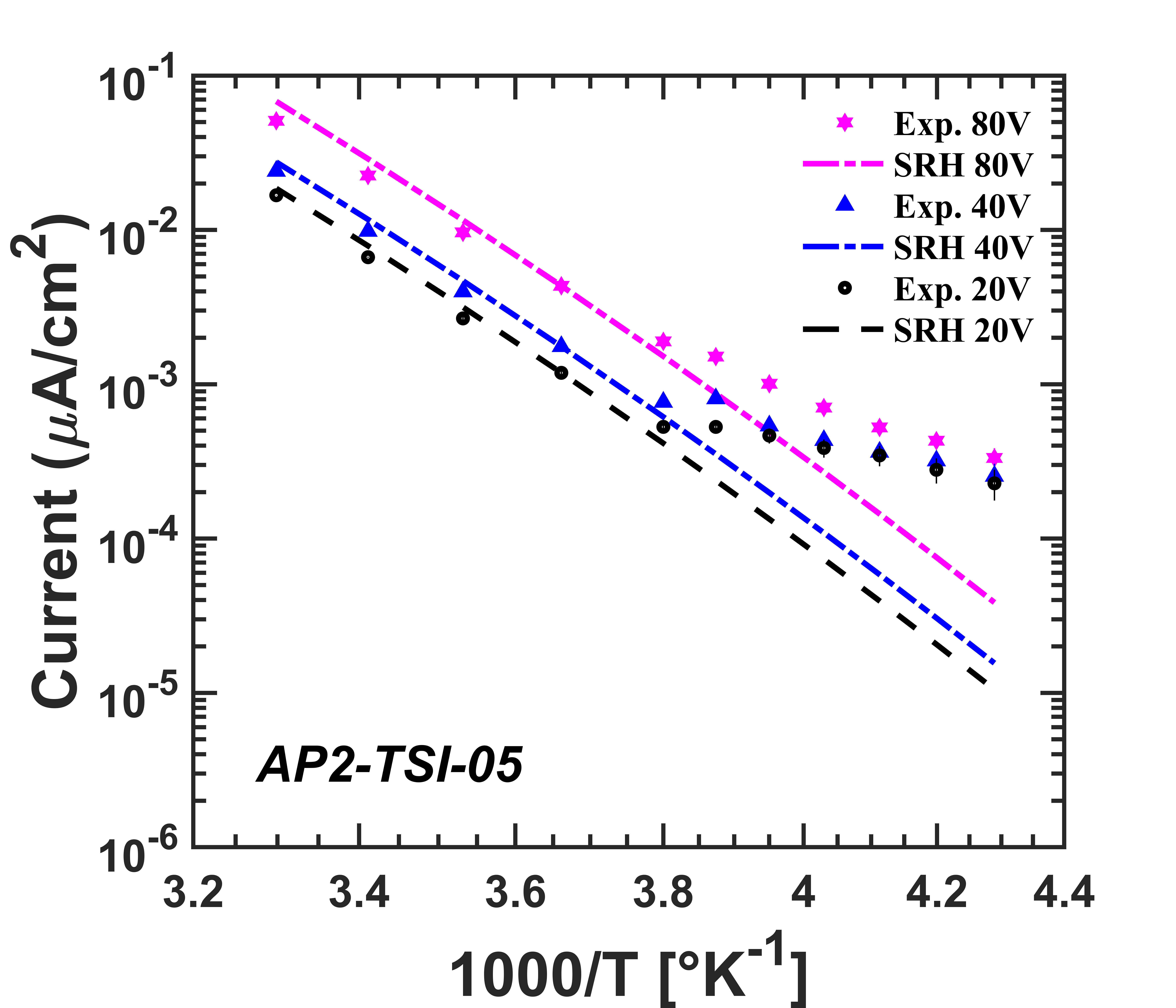}}\\
\caption{\label{f:fig4}Measured I-V curves (top) and corresponding Arrhenius plots (bottom) of non-irradiated ATLASPix2 chips from two different foundries: AP2-AMS-09 (left) and AP2-TSI-05 (right). See table~\ref{t:samples} for a description of the samples.}
\end{figure}

As previously reported in \cite{sultan1}, at near room temperature, the breakdown voltage for the ams ATLASPix2 was found to be $\sim$\SI{50}{\volt}, and $\sim$\SI{100}{\volt} for the TSI ATLASPix2 version. The ATLASPix2 version produced by TSI shows a breakdown voltage in good agreement with the numerical TCAD simulation, estimated to be $\sim$\SI{120}{\volt}. Selected samples of table \ref{t:samples} were probed by varying the chuck temperature between \SI{30}{\celsius} and \SI{-10}{\celsius} with a step of \SI{10}{\celsius} and between \SI{-15}{\celsius} to \SI{-40}{\celsius} with a step of \SI{5}{\celsius}. The acquired leakage current data at different temperatures were then compared with  the Shockley Read-Hall (SRH) model prediction at a reference temperature, using formula \ref{SRHeq},
%The die temperature  was varied between \SI{30}{\celsius} and \SI{10}{\celsius} with \SI{5}{\celsius} step.% 
\begin{equation}
    I(T)=I(T_R)\cdot\left(\frac{T}{T_R}\right)^2\cdot\exp\left[\frac{E_\text{eff}}{2k_B}\left(\frac{1}{T_R}-\frac{1}{T}\right)\right]
    \label{SRHeq}
\end{equation}
where $I(T_R)$ is the reference leakage at \SI{10}{\celsius} as in \cite{sultan1},  $E_\text{eff}$ the Si band gap (\SI{1.21}{\electronvolt})~\cite{alex} and $k_B$ the Boltzmann constant. The SRH model is the simplest approach to the modeling of the device leakage current as function of temperature, considering only the bulk generated current from carrier generation and recombination. Any disagreement between the measurements and the SRH prediction hints that leakage current from other sources than the bulk, such as surface leakage from the oxide-silicon interface due to stress, are present in the device.

Figure \ref{f:fig4} shows the I-V curves measured for both ams ATLASPix2 and TSI ATLASPix2. As expected, the leakage current decreases with a lower temperature. ATLASPix2 chips from ams production show a breakdown voltage of $\sim$\SI{50}{\volt}, similar to the earlier of HV-CMOS sensors, the ams ATLASPix1. Such an early breakdown in comparison to TCAD results could be the result of a large process-induced surface damage present only in the ams chip. Measured data have been fitted with the SRH model, using the reference temperature $T_R=$\SI{0}{\celsius} (equation~\ref{SRHeq}). Careful observation in figure \ref{f:fig4c} shows little deviation from the SRH prediction at higher temperatures, and supports the hypothesis that for the ams sample, surface damage is adding an extra leakage current contribution. Figure \ref{f:fig4b} shows the current-voltage (I-V) characteristics of TSI processed samples. While the leakage followed the decreasing trend with lower temperature, the sensor breakdown voltage remained at around 100 V, as it was seen in~\cite{sultan1}. It is worth noting that the breakdown voltage is reduced at lower temperature due to an increased ionization rate~\cite{crowell}.

%\begin{figure}[!tbp]
%\centering
%\includegraphics[width=0.6\textwidth]{f/Figure5.png}
%\caption{\label{f:fig5}Measured C-V curves of non-irradiated ATLASPix2 chips from two different foundries. See table~\ref{t:samples} for a description %of the samples.}
%\end{figure}

The leakage current of the TSI samples was found to be $\sim$6-7 times smaller than for the ams  ATLASPix2. Arrhenius plot of the TSI sample (as found in figure~\ref{f:fig4d}) shows a better agreement between measured leakage current and the SRH prediction even for a bias voltage reference a few volts before breakdown until \SI{0}{\celsius}. At temperatures lower than \SI{-10}{\celsius}, the measured data clearly deviated from the SRH predictions, and the same phenomenon is also observed for the ams samples. At lower ambient condition, the mean free path of charge carriers increases and triggers earlier the impact ionization process. Therefore, there has been an early avalanche-introduced breakdown voltage for TSI ATLASPix2 prototypes at lower temperatures.

%Figure~\ref{f:fig5} shows the C-V data for both non-irradiated ams ATLASPix2 and TSI ATLASPix2, measured at different temperatures with a 1 kHz small AC-signal. No significant temperature dependence on capacitance has been observed nor a full-depletion plateau been found. We know, irradiated samples require to be operated in a lower temperature to avoid the thermal runaway condition. Though temperature has a negligible effect on C-V measurements for a non-irradiated sample, sometimes C-V data can be also misleading at a higher temperature since the space charge density builds up exponentially with the reverse bias increase~\cite{campbel}. The capacitance of ATLASPix2 sensors, fabricated at low substrate resistivity (20~\ohmcm), showed comparatively the larger value for the TSI sample. It can be anticipated from additional metal lines used for TSI chips. The ATLASPix2 chip thickness is much larger than the depletion depth expected on both ams and TSI prototypes ($\sim$10-20~\micron). The depletion volume expansion remains continuous in this case as the reverse bias increases. Therefore, no depletion plateau has been seen in C-V measurements (Fig.~\ref{f:fig5}).  

%------------------------------------------
\subsection{Results after irradiation}
%------------------------------------------
\begin{figure}[!htbp]
\centering
\subfloat[]{\label{f:fig5a}\includegraphics[width=0.48\textwidth]{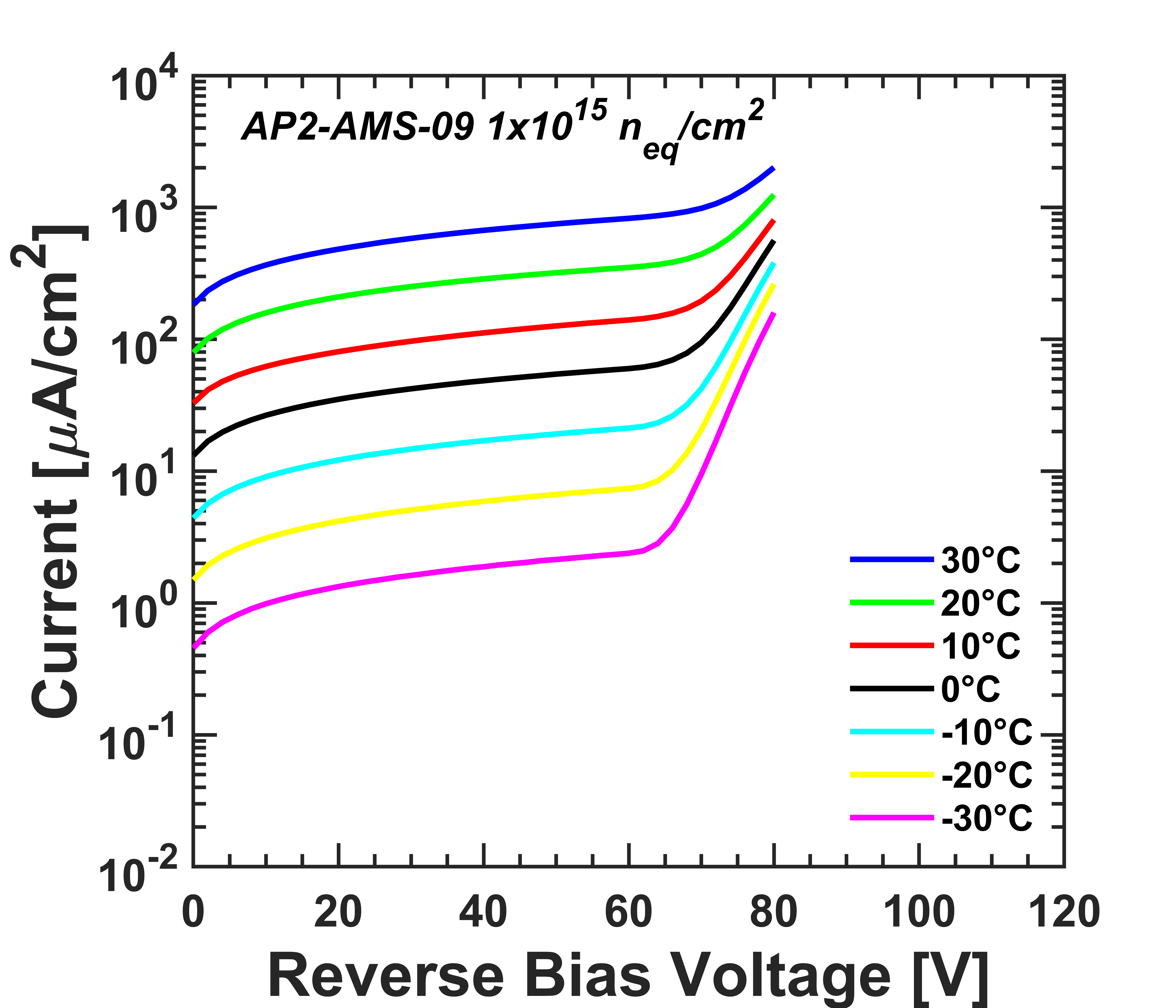}}
\subfloat[]{\label{f:fig5b}\includegraphics[width=0.48\textwidth]{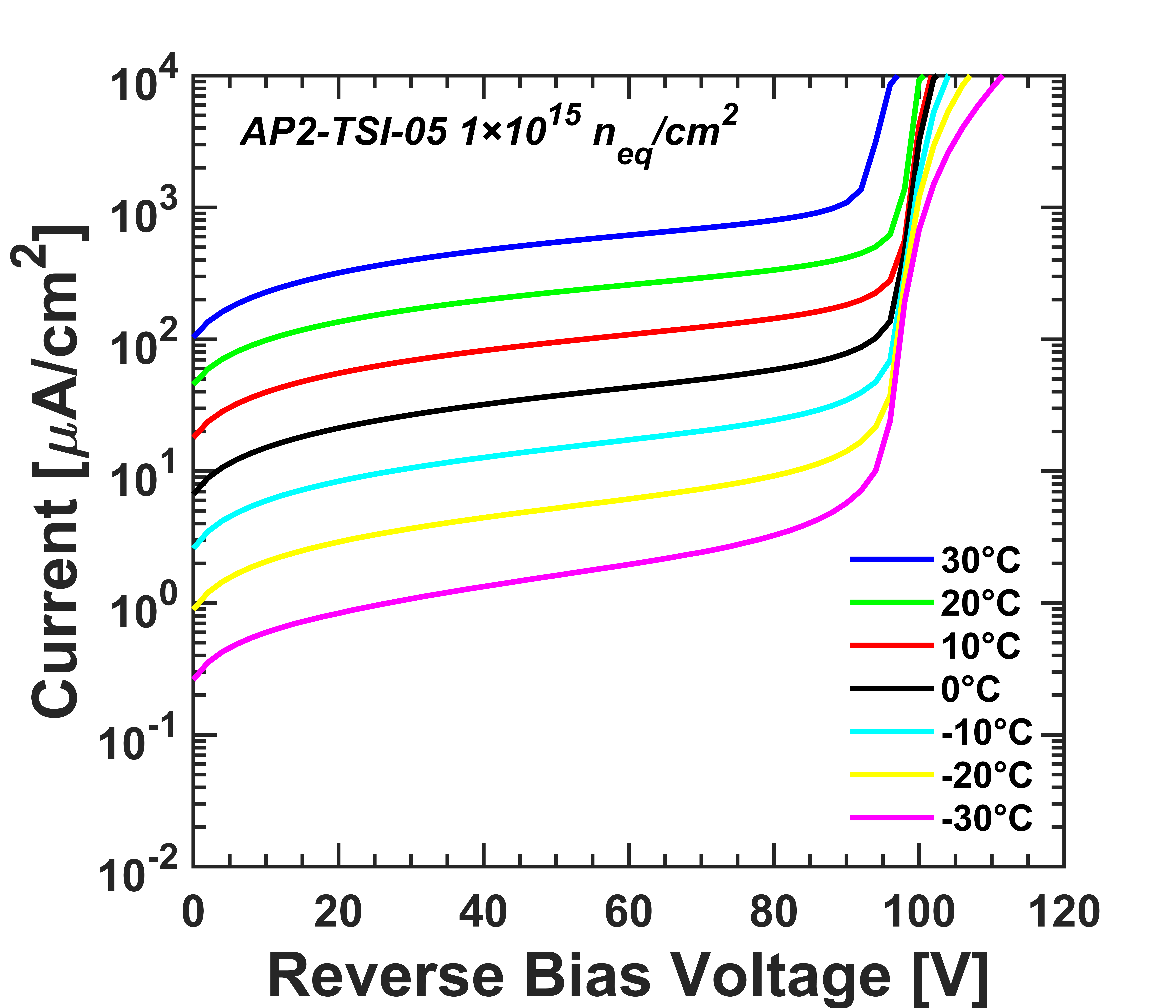}}\\ 
\subfloat[]{\label{f:fig5c}\includegraphics[width=0.48\textwidth]{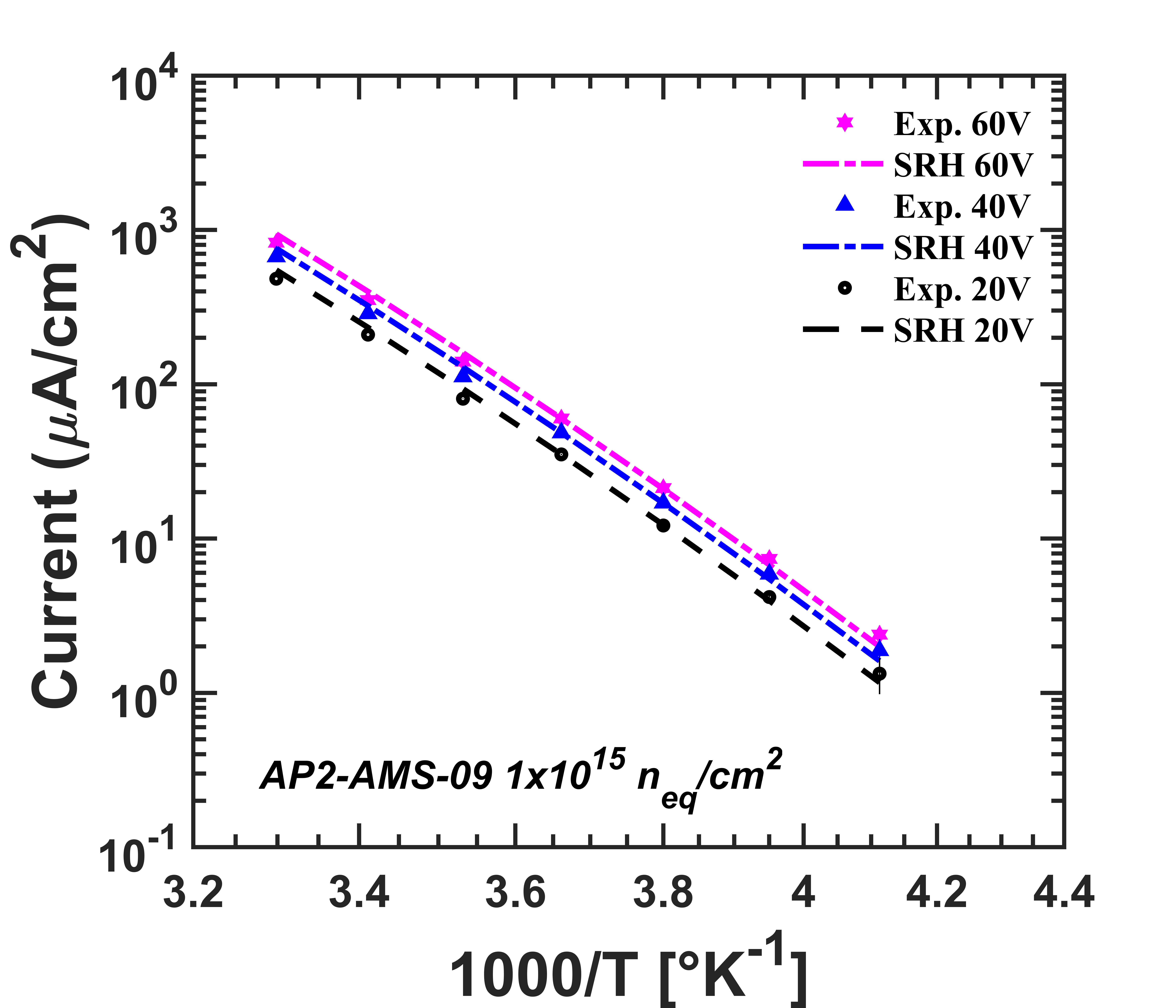}}
\subfloat[]{\label{f:fig5d}\includegraphics[width=0.48\textwidth]{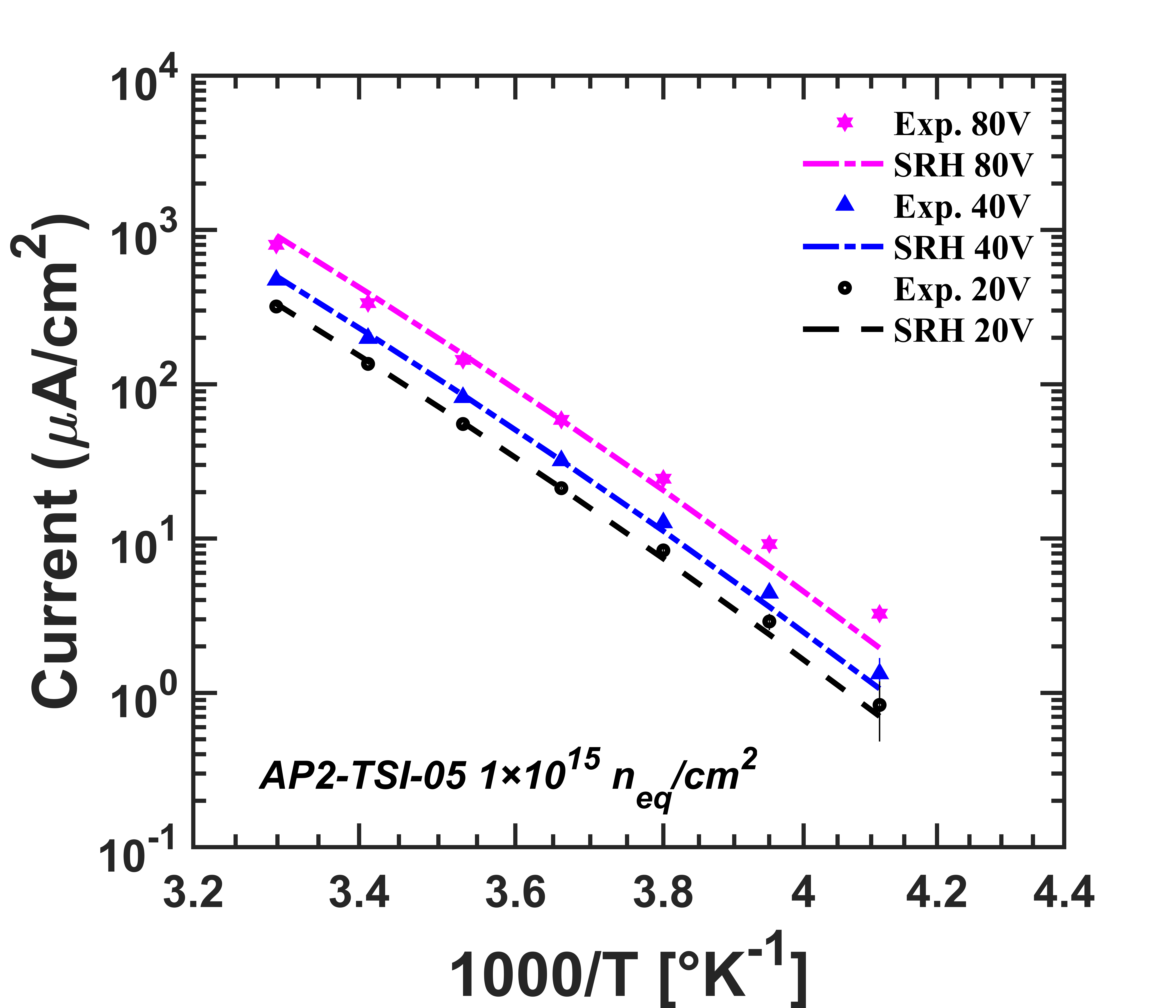}}\\
\caption{\label{f:fig5}Measured I-V curves (top) and corresponding Arrhenius plots (bottom) of ATLASPix2 chips from two different foundries irradiated to $1\times 10^{15}~\onemev$: AP2-AMS-09 (left) and AP2-TSI-05 (right). See table~\ref{t:samples} for a description of the samples.}
\end{figure}

All eight ATLASPix2 samples from table \ref{t:samples} were sent for neutron irradiation to JSI and were then measured in the clean room facility of the DPNC - University of Geneva under controlled ambient conditions. As an example, figure~\ref{f:fig5} shows I-V curves for two prototypes of ATLASPix2 fabricated in ams (AP2-AMS-09), and TSI (AP2-TSI-05), and irradiated to $1\times 10^{15}~\onemev$. In both cases, the leakage increased by four orders of magnitude  with respect to the data before irradiation. It can be explained by the introduction of electrically active traps, generated by crystalline defects produced by neutrons~\cite{betta, li, moll} and the surface damage increase due to background $\gamma$-rays, present at the JSI beam-line~\cite{ambrozic}. For irradiated ams chips, a breakdown voltage beyond $\SI{60}{\volt}$ is observed, slightly higher in comparison to the non-irradiated case. This can be explained by interface states (introduced by irradiation) fill up with carriers, where an additional reverse bias voltage is required to de-trap them before reaching the surface condition similar to a non-irradiated chip and triggering a delayed avalanche breakdown. On the contrary, the breakdown voltage of the TSI samples is measured to be $\SI{90}{\volt}$, $\SI{10}{\volt}$ lower than a non-irradiated case. Before irradiation the TSI chip has already shown a comparatively smaller leakage current than an ams prototype, hinting to fewer surface interface traps from the production phase (a better surface condition). Despite the irradiation-related uncertainties (fluence, self-heating), it is expected that the TSI samples would still have a lower number of interface states in comparison to the ams samples after irradiation. These comparatively fewer active interface-states would have led TSI chips to experience a lesser fill-up of carriers. Thus, a larger peripheral and bulk charge-carriers may cause an earlier avalanche breakdown, where they drive into a device high electric field area near deep $\rm{N^-}$-well and P-substrate (viz. figure~\ref{f:atlaspix2-layout}).    

After irradiation, the predictions from the Arrhenius model have been compared again with the measured leakage current of both irradiated ams and TSI samples (figure 5c and 5d). Interestingly, samples from both foundries show a good agreement for all measured temperatures, meaning the leakage current increase after irradiation is mostly driven by thermionic emission process in the bulk of the sensors.

%After irradiation, the surface current increases by means of an increase of both oxide charge and interface state. Despite the irradiation uncertainty, TSI samples would have still comparatively lower interface states after irradiation. The new surface current introduced for an irradiated TSI chip would have triggered an early breakdown by supplying sufficiently larger charge carriers (in this context, a lesser fill-up to the interface states) in a device high electric field area, that can lead to the impact ionization current. The predictions from the Arrhenius model have been compared with the measured leakage current of both irradiated ams and TSI samples (Figs.~\ref{f:fig5c} and~\ref{f:fig5d}). Interestingly, samples from both foundries show a good agreement for all measured temperatures, meaning the leakage current increase is mostly driven by thermionic emission process in the bulk of the sensors.

\begin{figure}[!htbp]
\centering
\subfloat[]{\label{f:fig6a}\includegraphics[width=0.48\textwidth]{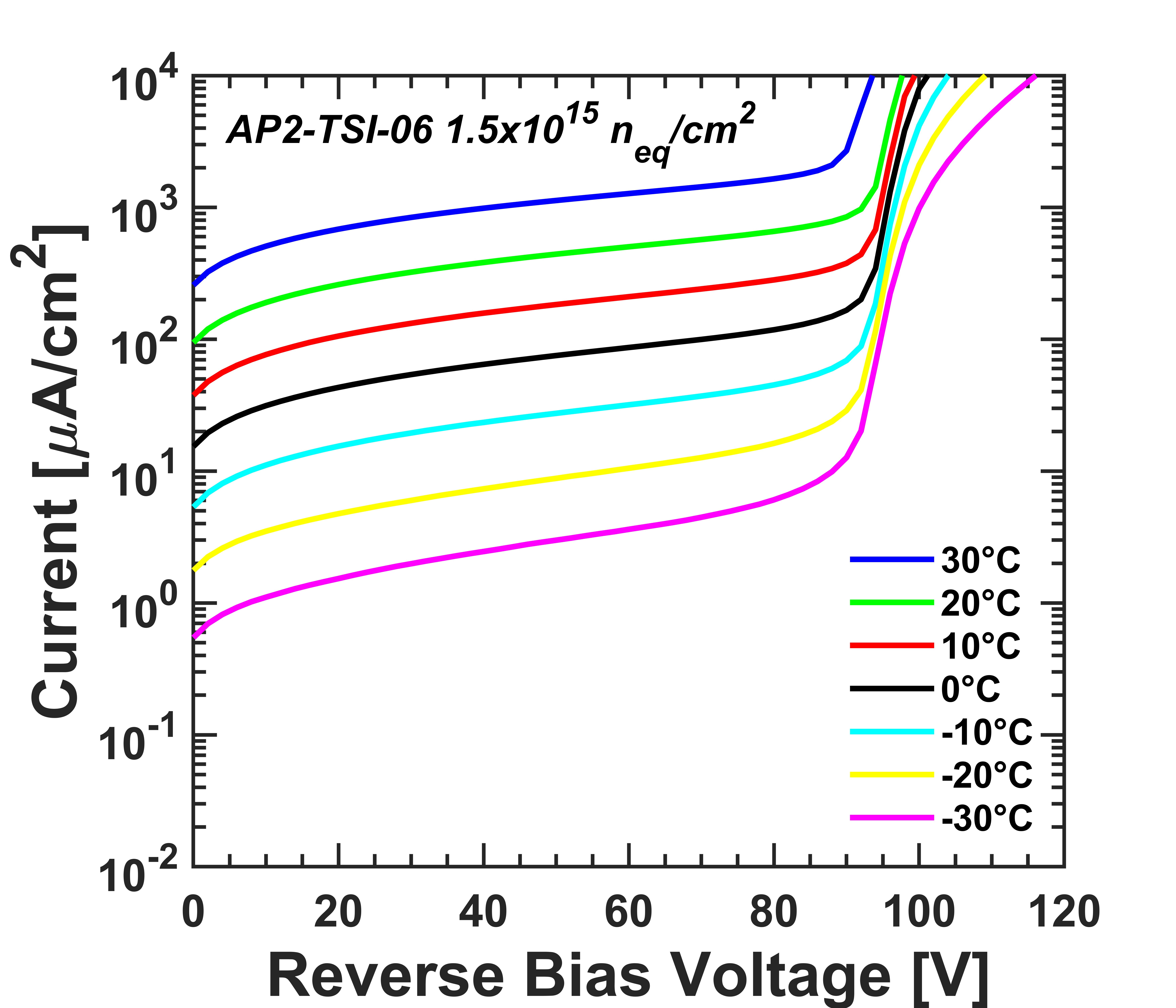}}
\subfloat[]{\label{f:fig6b}\includegraphics[width=0.48\textwidth]{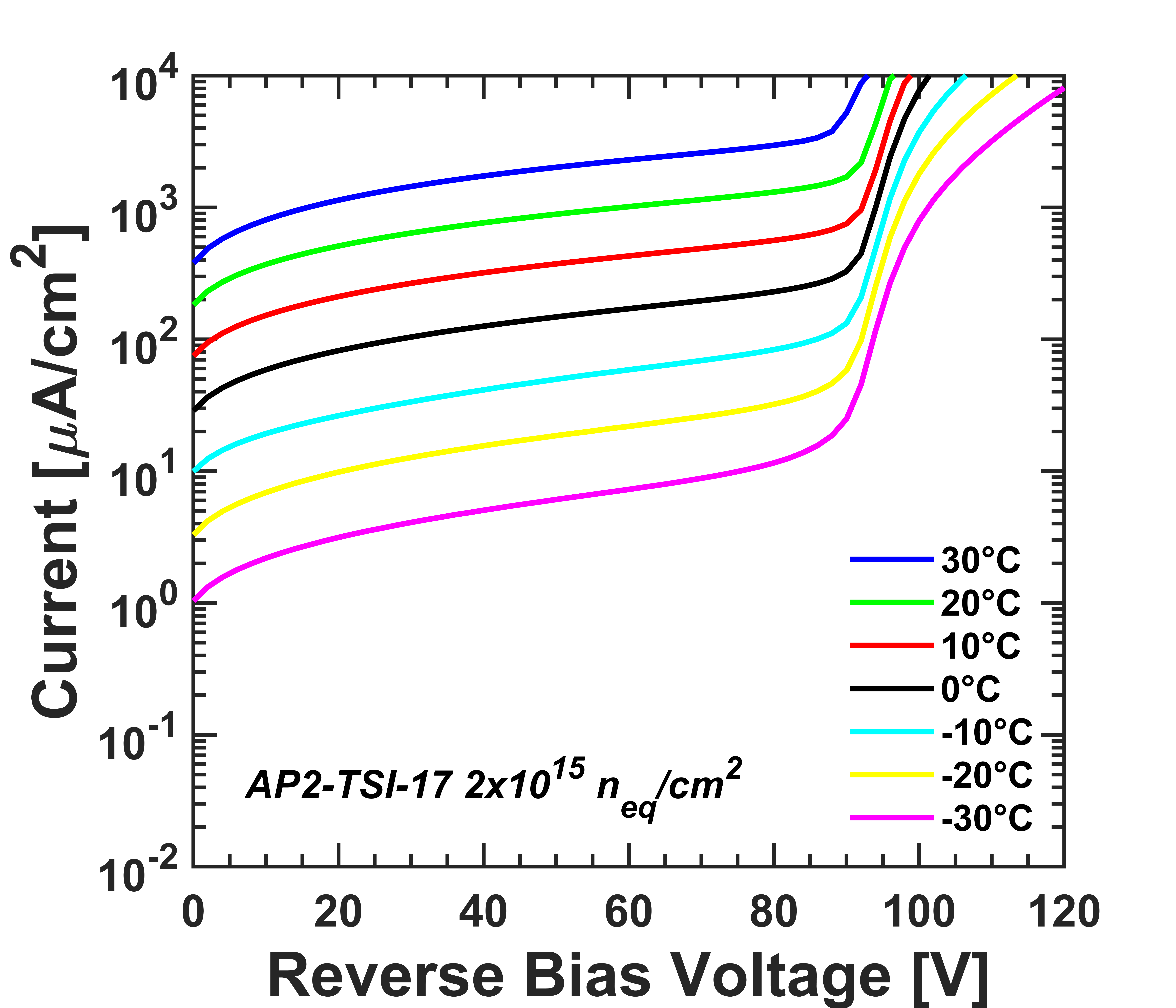}}\\ 
\subfloat[]{\label{f:fig6c}\includegraphics[width=0.48\textwidth]{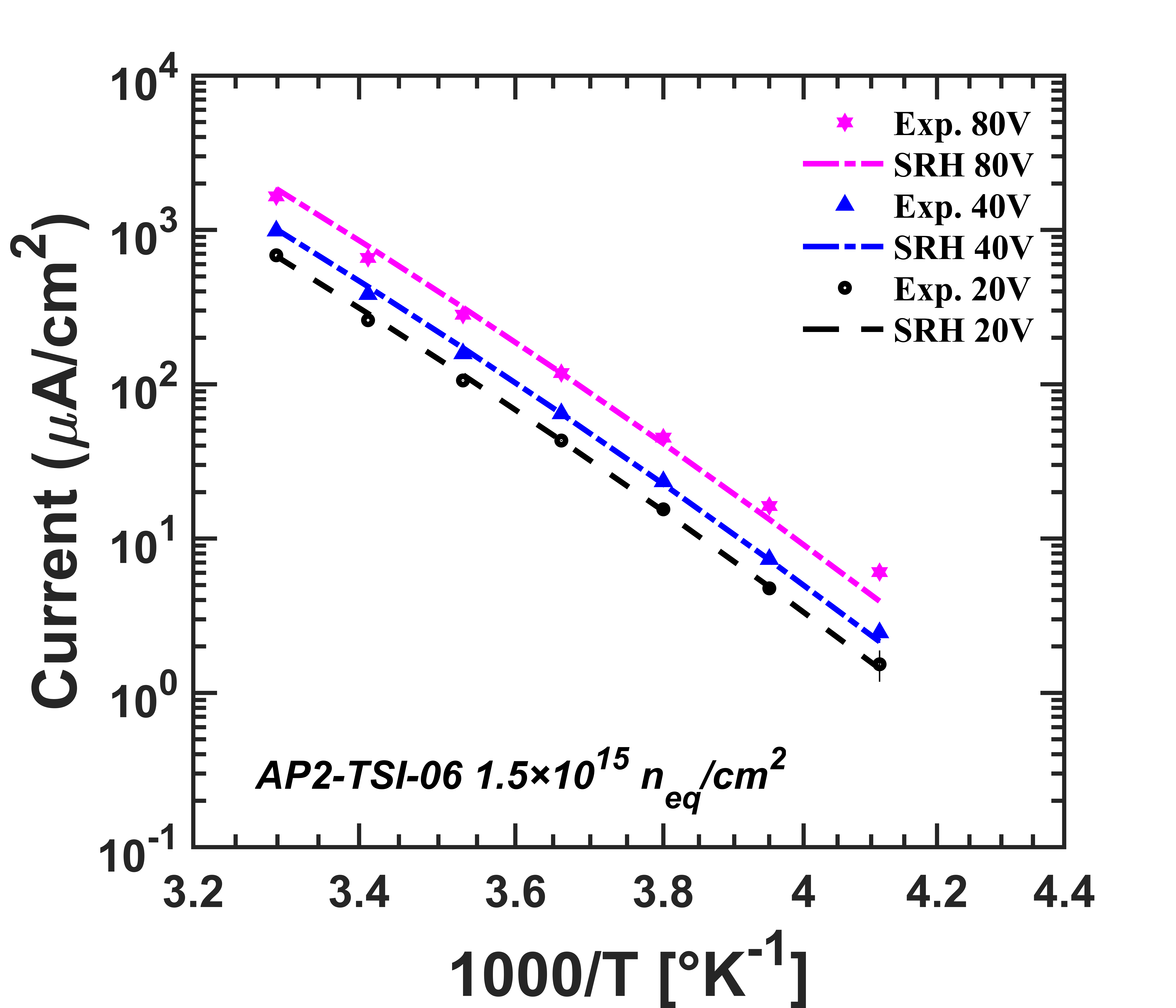}}
\subfloat[]{\label{f:fig6d}\includegraphics[width=0.48\textwidth]{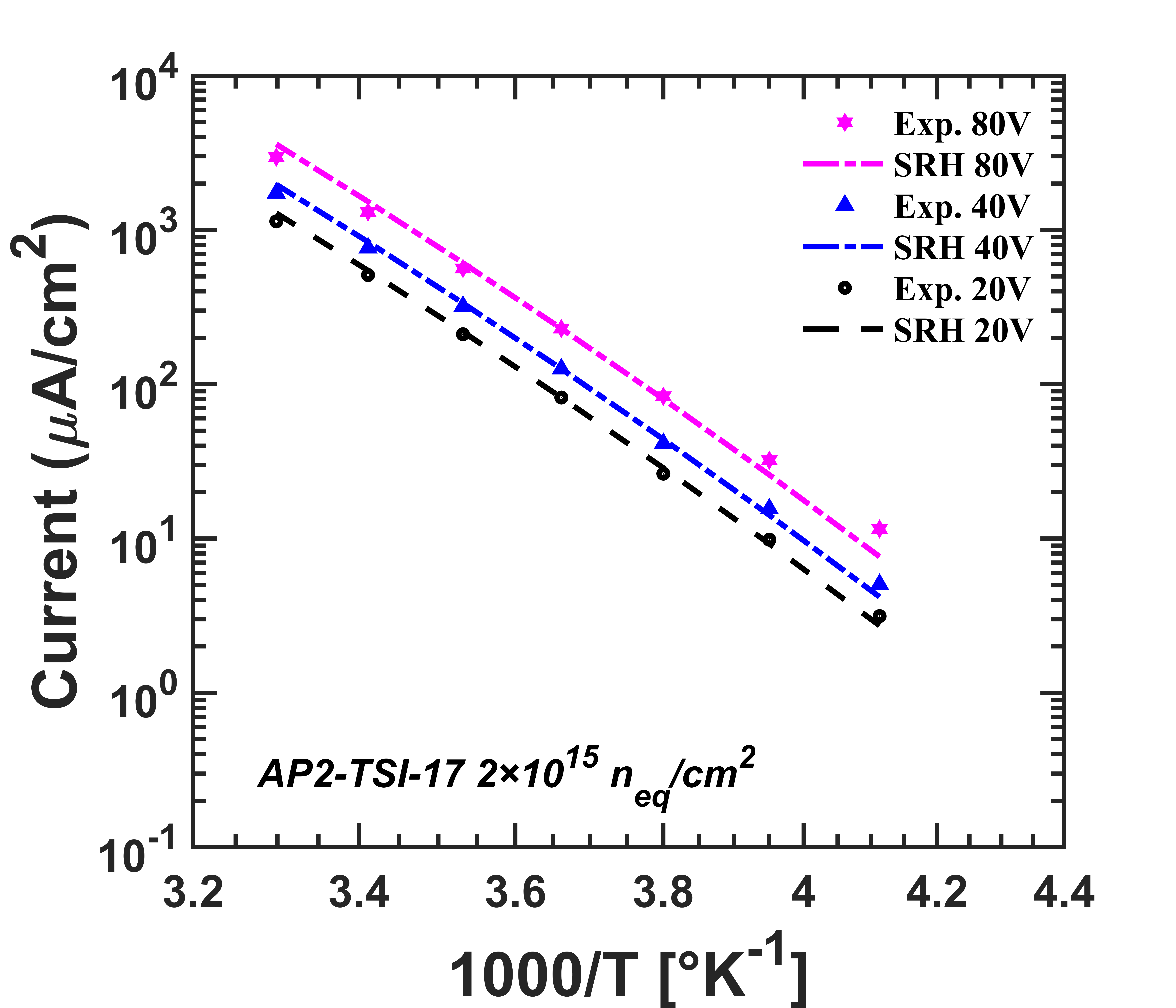}}\\
\caption{\label{f:fig6} Measured I-V curves (top) and corresponding Arrhenius plots (bottom) of two different ATLASPix2 chips from TSI: AP2-TSI-06, irradiated to $1.5\times 10^{15}~\onemev$ (left) and AP2-TSI-17, irradiated to $2\times 10^{15}~\onemev$ (right). See table~\ref{t:samples} for a description of the samples.}
\end{figure}

Figure~\ref{f:fig6a} and \ref{f:fig6b} show the I-V measurements of the TSI processed samples irradiated to $1.5\times 10^{15}$ and $2\times 10^{15}~\onemev$ neutrons at JSI. The breakdown voltage is reduced by a few volts ($\sim$\SI{85}{\volt}) with regard to the TSI samples irradiated to $1\times 10^{15}~\onemev$ neutrons. More interface states are formed with higher irradiation fluences (by $\gamma$-rays) and therefore leads to an early avalanche breakdown. The leakage current of the TSI sample increases by a factor of $4\times 10^{4}$ and $8\times 10^{4}$ respectively, regarding the non-irradiated case compared to the irradiated cases with $1.5\times 10^{15}$ and $2\times 10^{15}~\onemev$ neutrons. A clear agreement between the Arrhenius model and the measured leakage data at different temperatures (as seen in figure~\ref{f:fig6c} and~\ref{f:fig6d}) is observed, again showing the bulk generated current remain dominant after irradiation.

\begin{table}[htbp]
\centering
\caption{\label{t:results-irrad} Summary of the electrical characteristics of irradiated ATLASPix2 with JSI neutron at \SI{-10}{\celsius}.}
\smallskip
\begin{tabular}{|c|c|l|r|c|c|}
\hline
{\bf Fluence} & {\bf TID} & \multicolumn{1}{c|}{\bf Device ID} & \multicolumn{1}{c|}{\bf $J_\text{lk}$} & {\bf $\alpha^*$} & {\bf $V_\text{bd}$} \\ 
 $[\onemev]$ & [Mrad] & & \multicolumn{1}{c|}{[\SI{}{\micro\ampere\per\centi\meter\squared}]} & $[\times 10^{-17}\,\rm{A/cm}]$ & [V] \\ \hline
\multirow{4}{*}{$1\times 10^{15}$} & \multirow{4}{*}{1.1} 
 & AP2-AMS-08 & $(35.12\pm 0.09)$ @64 V & $(2.30\pm 0.35)$ @50V & $66\pm 2$ \\ \cline{3-6}
&& AP2-AMS-09 & $(26.22\pm 0.08)$ @66 V & $(1.62\pm 0.25)$ @50V & $68\pm 2$ \\ \cline{3-6}
&& AP2-TSI-05 & $(31.31\pm 0.08)$ @88 V & $(1.74\pm 0.50)$ @80V & $90\pm 2$ \\ \cline{3-6}
&& AP2-TSI-11 & $(38.77\pm 0.09)$ @88 V & $(2.31\pm 0.57)$ @80V & $90\pm 2$ \\ \hline
\multirow{2}{*}{$1.5\times 10^{15}$} & \multirow{2}{*}{1.65} 
 & AP2-TSI-06 & $(54.44\pm 0.09)$ @86 V & $(2.20\pm 0.56)$ @80V & $88\pm 2$ \\ \cline{3-6}
&& AP2-TSI-24 & $(60.34\pm 0.09)$ @88 V & $(2.21\pm 0.53)$ @80V & $90\pm 2$ \\ \hline
\multirow{2}{*}{$2\times 10^{15}$} & \multirow{2}{*}{2.2} 
 & AP2-TSI-08 & $(72.35\pm 0.10)$ @86 V & $(2.24\pm 0.58)$ @80V & $88\pm 2$ \\ \cline{3-6}
&& AP2-TSI-17 & $(100.51\pm 0.10)$ @86 V & $(3.18\pm 0.80)$ @80V & $88\pm 2$ \\ \hline
\end{tabular}
\end{table}

\begin{figure}[htbp]
\centering
\subfloat[]{\label{f:7a}\includegraphics[width=0.48\textwidth]{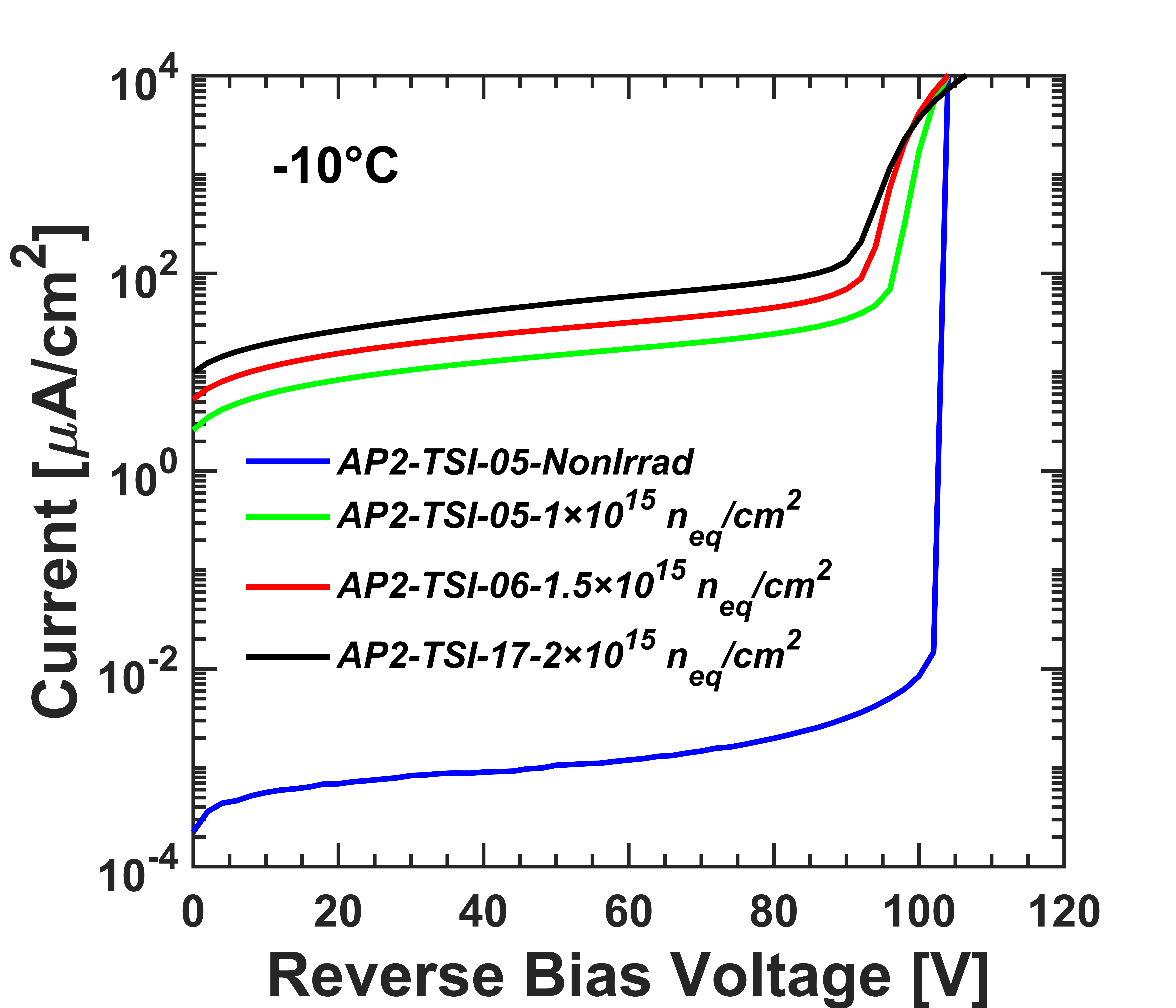}}
\subfloat[]{\label{f:7b}\includegraphics[width=0.48\textwidth]{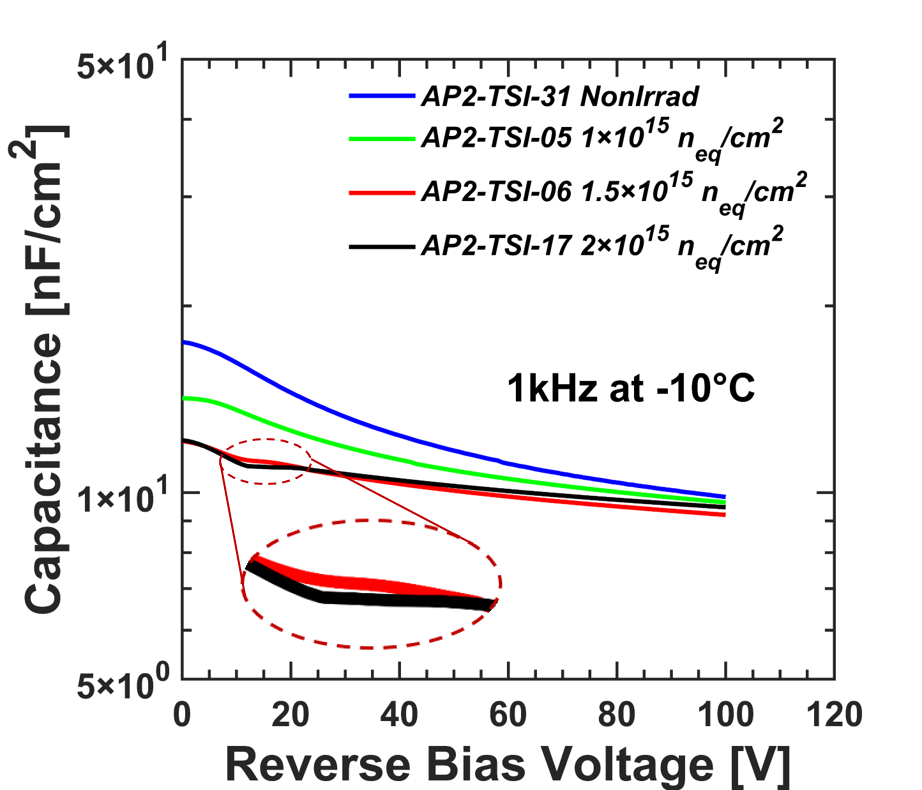}}
\caption{\label{f:fig7}Measured I-V curves of the TSI ATLASPix2 samples, irradiated with JSI neutron at the different fluences (left) and measured C-V curves of same samples at different fluences (right). Zoomed inset of the C-V plot shows the plateau seen due to lateral depletion.}
\end{figure}

Table~\ref{t:results-irrad} summarizes the electrical characterization data of all irradiated samples. While the leakage current density ($J_\text{lk}$) includes contributions from both the bulk and the peripheral current, the damage constant $\alpha^*$ provides information about the irradiation effects in the active pixel volume. The damage constant $\alpha^*$, as defined in~\cite{wonsak} to deal with highly irradiated sensors where full depletion cannot be achieved, is just dependent on the chip thickness. In our case the usage of $\alpha^*$ is considered to be more appropriate than $\alpha$ due to the absence of dedicated edge-TCT measurements for the estimation of the depletion depth (needed for the calculation of $\alpha$). The values shown in Table~\ref{t:results-irrad} are in reasonable agreement with those observed in other irradiated devices not subject to a high temperature annealing process~\cite{wonsak, moll}.

%Table$~\ref{t:results-irrad}$ summarizes the electrical characterization data of all  irradiated samples. It is worth highlighting that the reported damage constant rate $\alpha^*$, dependent on the chip thickness, differs from $\alpha$. Considering the time and required cost, no post-sintering step (a high temperature annealing treatment at the wafer-level) has been applied for ATLASPix2 of any foundry. The additional non-idealities i.e. the uncertainties in the irradiation fluences, the self-heating effects experienced by the devices during irradiation, and the lacking of dedicated edge-TCT measurements to estimate the depletion depth of investigating samples also make the extraction of the current damage constant ($\alpha$) difficult. However, given several fluences considered in this irradiation study, the leakage current density ($J_\text{lk}$) would not be enough to justify the order of magnitude, so the geometric current related damage rate ($\alpha^*$) has been introduced, as defined in$~\cite{wonsak}$ to deal with highly irradiated sensors where full depletion cannot be achieved. So, the $\alpha^*$ parameter is thought to be enough to demonstrate the leakage current trend after irradiation, and the $\alpha^*$ value are also in reasonable agreement with those observed in irradiated devices not subjected to a high temperature annealing$~\cite{wonsak, moll}$.

Figure \ref{f:7a} shows the comparative I-V plots of the TSI ATLASPix2 sensors, irradiated at different accumulated fluences and measured in -\SI{10}{\degreeCelsius} steps. The leakage current increased with a higher fluence received. Again, the $\gamma$-rays present at JSI neutron beam increase the density of interface states. Bulk trap-density also increases with exposure to neutrons and modifies the effective doping concentrations. Consequently, as previously explained, a smaller avalanche-driven breakdown voltage was observed for samples exposed to higher fluences.

The introduction rate of traps is proportional to the received fluence and affects the substrate effective doping concentration and therefore the electric field distribution under bias, thus affecting the depletion voltage. In this context, C-V measurements were made for different TSI ATLASPix2 samples. Figure~\ref{f:7b} comprises the measured C-V data for TSI samples, irradiated with different fluences. Because of the larger sensor thickness (\SIrange{220}{254}{\micro\meter}) in comparison to the expected depletion depths of tens of microns ~\cite{sultan1} for our samples, no plateau after full depletion can be observed. For higher neutron fluences ($1.5\times 10^{15}$ and $2\times 10^{15}~\onemev$), a sort of plateau observed in the \SIrange{10}{15}{\volt} reverse bias range that perhaps denotes the lateral depletion (figure~\ref{f:7b}). Per-pixel capacitance is found to be approximately \SI{100}{\femto\farad}.

%------------------------------------------
\subsection{Annealing}
%------------------------------------------

It is foreseen that the installed detectors at the HL-LHC will experience annealing during the required periodic shutdowns. To understand the annealing-driven effects on these sensors, a systematic study by accelerated annealing was performed. %From irradiation, interstitials (I), and vacancies (V) defects are formed within the sensors and become very mobile beyond \SI{150}{\kelvin} operating condition. During the annealing process, several kinds of recombination happen: Frenkel pair recombination ($\rm{I + V \rightarrow Si}$), multi vacancy and multi interstitial combination (i.e., $\rm{I+I\rightarrow I2}$ or $\rm{V+V\rightarrow V2}$) and combination of more complex (C) defects (i.e., $\rm{C_i + O_i\rightarrow C_i O_i}$, or, $\rm{V + P\rightarrow VP}$), where O denotes oxygen vacancies and P denotes acceptor in P-type bulk~\cite{moll}. 
From irradiation, interstitial and vacancy defects are typically formed within the sensor. During the annealing process, several kinds of recombination happen: Frenkel pair recombination, multi vacancy and multi interstitial combination and combination of more complex defects~\cite{moll}.
The first two types are typically very mobile and can be cured at room temperature in a short period. The recombination of complex defect rather requires a longer time and affects adversely the sensor’s leakage current, the depletion voltage, and Charge Collection Efficiency (CCE). The accelerated annealing process at an elevated temperature helps to mitigate vacancies and defects, acts as the beneficial annealing initially by reducing the leakage current and the depletion voltage. However, acceptor removal is also dependent on annealing time and temperature and may trigger the reverse effects, as detailed in \cite{wiik,mandic2}. The damage constant rate, $\alpha$, can be corrected for room temperature annealing from accelerated annealing using the model proposed in \cite{moll},
\begin{equation}
\label{eq:annealing}
\alpha(t,T_a) = \alpha_1\exp\left(-\frac{t}{\tau_1(T_a)}\right) + \alpha_0^* - \beta\ln\left(\Theta(T_a)\frac{t}{t_0}\right)
\end{equation}
\noindent where $\alpha(t,T_a)$ is evaluated at a time $t$ and annealing temperature $T_a$, and $t_0$ is the initial annealing period, fixed to \SI{1}{\minute}. The model consists of an exponential term related to bulk effect and a logarithmic term related to surface effects. The correction factor $\Theta(T_a)$ is used to quantify the damage constant rate $\alpha$ below \SI{21}{\celsius}, and is computed from the following equation:
\begin{equation}
\label{eq:annealingcorr}
    \Theta(T_a) = \exp\left[-\frac{E_\text{eff}}{k_B}\left(\frac{1}{T_a}-\frac{1}{T_\text{ref}}\right)\right]
\end{equation}
\begin{table}[!htbp]
\centering
\caption{\label{t:annealing-fit}Table 3: Fit parameters of current annealing at different annealing temperatures.}
\smallskip
\begin{tabular}{|c|c|c|c|c|}
\hline
{\bf $T_a$} & {\bf $\alpha_1$} & {\bf $\tau_1$} & {\bf $\alpha_0$} & {\bf $\beta$} \\ 
$[~\SI{}{\celsius}]$ & $[\times 10^{-17}\,\rm{A/cm}]$ & [min] & $[\times 10^{-17}\,\rm{A/cm}]$ & $[\times 10^{-18}\,\rm{A/cm}]$\\ \hline
21 & 1.23 & $1.4\times 10^4$ & 7.07 & 3.29 \\ \hline
60 & 1.26 & 94 & 4.87 & 3.16 \\ \hline
80 & 1.13 &  9 & 4.23 & 2.83 \\ \hline
\end{tabular}
\end{table}

\begin{table}[!htbp]
\centering
\caption{\label{t:annealing-steps} Summary of the four applied accelerated annealing steps and corresponding equivalent annealing times at \SI{20}{\celsius}.}
\smallskip
\begin{tabular}{|c|c|c|c|}
\hline
{\bf Annealing stage} & {\bf Temperature} & {\bf Duration} & {\bf Equivalent annealing at \SI{20}{\celsius}} \\ \hline
1st stage & \SI{60}{\celsius} & 80 min & $\sim$18 days \\ \hline
2nd stage & \SI{80}{\celsius} & 30 min & $\sim$53 days \\ \hline
3rd stage & \SI{80}{\celsius} & 120 min & $\sim$288 days \\ \hline
4th stage & \SI{80}{\celsius} & 300 min & $\sim$634 days \\ \hline
\end{tabular}
\end{table}
An effective activation energy $E_\text{eff}$ of \SI{1.21}{\electronvolt} for irradiated Si was used, following \cite{alex}. The temperature steps for the study were carefully chosen since it has been observed that MCz P-type substrate (used in our prototypes) experiences reverse annealing for an annealing treatment beyond \SI{400}{\celsius} \cite{li}. The first annealing step was made at \SI{60}{\celsius} for \SI{80}{\minute} for all irradiated samples of table~\ref{t:samples} and in later steps, selected irradiated samples from the 1st stage annealing were further treated with accelerated annealing process at \SI{80}{\celsius} for \SI{30}{\minute}, \SI{120}{\minute} and \SI{300}{\minute} successively. All necessary values of $\alpha_1$, $\alpha_0^*$, $\beta$ and time constant, $\tau_1$ for equation~\ref{eq:annealing} were taken for the corresponding annealing temperature, as can be found in table~\ref{t:annealing-fit}. These values were selected because of a better agreement with years-long annealing data of silicon sensors at the LHC \cite{atlas-modelling} than the recommended values proposed in \cite{moll}. The equivalent annealing time at near room temperature (\SI{20}{\celsius}) from the applied accelerated annealing has been shown in table \ref{t:annealing-steps} by numerically solving the equations~\ref{eq:annealing} and~\ref{eq:annealingcorr}.

%%%%%%%%%%%%%%%%%%%%%%%%%%%%%%%%%%%%%%%%%%%%%%%%%%%%%%%%%%%%%%%%%%%%%%%
\subsubsection{First stage annealing (\SI[detect-weight]{60}{\celsius} for 80 min)}
\begin{figure}[!htbp]
\centering
\subfloat[]{\label{f:fig8a}\includegraphics[width=0.48\textwidth]{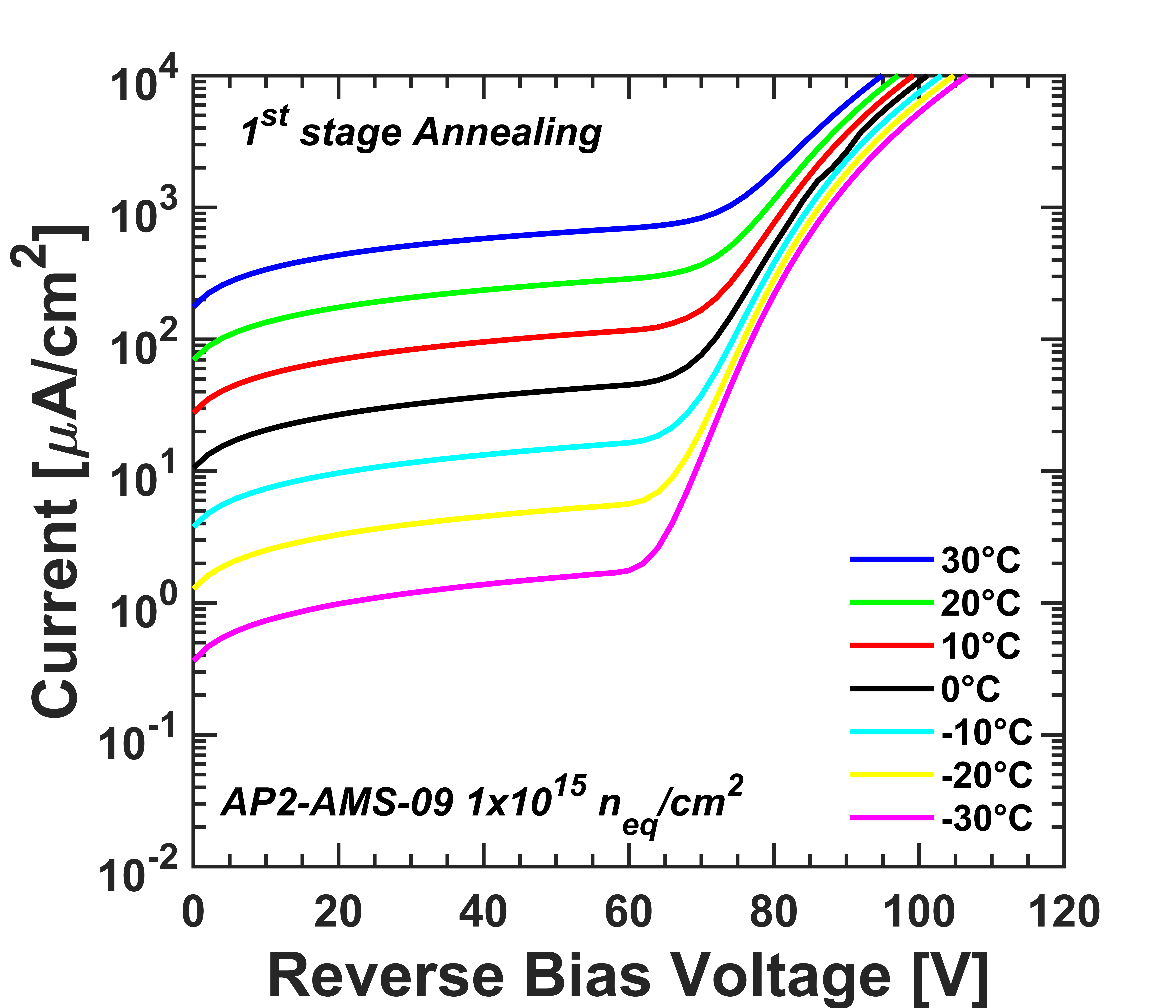}}
\subfloat[]{\label{f:fig8b}\includegraphics[width=0.48\textwidth]{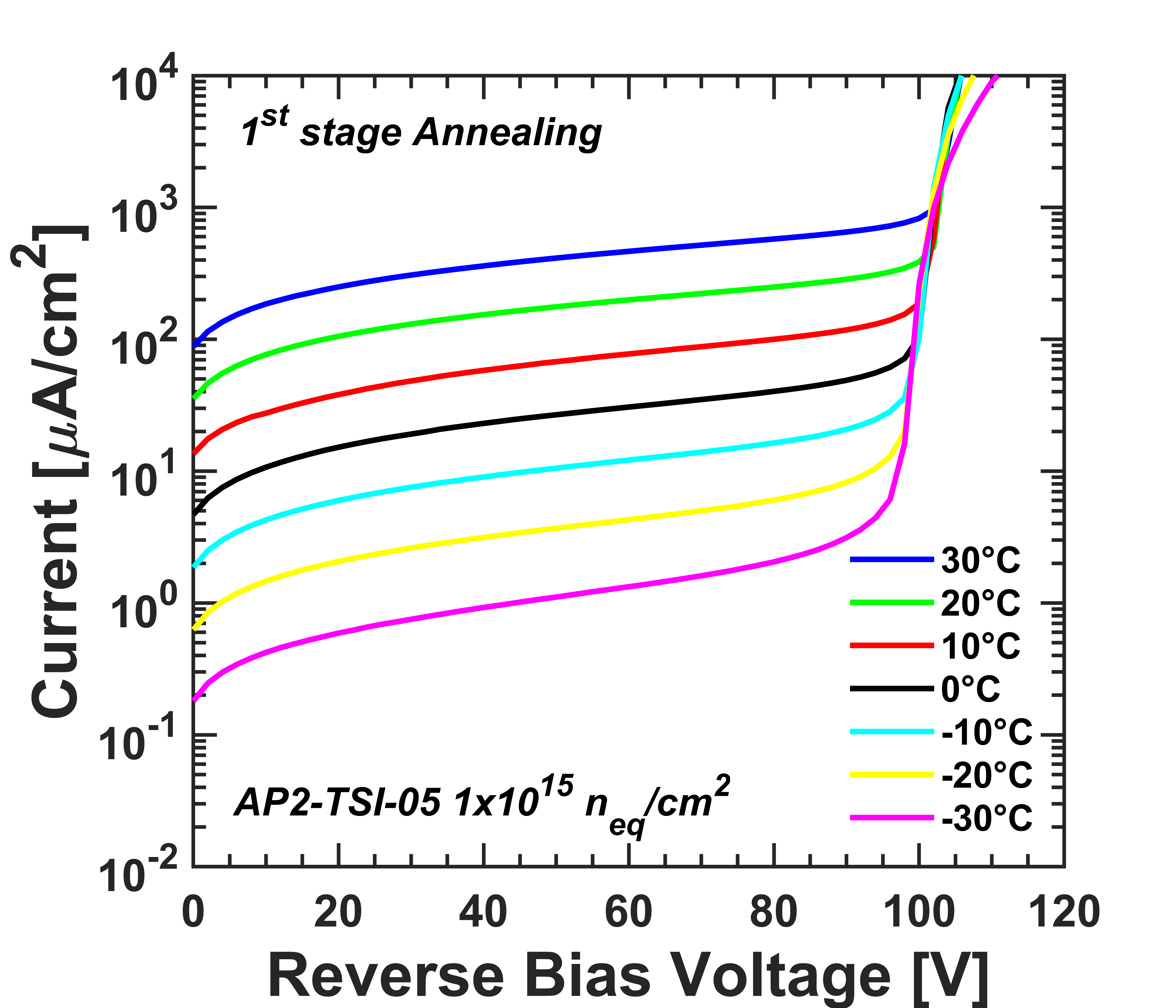}}\\ 
\subfloat[]{\label{f:fig8c}\includegraphics[width=0.48\textwidth]{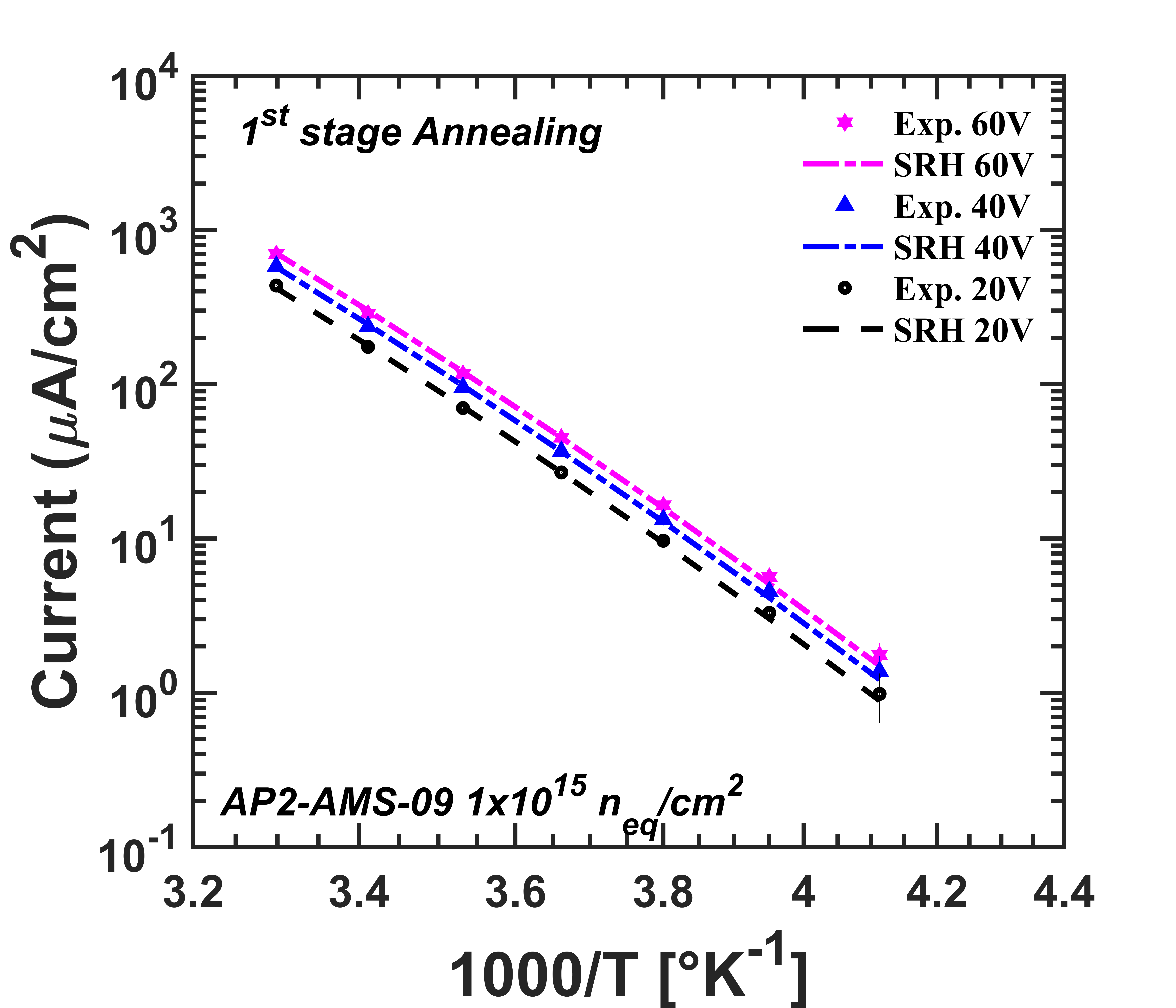}}
\subfloat[]{\label{f:fig8d}\includegraphics[width=0.48\textwidth]{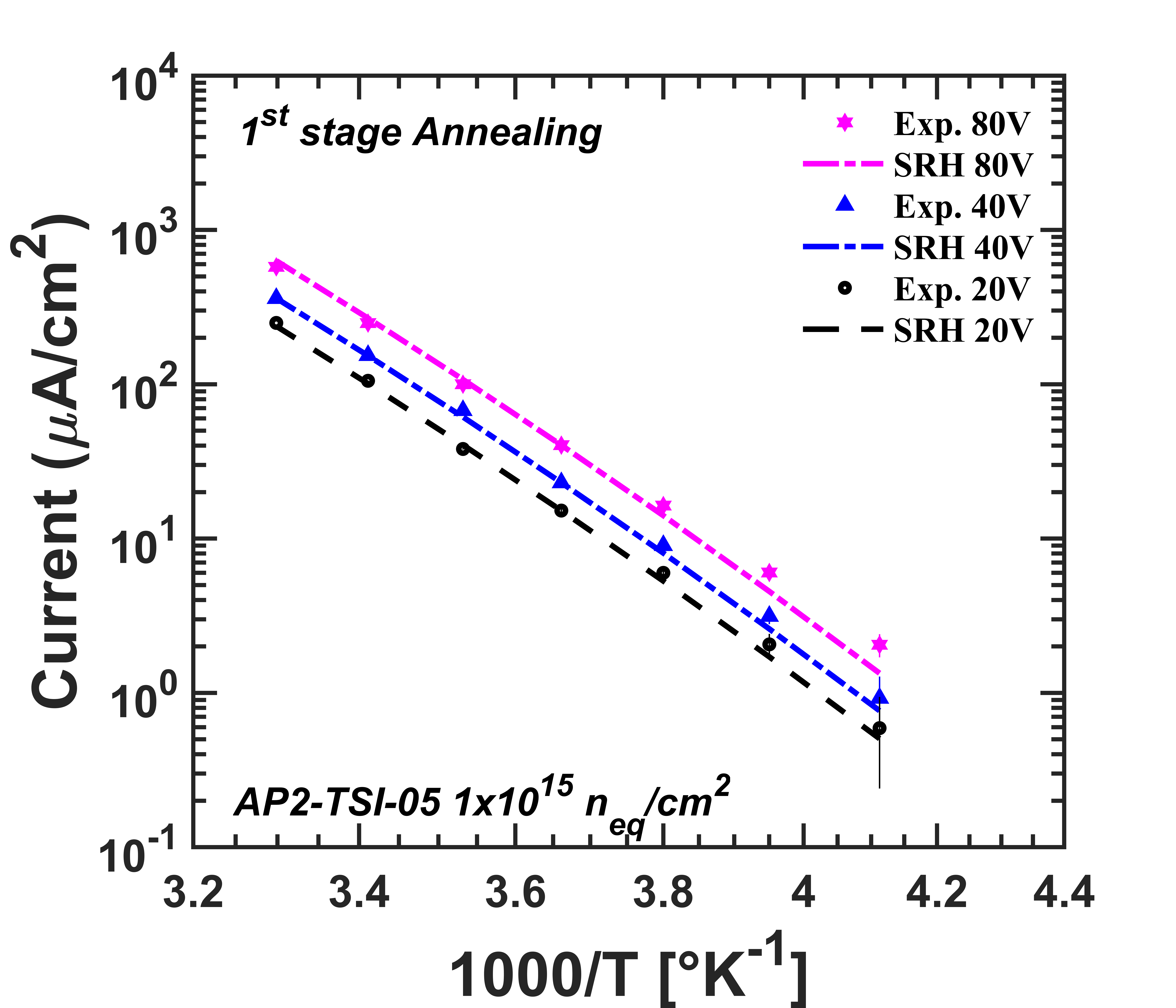}}\\
\caption{\label{f:fig8}Measured I-V curves (top) and corresponding Arrhenius plots (bottom) of ATLASPix2 chips from two different foundries, both irradiated to $1\times 10^{15}~\onemev$ and then annealed at \SI{60}{\celsius} for 80 min: AP2-AMS-09 (left) and AP2-TSI-05 (right). See table~\ref{t:samples} for a description of the samples.}
\end{figure}

%%%%%%%%%%%%%%%%%%%%%%%%%%%%%%%%%%%%%%%%%%%%%%%%%%%%%%%%%%%%%%%%%%%%%%%
All the neutron-irradiated samples were annealed on the thermal chuck of CM300 at \SI{60}{\celsius} for 80 min in the first stage. Table \ref{t:results-annealing-1} summarizes the data obtained from the I-V characterizations after annealing. Figure \ref{f:fig8} shows the I-V behavior of ams (AP2-AMS-09) and TSI (AP2-TSI-05) prototypes irradiated up to $1\times 10^{15}\ \onemev$ neutrons and treated with first stage accelerated annealing. Interface states and traps recovery by accelerated annealing led to a leakage current decrease and a lower value of the damage constant rate $\alpha^*$ is obtained. The interface states recovery and the effective doping concentration reduction by the first stage of accelerated annealing possibly helped to delay the avalanche process for TSI sample and improve the breakdown voltage by a few volts.
\begin{table}[!htbp]
\small
\centering
\caption{\label{t:results-annealing-1} Electrical characteristics of irradiated ATLASPix2 samples after the 1st annealing stage (\SI{60}{\celsius} for 80 min). The last column indicates the total cumulated equivalent annealing time at \SI{20}{\celsius}.}
\smallskip
\begin{tabular}{|c|l|c|c|c|c|}
\hline
{\bf Fluence} & \multicolumn{1}{c|}{\bf Device ID} & \multicolumn{1}{c|}{\bf $J_\text{lk}$} & \multicolumn{1}{c|}{\bf $\alpha^*$} & {\bf $V_\text{bd}$} & {\bf Equiv. cum.} \\ 
 $[\onemev]$ && \multicolumn{1}{c|}{[\SI{}{\micro\ampere\per\centi\meter\squared}]} & $[\times 10^{-17}\,\rm{A/cm}]$ & [V] & {\bf ann. at \SI{20}{\celsius}} \\ \hline
\multirow{4}{*}{$1\times 10^{15}$} 
 & AP2-AMS-08 & $(30.14\pm 0.09)$ @62 V & $(1.85\pm 0.25)$ @50V & $64\pm 2$ & \multirow{8}{*}{$\sim$18 days}\\ \cline{2-5}
& AP2-AMS-09 & $(18.52\pm 0.08)$ @64 V & $(1.28\pm 0.15)$ @80V & $64\pm 2$ &\\ \cline{2-5}
& AP2-TSI-05 & $(22.35\pm 0.08)$ @92 V & $(1.17\pm 0.28)$ @80V & $94\pm 2$ &\\ \cline{2-5}
& AP2-TSI-11 & $(29.94\pm 0.08)$ @92 V & $(1.60\pm 0.37)$ @80V & $94\pm 2$ &\\ \cline{1-5}
\multirow{2}{*}{$1.5\times 10^{15}$} 
 & AP2-TSI-06 & $(45.37\pm 0.09)$ @92 V & $(1.57\pm 0.29)$ @80V & $94\pm 2$ &\\ \cline{2-5}
 & AP2-TSI-24 & $(44.30\pm 0.09)$ @92 V & $(1.57\pm 0.30)$ @80V & $94\pm 2$ &\\ \cline{1-5}
\multirow{2}{*}{$2\times 10^{15}$} 
 & AP2-TSI-08 & $(67.20\pm 0.10)$ @92 V & $(1.81\pm 0.35)$ @80V & $94\pm 2$ &\\ \cline{3-5}
 & AP2-TSI-17 & $(97.59\pm 0.10)$ @90 V & $(2.91\pm 0.60)$ @80V & $92\pm 2$ &\\ \hline
\end{tabular}
\end{table}

In contrast, the annealing treatment of the ams prototype enhanced the already present surface leakage by recovering the interface states. These carriers were triggering the ionization process at a relatively low reverse bias and so, the breakdown voltage of ams candidate is reduced by a few volts in comparison to data reported before annealing. The good agreement of measured data for annealed samples to the Arrhenius predictions is visible though in figure~\ref{f:fig8c} and \ref{f:fig8d}. 

\begin{figure}[htbp]
\centering
\subfloat[]{\label{f:fig9a}\includegraphics[width=0.48\textwidth]{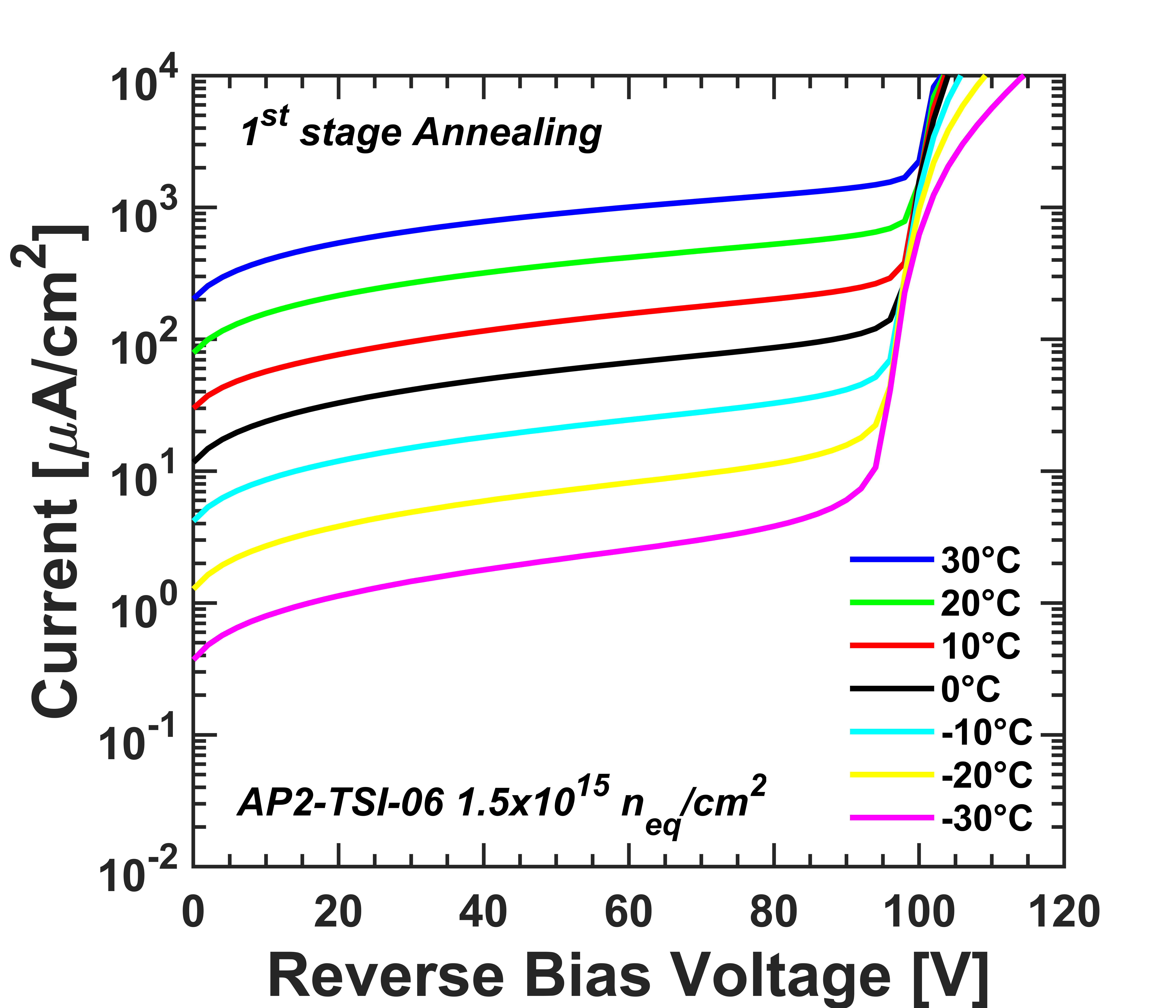}}
\subfloat[]{\label{f:fig9b}\includegraphics[width=0.48\textwidth]{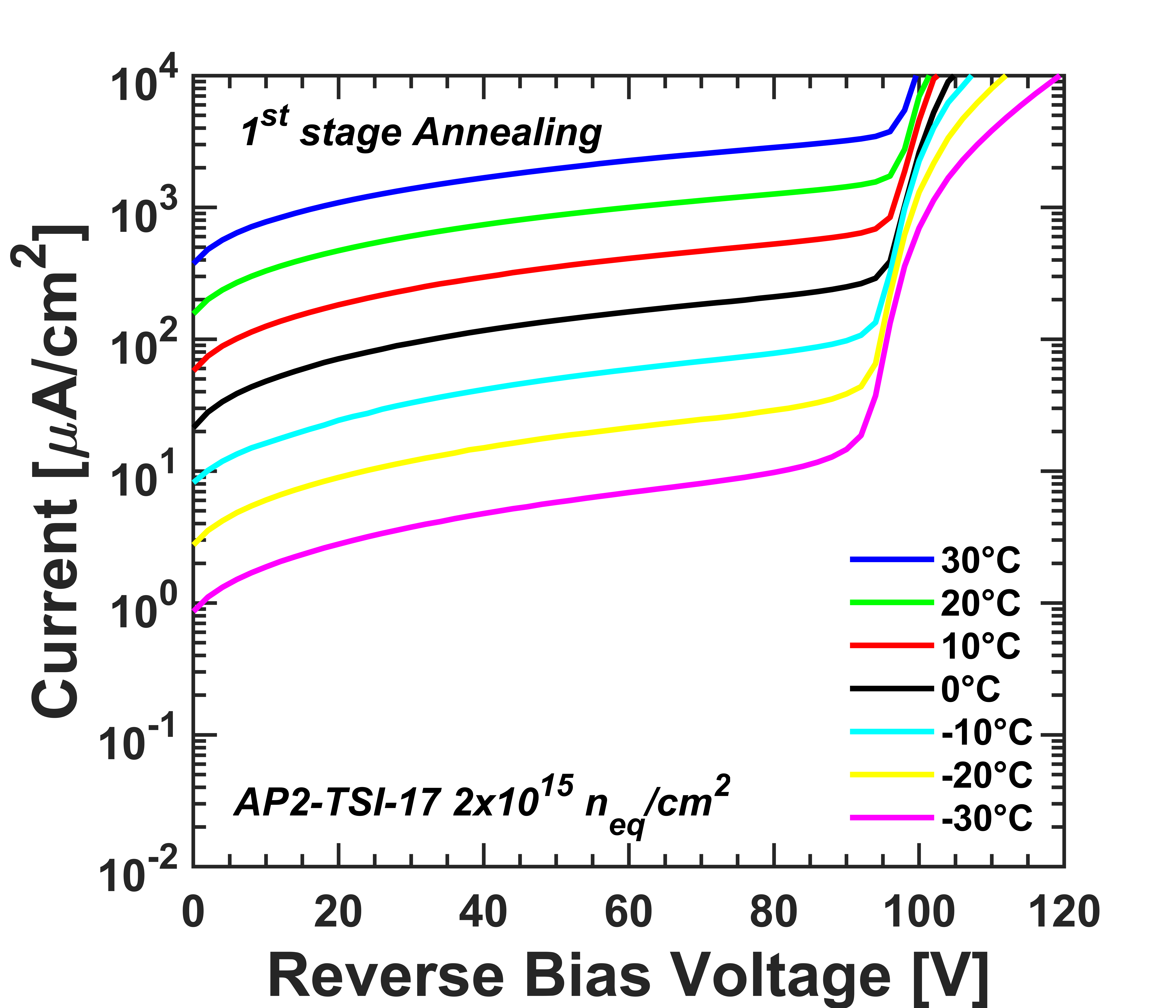}}\\ 
\subfloat[]{\label{f:fig9c}\includegraphics[width=0.48\textwidth]{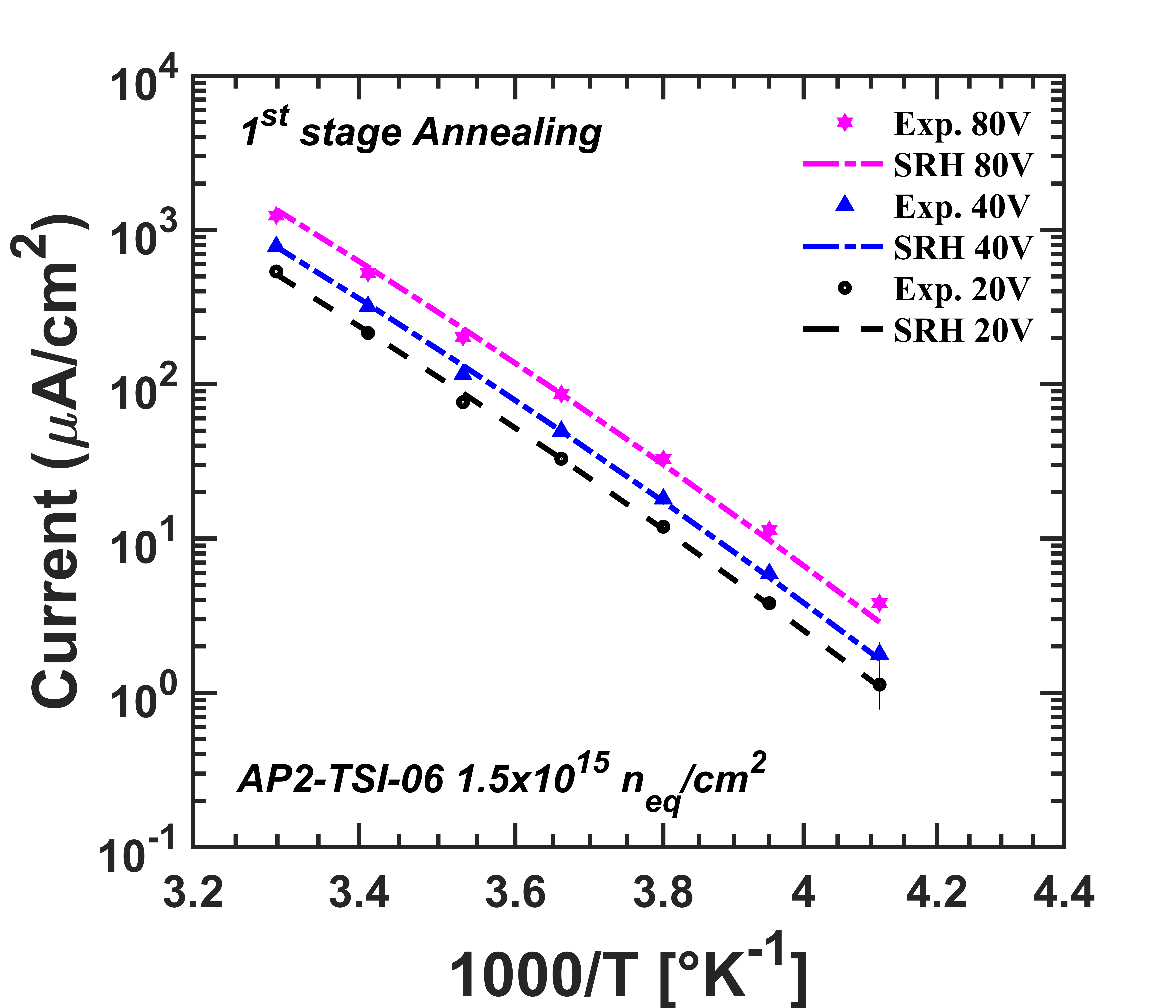}}
\subfloat[]{\label{f:fig9d}\includegraphics[width=0.48\textwidth]{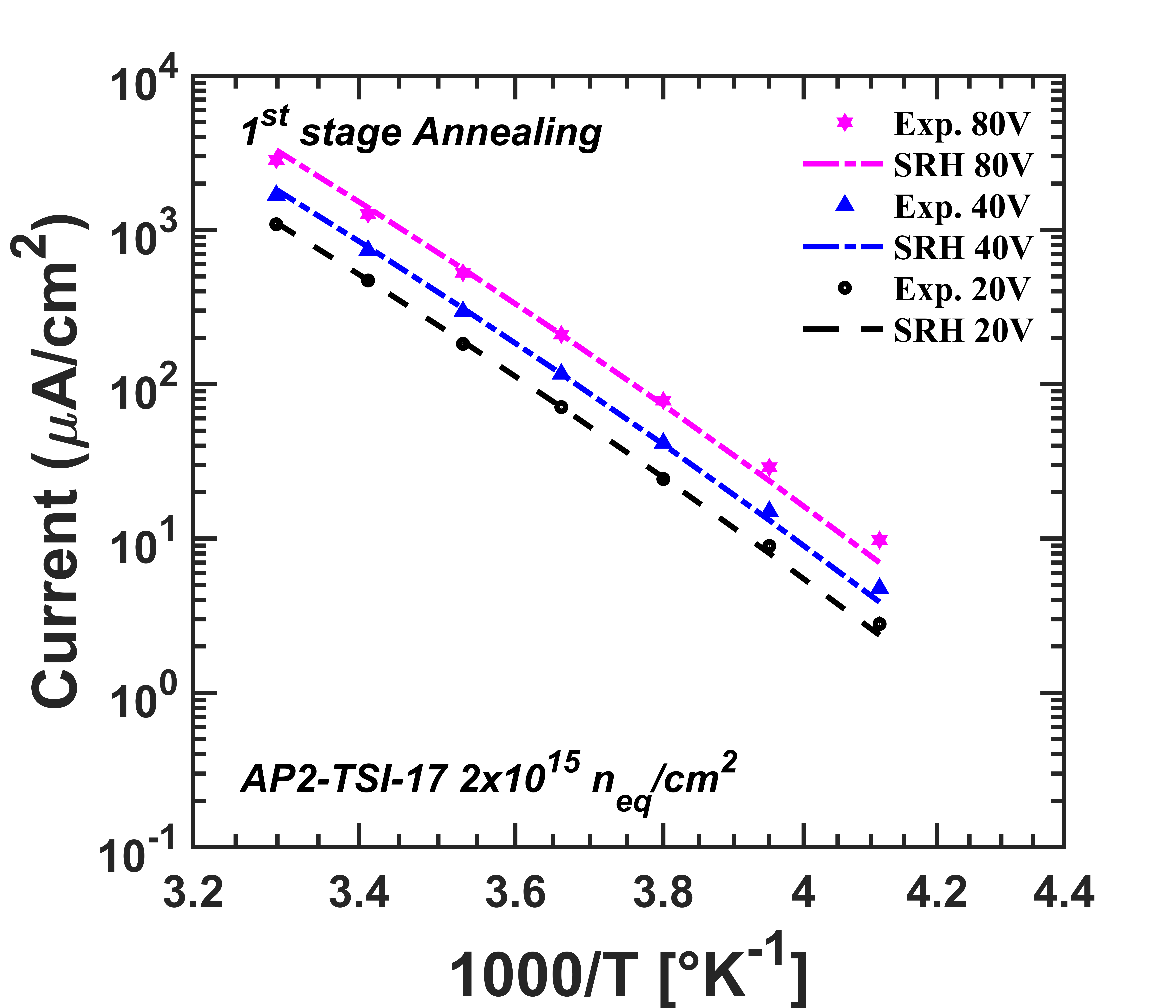}}\\
\caption{\label{f:fig9}Measured I-V curves (top) and corresponding Arrhenius plots (bottom) of irradiated ATLASPix2 chips: AP2-TSI-06 (irradiated to $1.5\times 10^{15}~\onemev$, left) and AP2-TSI-17 (irradiated to $2\times 10^{15}~\onemev$, right). Both chips were annealed at \SI{60}{\celsius} for \SI{80}{\minute} after irradiation. See table~\ref{t:samples} for a description of the samples.}
\end{figure}

I-V curves of TSI samples (AP2-TSI-06 and AP2-TSI-17, irradiated with $1.5\times 10^{15}$ and $2\times 10^{15}~\onemev$ JSI neutrons respectively) after the 1st stage annealing treatment are shown in figure~\ref{f:fig9a} and~\ref{f:fig9b}. The leakage current has been improved by 5 to 10\% after the first stage annealing treatment and thus, lower values for $\alpha^*$  are reported (table \ref{t:results-annealing-1}). The breakdown voltage has also been improved by a few volts for these samples. In reference to the section 5.2, the Arrhenius predictions have already been in good agreement with the measurements just after irradiation (before annealing) because the peripheral current becomes negligible in comparison with the increased bulk current. The expected recovery of some radiation-induced defects (both surface and bulk-damage) from the accelerated annealing leads to an improved agreement between the Arrhenius predictions and the measured data, as seen in figure~\ref{f:fig9c} and~\ref{f:fig9d}.

%%%%%%%%%%%%%%%%%%%%%%%%%%%%%%%%%%%%%%%%%%%%%%%%%%%%%%%%%%%%%%%%%%%%%%%
\subsubsection{Second stage annealing (\SI[detect-weight]{80}{\celsius} for 30 min)}

The second stage accelerated annealing was performed on selected candidates from the samples of first stage annealing. Candidates were chosen by cumulative fluence and processing foundry (ams and TSI). Samples were treated for \SI{30}{\minute} at \SI{80}{\celsius} in the second stage, and a summary of the I-V characteristics can be found in table \ref{t:results-annealing-2}. The leakage current improved by $\sim$30\% at this stage with respect to the first stage accelerated annealing. 

\begin{table}[htbp]
\small
\centering
\caption{\label{t:results-annealing-2} Electrical characteristics of irradiated ATLASPix2 samples after the second annealing stage (\SI{80}{\celsius} for 30 min). The last column indicates the total cumulative equivalent annealing time at \SI{20}{\celsius}.}
\smallskip
\begin{tabular}{|c|l|c|c|c|c|}
\hline
{\bf Fluence} & \multicolumn{1}{c|}{\bf Device ID} & \multicolumn{1}{c|}{\bf $J_\text{lk}$} & \multicolumn{1}{c|}{\bf $\alpha^*$} & {\bf $V_\text{bd}$} & {\bf Equiv. cum.} \\ 
 $[\onemev]$ && \multicolumn{1}{c|}{[\SI{}{\micro\ampere\per\centi\meter\squared}]} & $[\times 10^{-17}\,\rm{A/cm}]$ & [V] & {\bf ann. at \SI{20}{\celsius}} \\ \hline
\multirow{2}{*}{$1\times 10^{15}$} 
 & AP2-AMS-09 & $(12.17\pm 0.08)$ @60 V & $(0.92\pm 0.11)$ @50 V & $62\pm 2$ & \multirow{4}{*}{$\sim$71 days}\\ \cline{2-5}
& AP2-TSI-05 & $(16.32\pm 0.08)$ @94 V & $(0.86\pm 0.17)$ @80 V & $96\pm 2$ &\\ \cline{1-5}
$1.5\times 10^{15}$ & AP2-TSI-06 & $(30.05\pm 0.08)$ @92 V & $(1.19\pm 0.18)$ @80 V & $94\pm 2$ &\\ \cline{1-5}
 $2\times 10^{15}$ & AP2-TSI-17 & $(86.99\pm 0.10)$ @94 V & $(2.36\pm 0.38)$ @80 V & $96\pm 2$ &\\ \hline
\end{tabular}
\end{table}
\begin{figure}[htbp]
\centering
\subfloat[]{\label{f:fig10a}\includegraphics[width=0.33\textwidth]{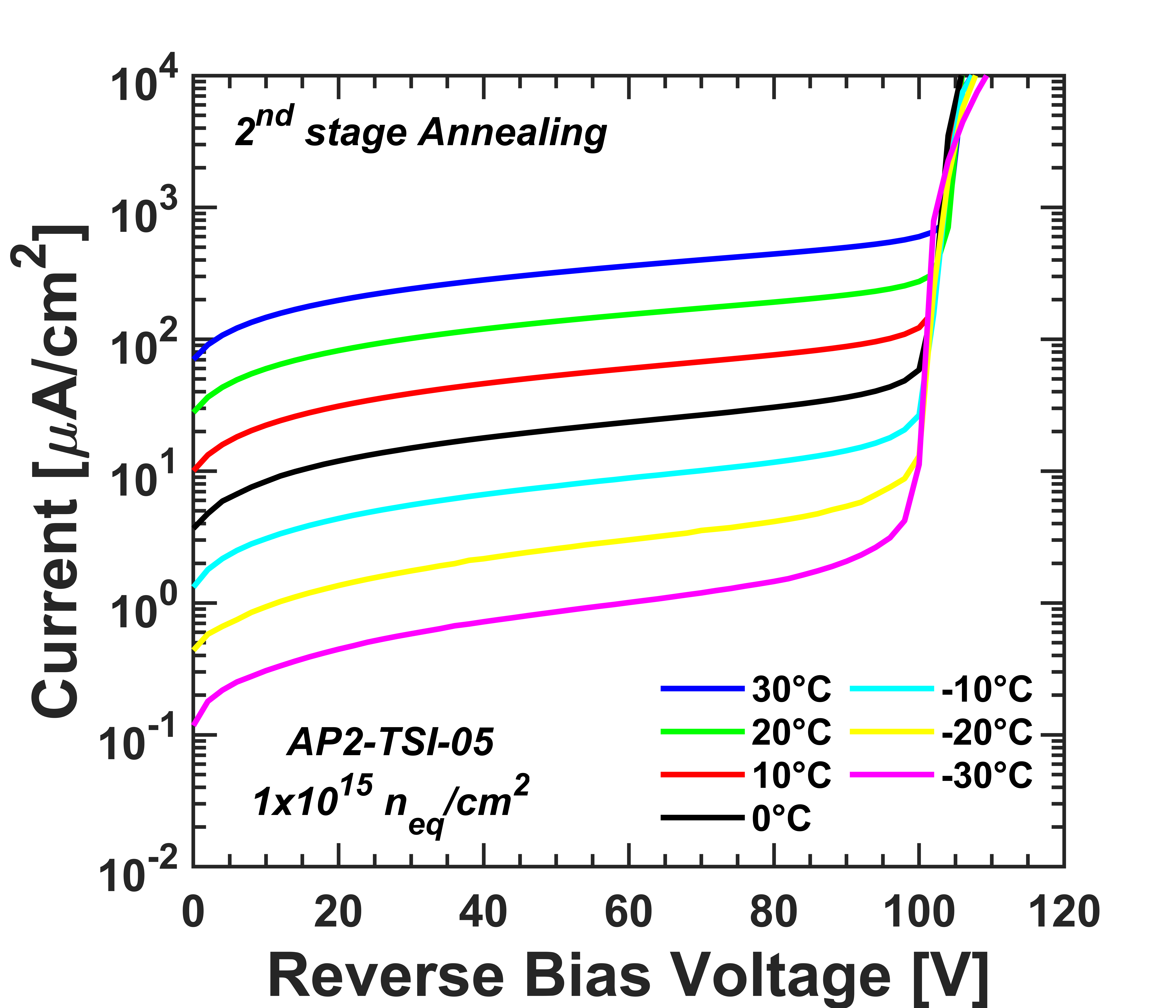}}
\subfloat[]{\label{f:fig10b}\includegraphics[width=0.33\textwidth]{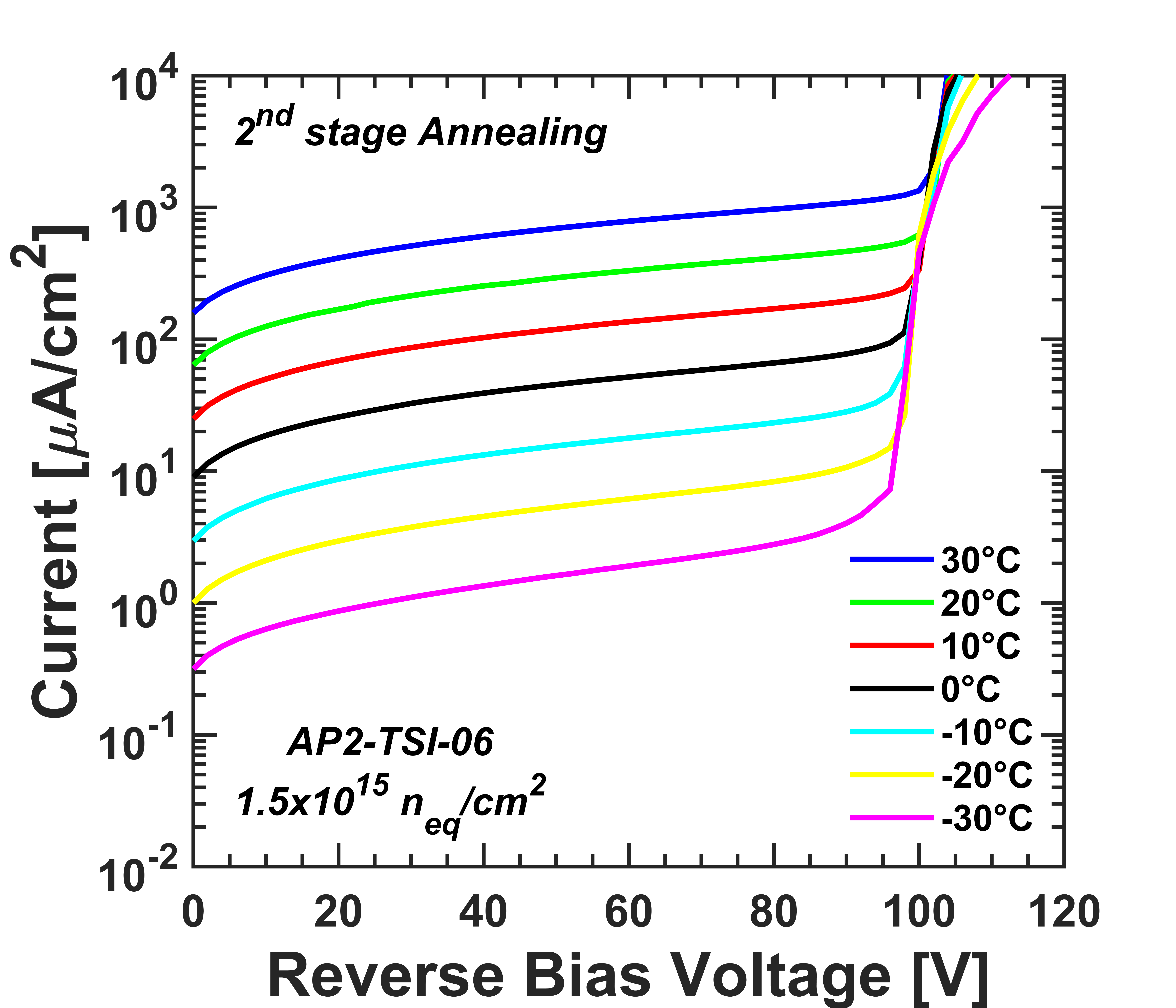}}
\subfloat[]{\label{f:fig10c}\includegraphics[width=0.33\textwidth]{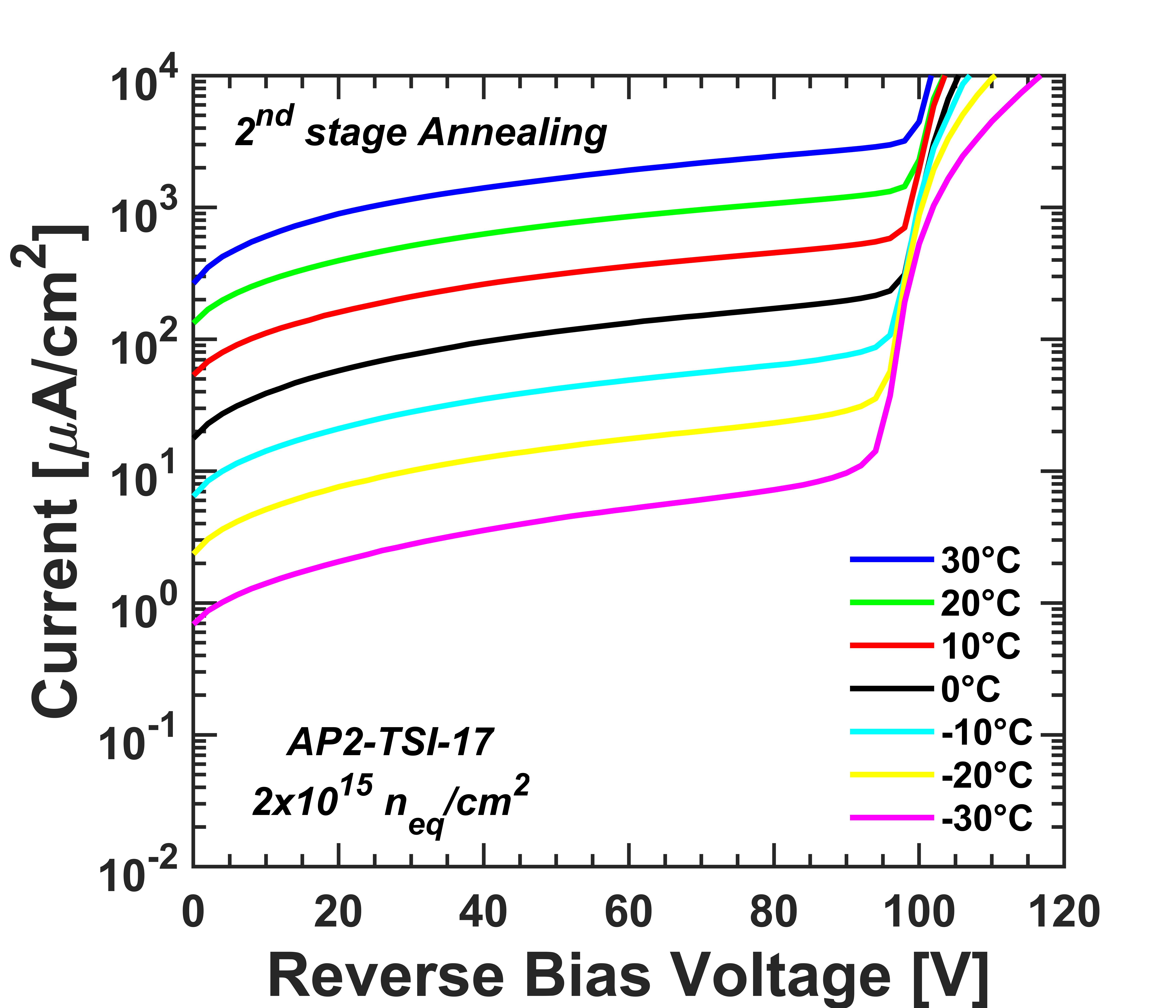}}\\
\subfloat[]{\label{f:fig10d}\includegraphics[width=0.33\textwidth]{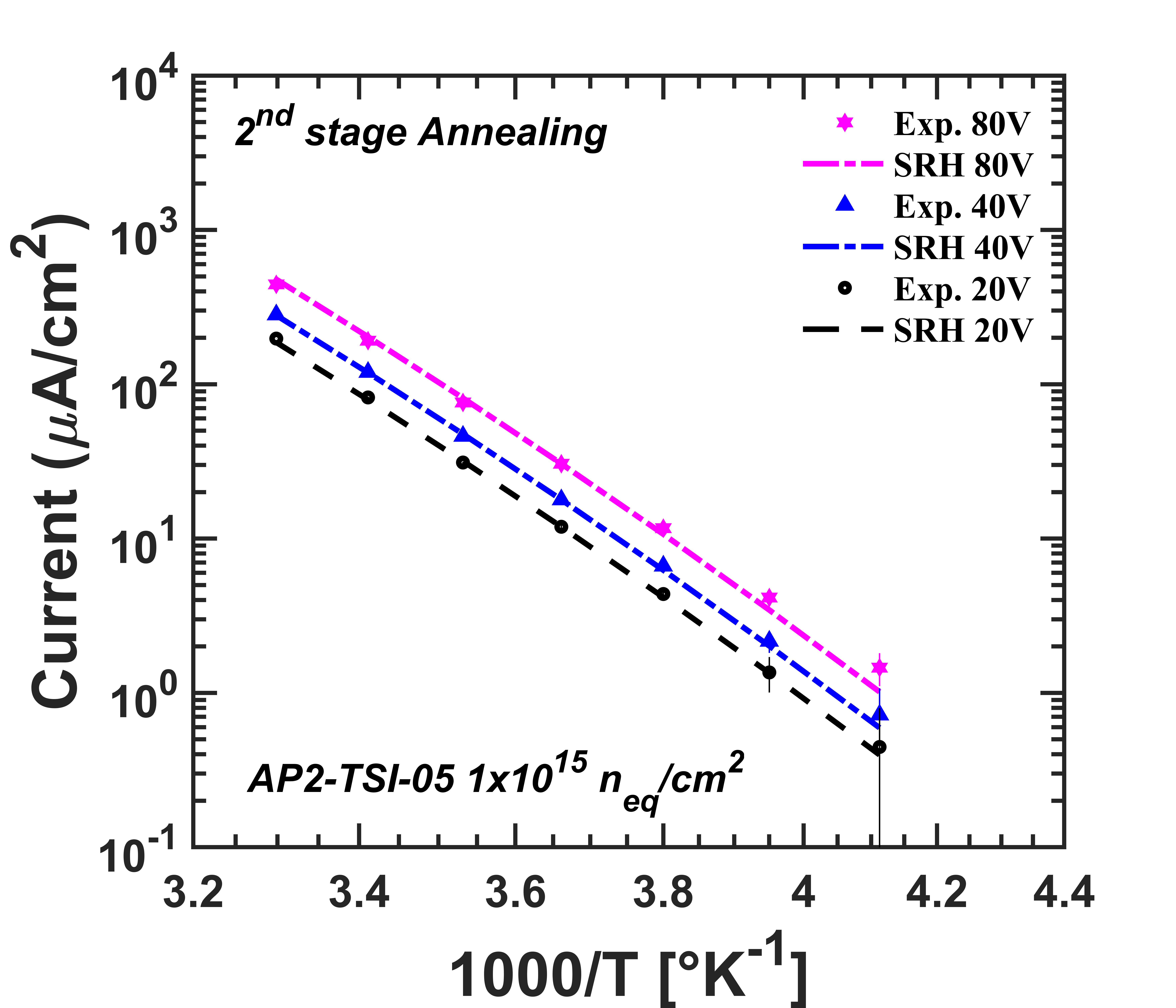}}
\subfloat[]{\label{f:fig10e}\includegraphics[width=0.33\textwidth]{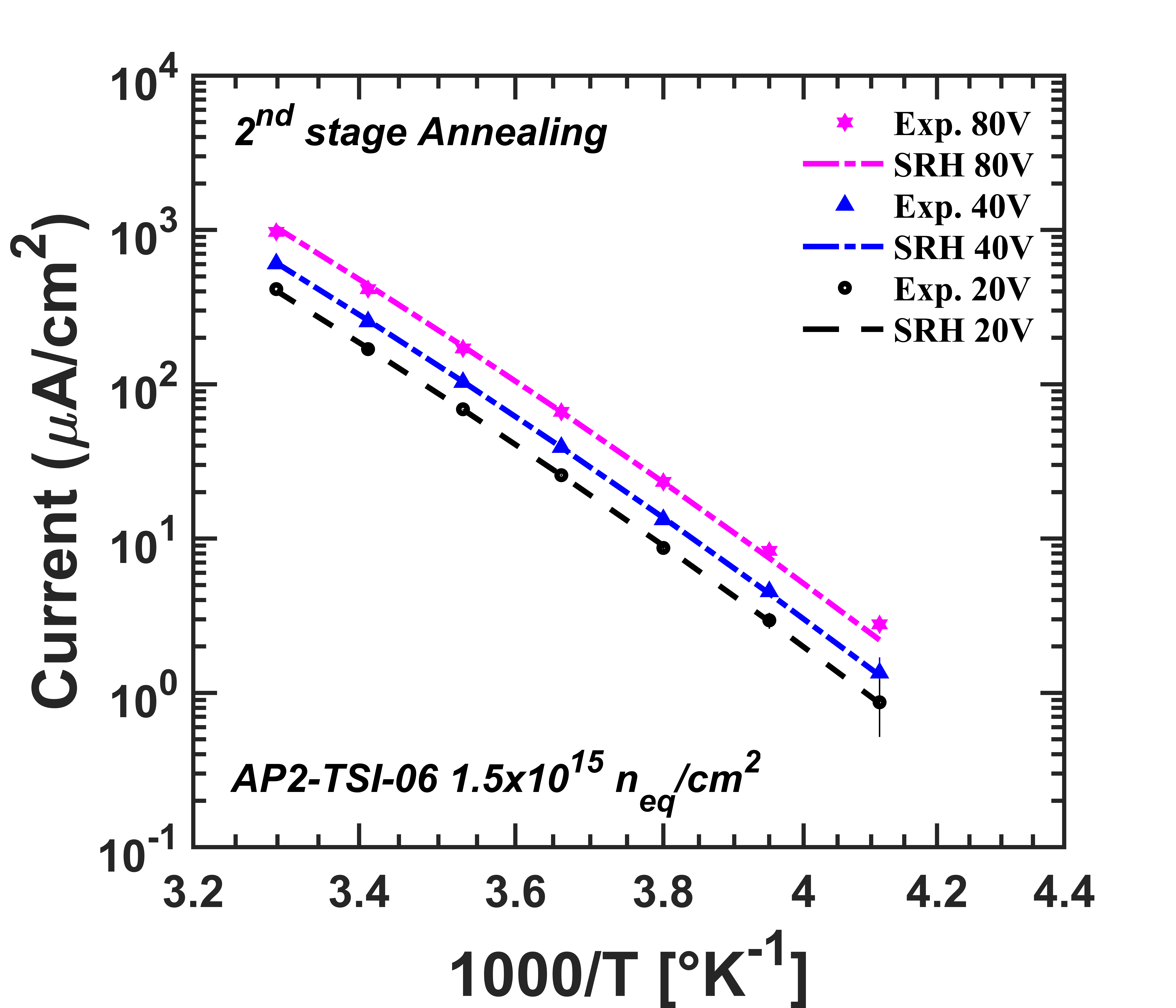}}
\subfloat[]{\label{f:fig10f}\includegraphics[width=0.33\textwidth]{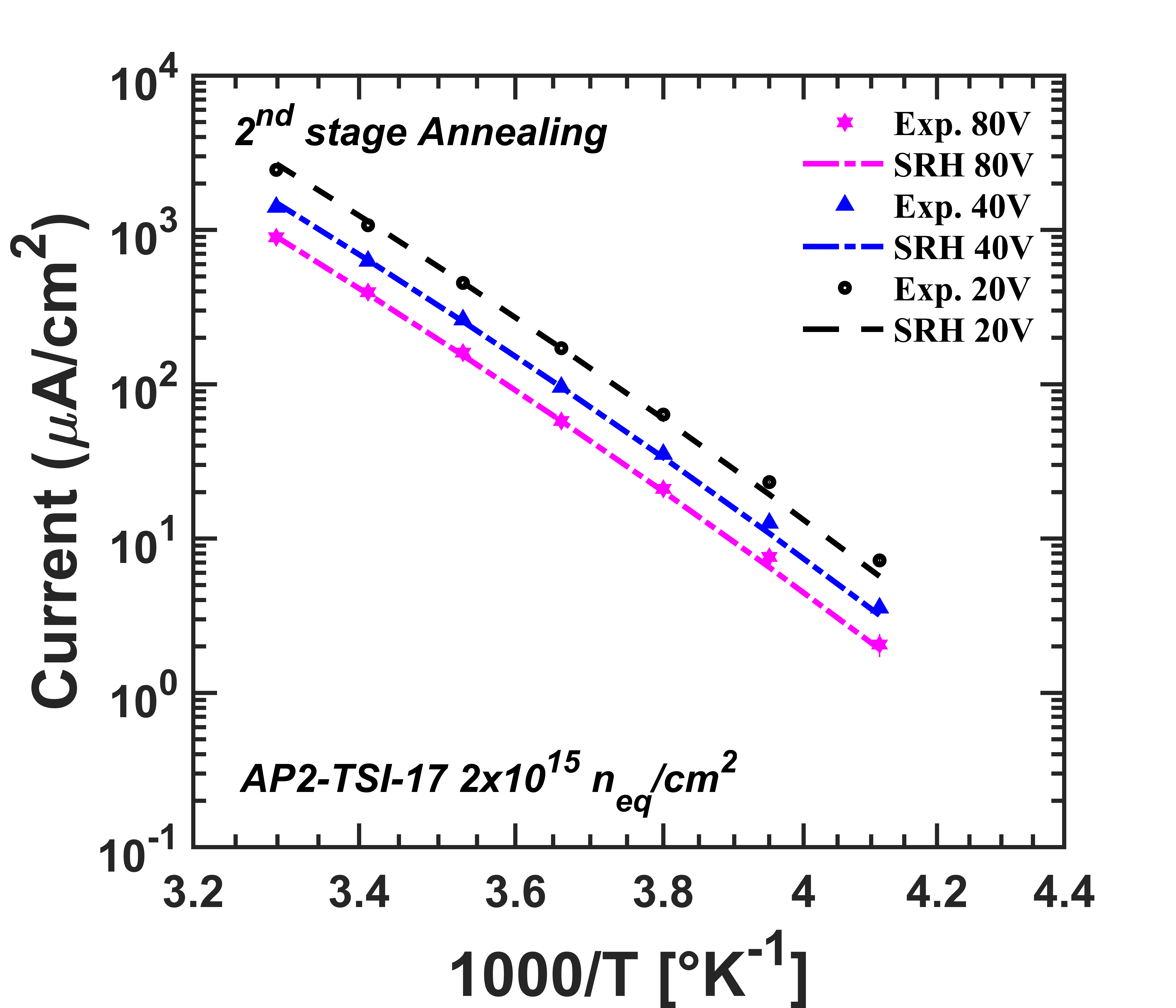}}
\caption{\label{f:fig10}Measured I-V curves (top) and corresponding Arrhenius plots (bottom) of irradiated ATLASPix2 chips: AP2-TSI-05 (irradiated to $1\times 10^{15}~\onemev$, left), AP2-TSI-06 (irradiated to $1.5\times 10^{15}~\onemev$, center) and AP-TSI-17 (irradiated to $2\times 10^{15}~\onemev$, right). All chips were annealed at \SI{60}{\celsius} for 80 min and \SI{80}{\celsius} for 30 min successively after irradiation. See table~\ref{t:samples} for a description of the samples.}
\end{figure}

Figure \ref{f:fig10} shows the I-V characteristics measured on TSI samples (irradiated with different fluences of neutron) after the second stage of accelerated annealing. The breakdown voltage tends to improve by a few volts due to the same reason of recovering the interface states and traps by annealing treatments. The surface conditions for TSI samples helped to trigger a delayed impact ionization. The corresponding Arrhenius plots also shows a good agreement for the measured data with the thermionic generation model.

\begin{figure}[!tbp]
\centering
\includegraphics[width=0.6\textwidth]{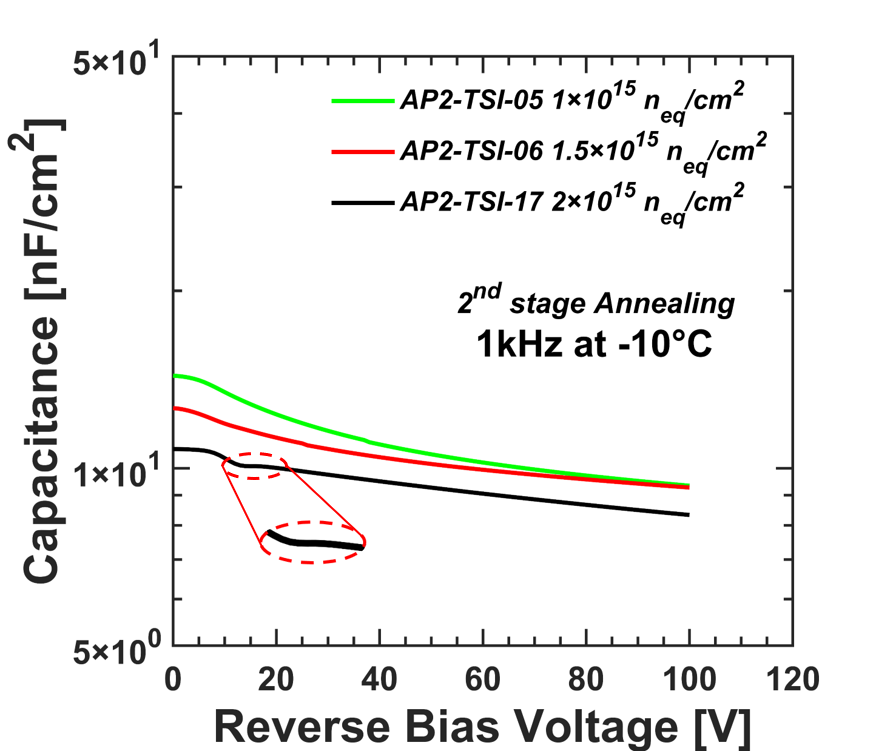}
\caption{\label{f:fig11}Measured C-V curves of several ATLASPix2 chips, irradiated to different fluences and then annealed at \SI{60}{\celsius} for 80 min and \SI{80}{\celsius} for 30 min successively. Measurements were taken at \SI{-10}{\celsius}. See table~\ref{t:samples} for a description of the samples.}
\end{figure}

C-V data shown in figure \ref{f:fig11} measured after the second stage annealing for an AC signal of \SI{100}{\milli\volt} and a \SI{1}{\kilo\hertz} frequency. They are quite comparable to the data seen before annealing, meaning there is no additional effect from these annealing treatments up to this second stage. 

%%%%%%%%%%%%%%%%%%%%%%%%%%%%%%%%%%%%%%%%%%%%%%%%%%%%%%%%%%%%%%%%%%%%%%%
\subsubsection{Third stage annealing (\SI[detect-weight]{80}{\celsius} for 120 min)}

\begin{figure}[htbp]
\centering
\subfloat[]{\label{f:fig12a}\includegraphics[width=0.33\textwidth]{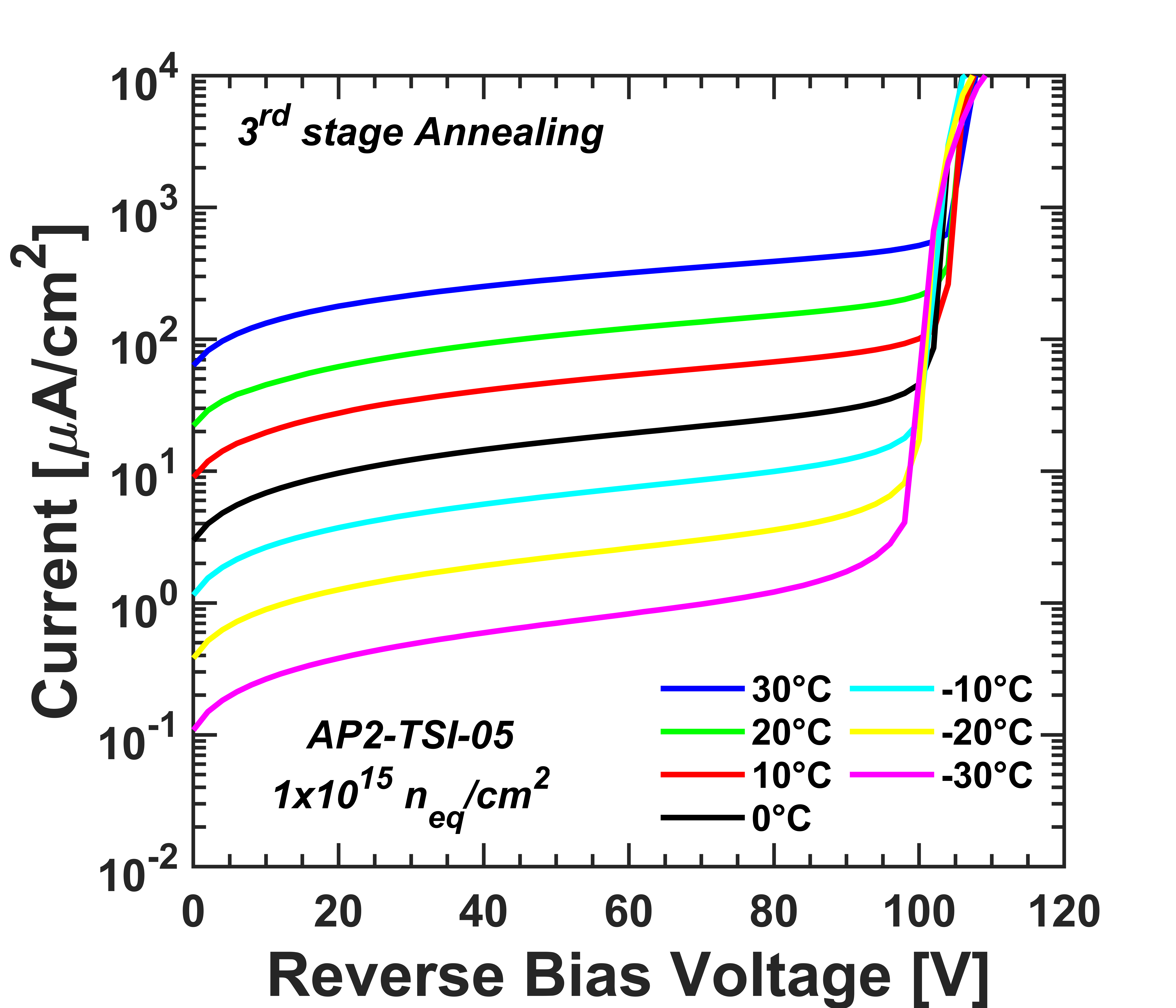}}
\subfloat[]{\label{f:fig12b}\includegraphics[width=0.33\textwidth]{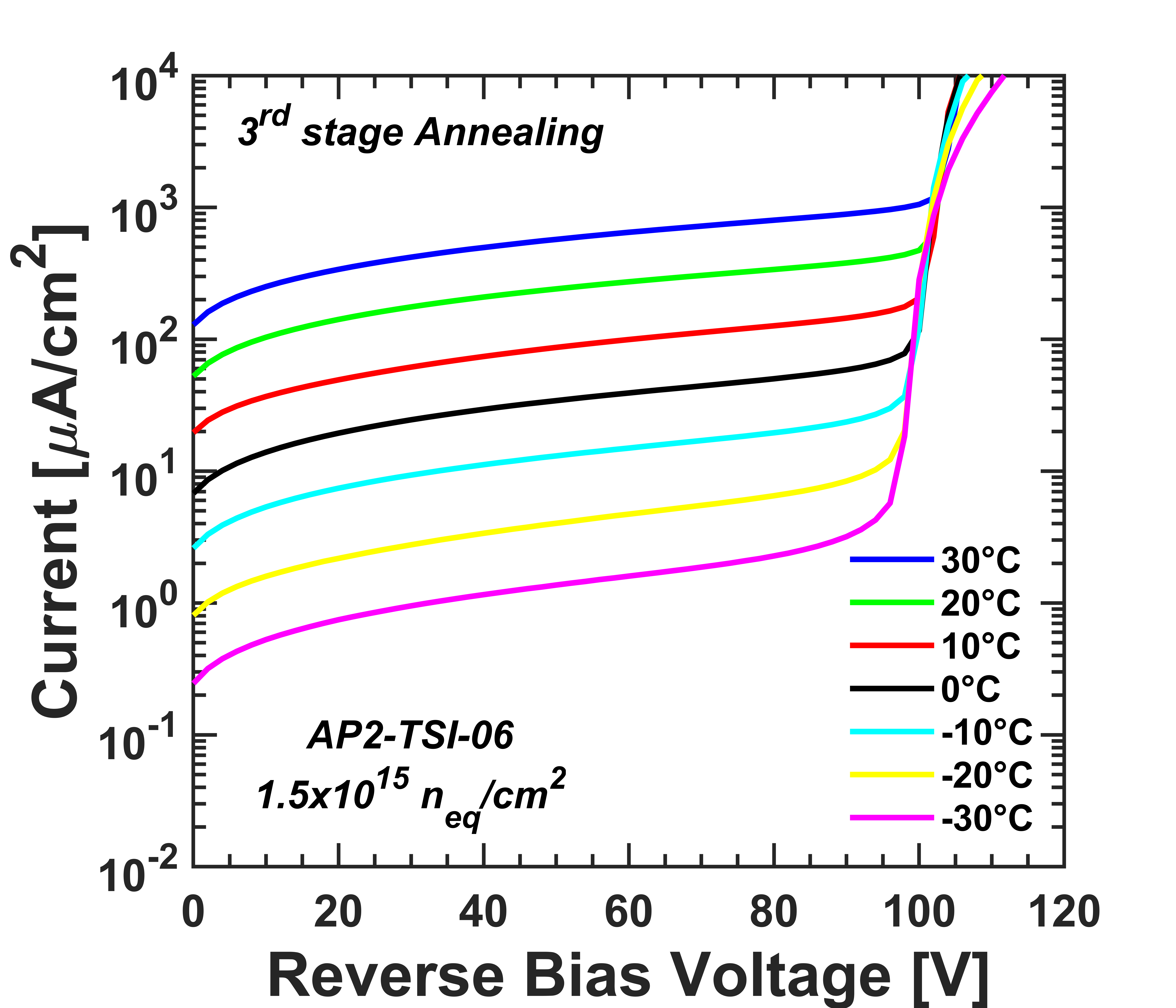}}
\subfloat[]{\label{f:fig12c}\includegraphics[width=0.33\textwidth]{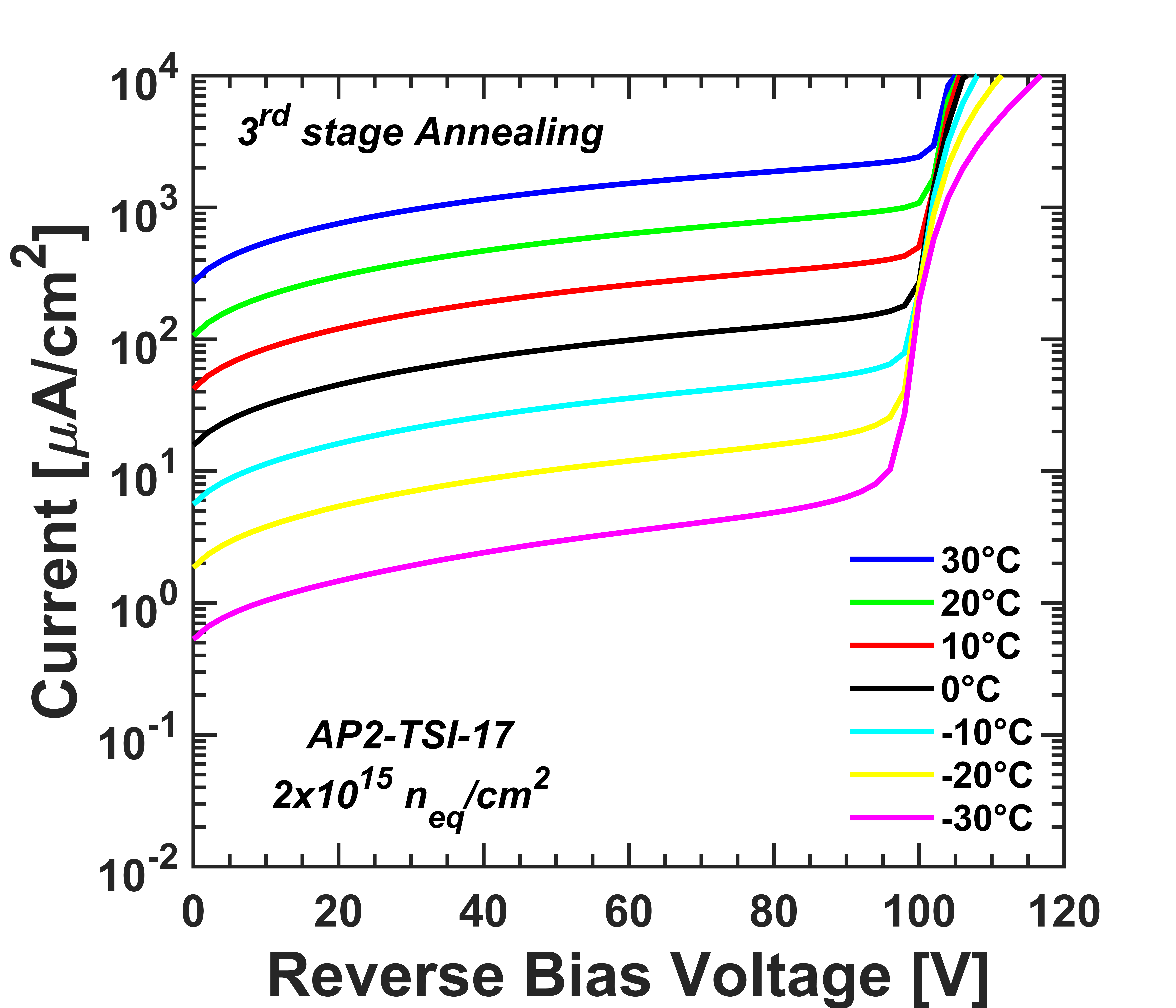}}\\
\subfloat[]{\label{f:fig12d}\includegraphics[width=0.33\textwidth]{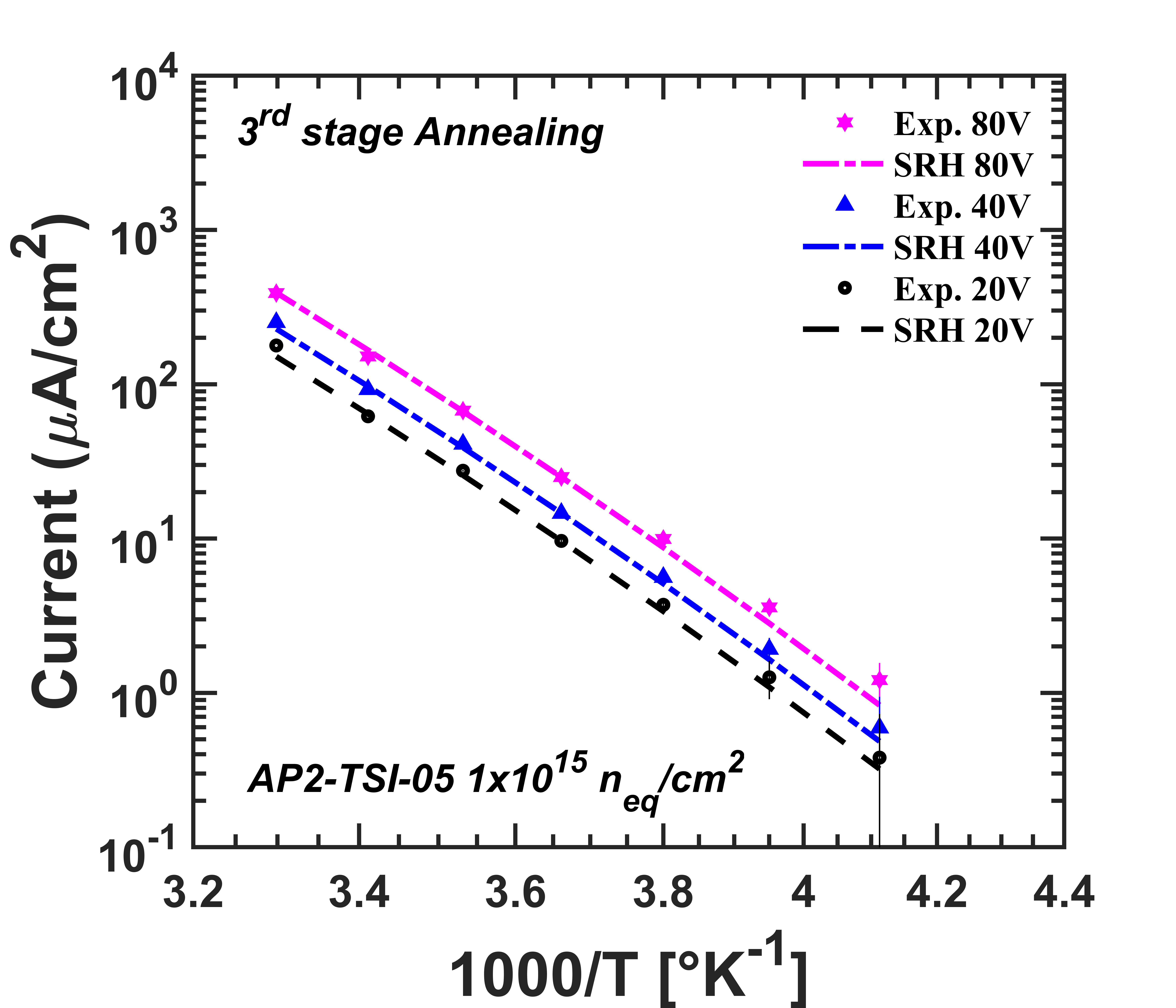}}
\subfloat[]{\label{f:fig12e}\includegraphics[width=0.33\textwidth]{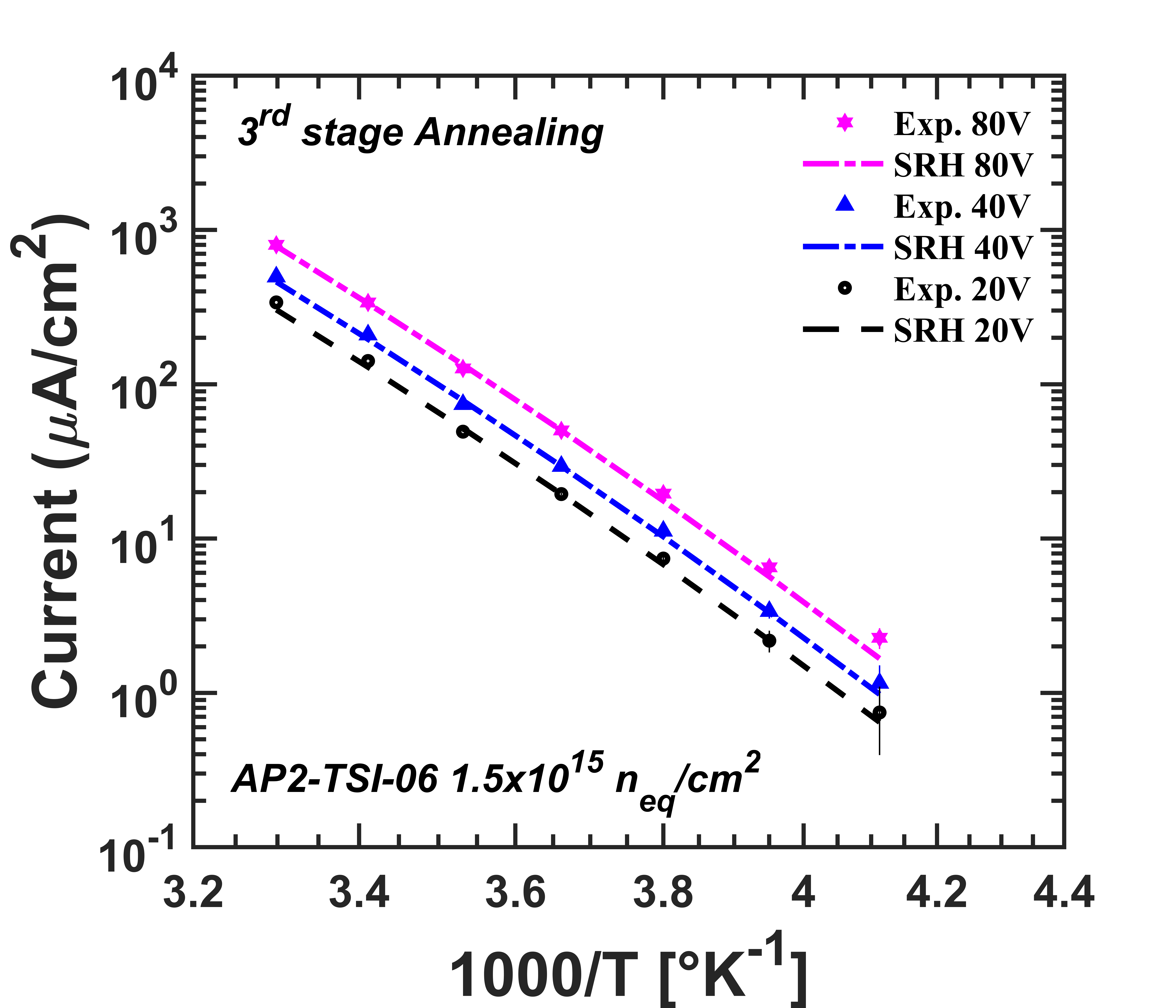}}
\subfloat[]{\label{f:fig12f}\includegraphics[width=0.33\textwidth]{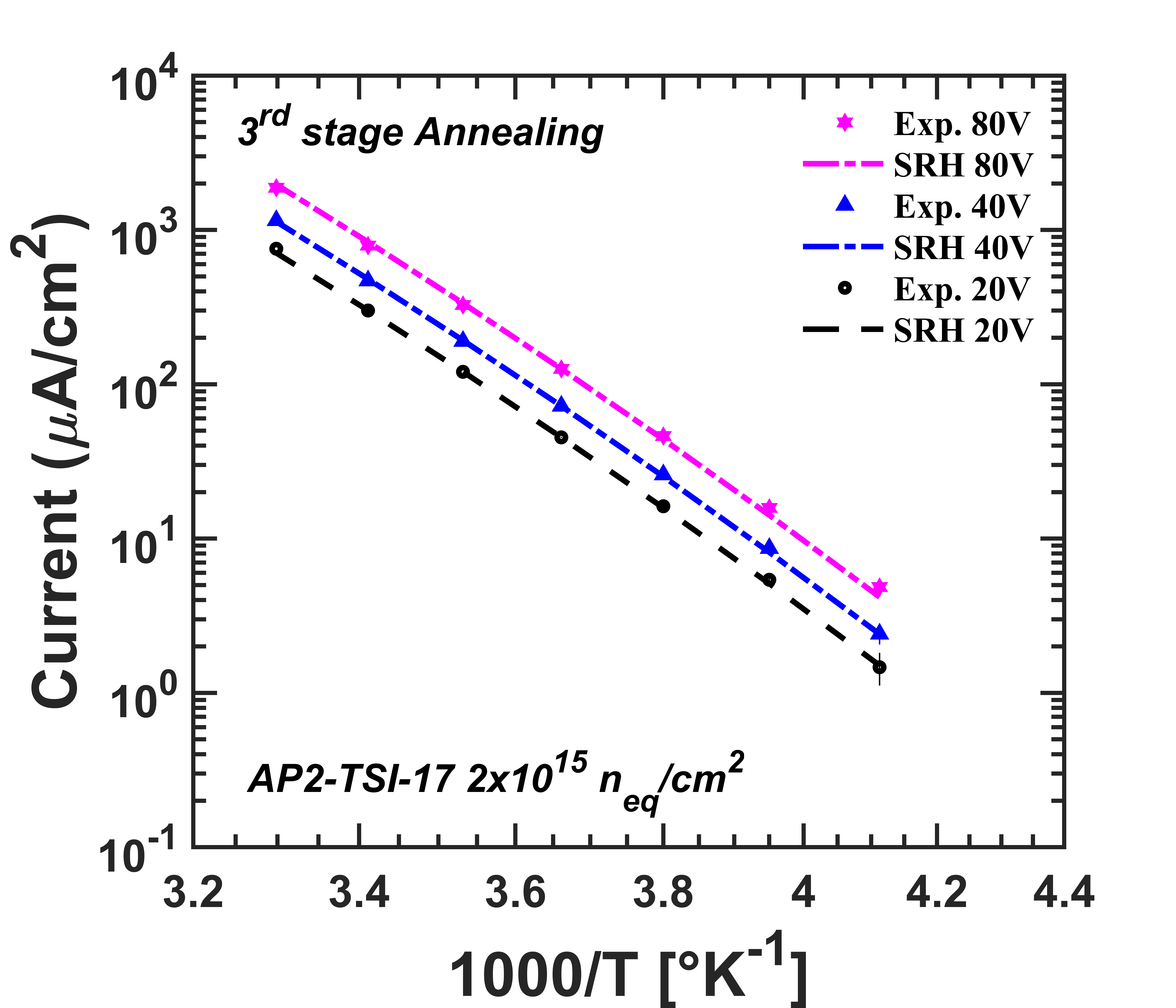}}
\caption{\label{f:fig12}Measured I-V curves (top) and corresponding Arrhenius plots (bottom) of irradiated ATLASPix2 chips: AP2-TSI-05 (irradiated to $1\times 10^{15}~\onemev$, left), AP2-TSI-06 (irradiated to $1.5\times 10^{15}~\onemev$, center) and AP2-TSI-17 (irradiated to $2\times 10^{15}~\onemev$, right). All chips were further annealed at \SI{80}{\celsius} for 120 min after second stage accelerated annealing. See table~\ref{t:samples} for a description of the samples.}
\end{figure}

Same selected samples used for second stage of accelerated annealing were further annealed for \SI{120}{\minute} at \SI{80}{\celsius} in the third stage. The summary of I-V characterizations is reported in table \ref{t:results-annealing-3}. The breakdown voltage increased by few volts for sample AP2-TSI-17 after the third stage annealing, as expected. Figure \ref{f:fig12} shows the measured I-V data of the TSI samples after the third stage annealing, where Arrhenius predictions are still in good agreement (irrespective of the fluence) with the measured data.

\begin{table}[htbp]
\small
\centering
\caption{\label{t:results-annealing-3} Electrical characteristics of irradiated ATLASPix2 samples after the third annealing stage (\SI{80}{\celsius} for 120 min). The last column indicates the total cumulative equivalent annealing time at \SI{20}{\celsius}.}
\smallskip
\begin{tabular}{|c|l|c|c|c|c|}
\hline
{\bf Fluence} & \multicolumn{1}{c|}{\bf Device ID} & \multicolumn{1}{c|}{\bf $J_\text{lk}$} & \multicolumn{1}{c|}{\bf $\alpha^*$} & {\bf $V_\text{bd}$} & {\bf Equiv. cum.} \\ 
 $[\onemev]$ && \multicolumn{1}{c|}{[\SI{}{\micro\ampere\per\centi\meter\squared}]} & $[\times 10^{-17}\,\rm{A/cm}]$ & [V] & {\bf ann. at \SI{20}{\celsius}} \\ \hline
\multirow{2}{*}{$1\times 10^{15}$} 
 & AP2-AMS-09 & $(11.07\pm 0.08)$ @60 V & $(0.80\pm 0.09)$ @50 V & $62\pm 2$ & \multirow{4}{*}{$\sim$359 days}\\ \cline{2-5}
& AP2-TSI-05 & $(13.98\pm 0.08)$ @94 V & $(0.73\pm 0.15)$ @80 V & $96\pm 2$ &\\ \cline{1-5}
$1.5\times 10^{15}$ & AP2-TSI-06 & $(26.96\pm 0.08)$ @94 V & $(0.95\pm 0.16)$ @80 V & $96\pm 2$ &\\ \cline{1-5}
 $2\times 10^{15}$ & AP2-TSI-17 & $(64.89\pm 0.09)$ @96 V & $(1.69\pm 0.22)$ @80 V & $98\pm 2$ &\\ \hline
\end{tabular}
\end{table}

%%%%%%%%%%%%%%%%%%%%%%%%%%%%%%%%%%%%%%%%%%%%%%%%%%%%%%%%%%%%%%%%%%%%%%%
\subsubsection{Fourth stage accelerated annealing (\SI[detect-weight]{80}{\celsius} for 300 min)}

\begin{figure}[!tbp]
\centering
\subfloat[]{\label{f:fig13a}\includegraphics[width=0.48\textwidth]{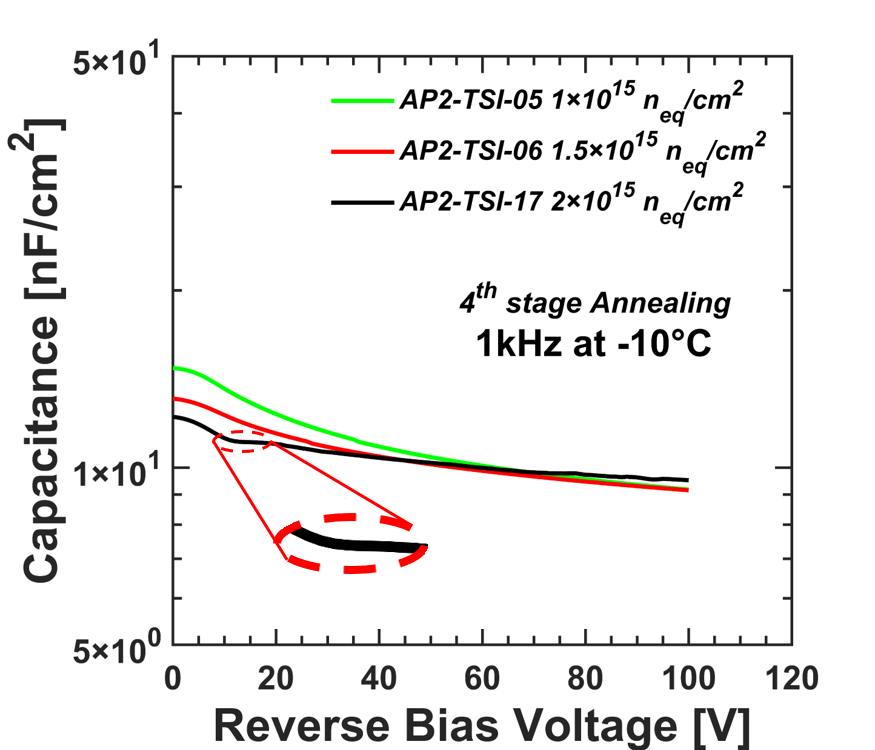}}
\subfloat[]{\label{f:fig13b}\includegraphics[width=0.48\textwidth]{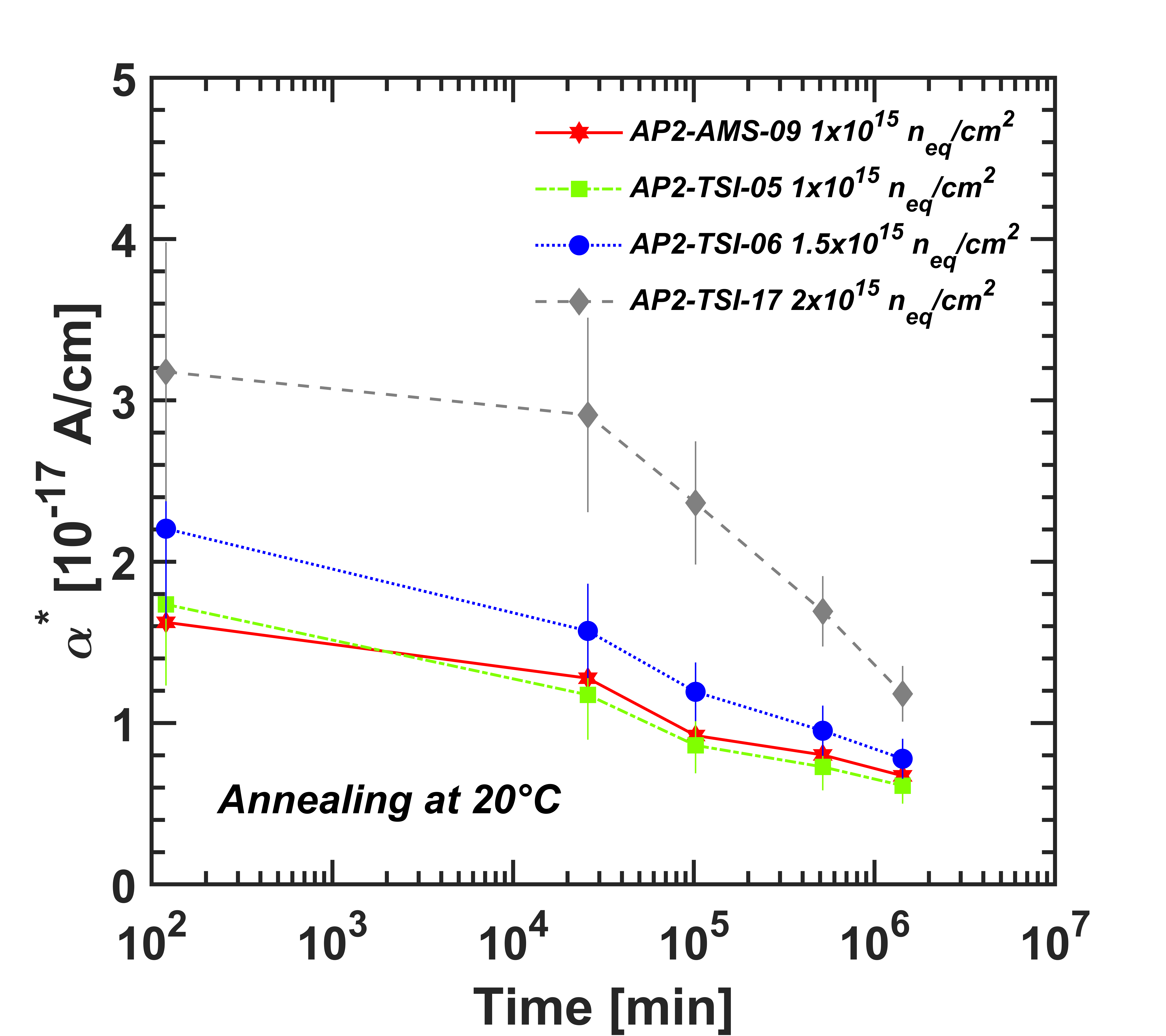}}
\caption{\label{f:fig13}Measured C-V surves of JSI neutron irradiated TSI ATLASPix2 samples of different fluences after fourth stage accelerated annealing. Data measured with \SI{1}{\kilo\hertz} small AC signal at \SI{-10}{\celsius} (left). Cumulative annealing effects for the damage constant rate $\alpha^*$ at \SI{20}{\celsius} (right). Data measured for a reverse bias of a \SI{10}{\volt} before than $V_\text{bd}$: ams sample at \SI{50}{\volt} and TSI prototypes at \SI{80}{\volt}.}
\end{figure}

Samples treated with third stage annealing were again treated for a  \SI{300}{\minute} long accelerated annealing at \SI{80}{\celsius} in the fourth stage. There was no sign of full depletion for the C-V data of TSI prototypes, reported in figure \ref{f:fig13}. However, the lateral depletion has still been seen at $\sim$\SI{14}{\volt} for the sample AP2-TSI-17 (irradiated with $2\times 10^{15}~\onemev$ neutrons). 

\begin{figure}[htbp]
\centering
\subfloat[]{\label{f:fig15a}\includegraphics[width=0.33\textwidth]{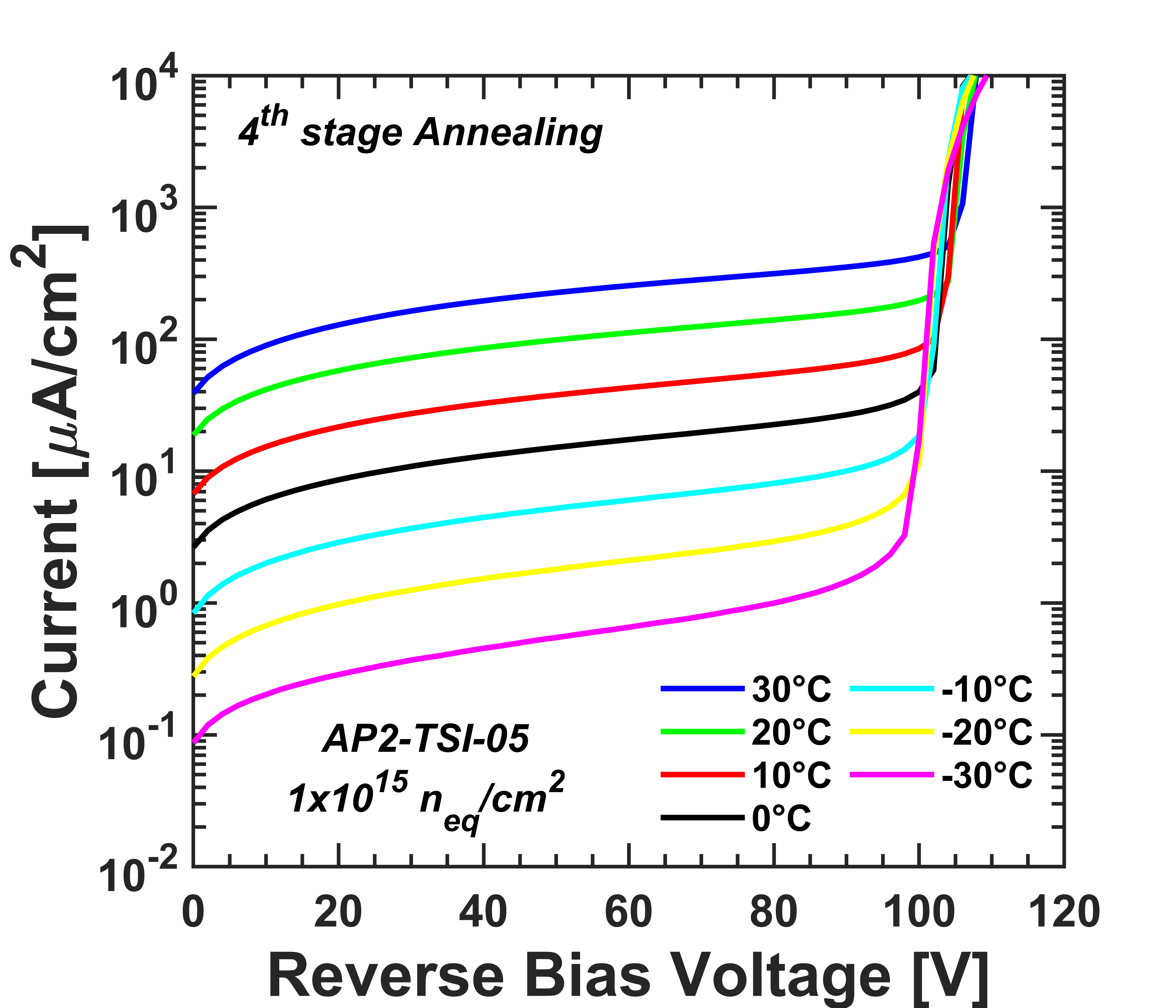}}
\subfloat[]{\label{f:fig15b}\includegraphics[width=0.33\textwidth]{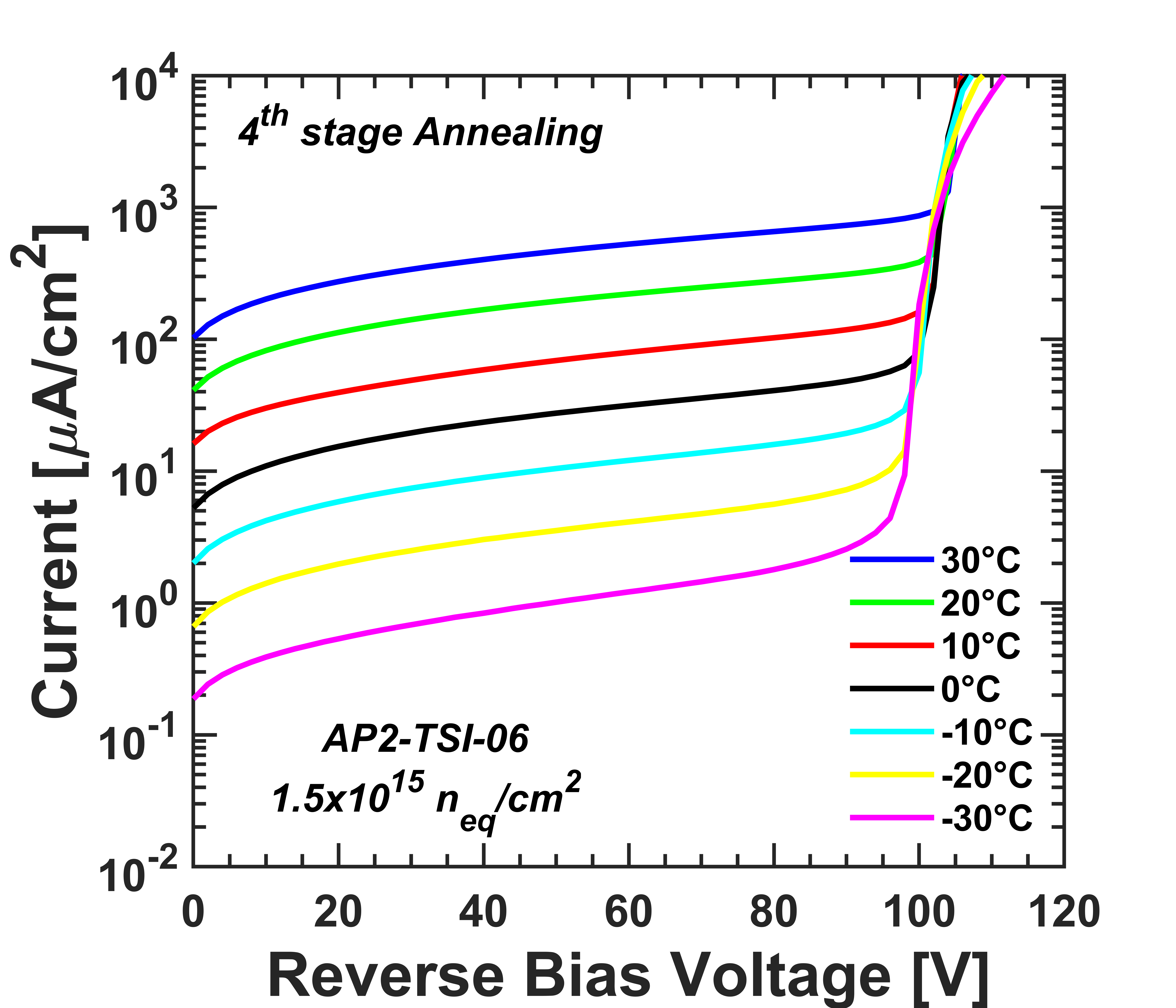}}
\subfloat[]{\label{f:fig15c}\includegraphics[width=0.33\textwidth]{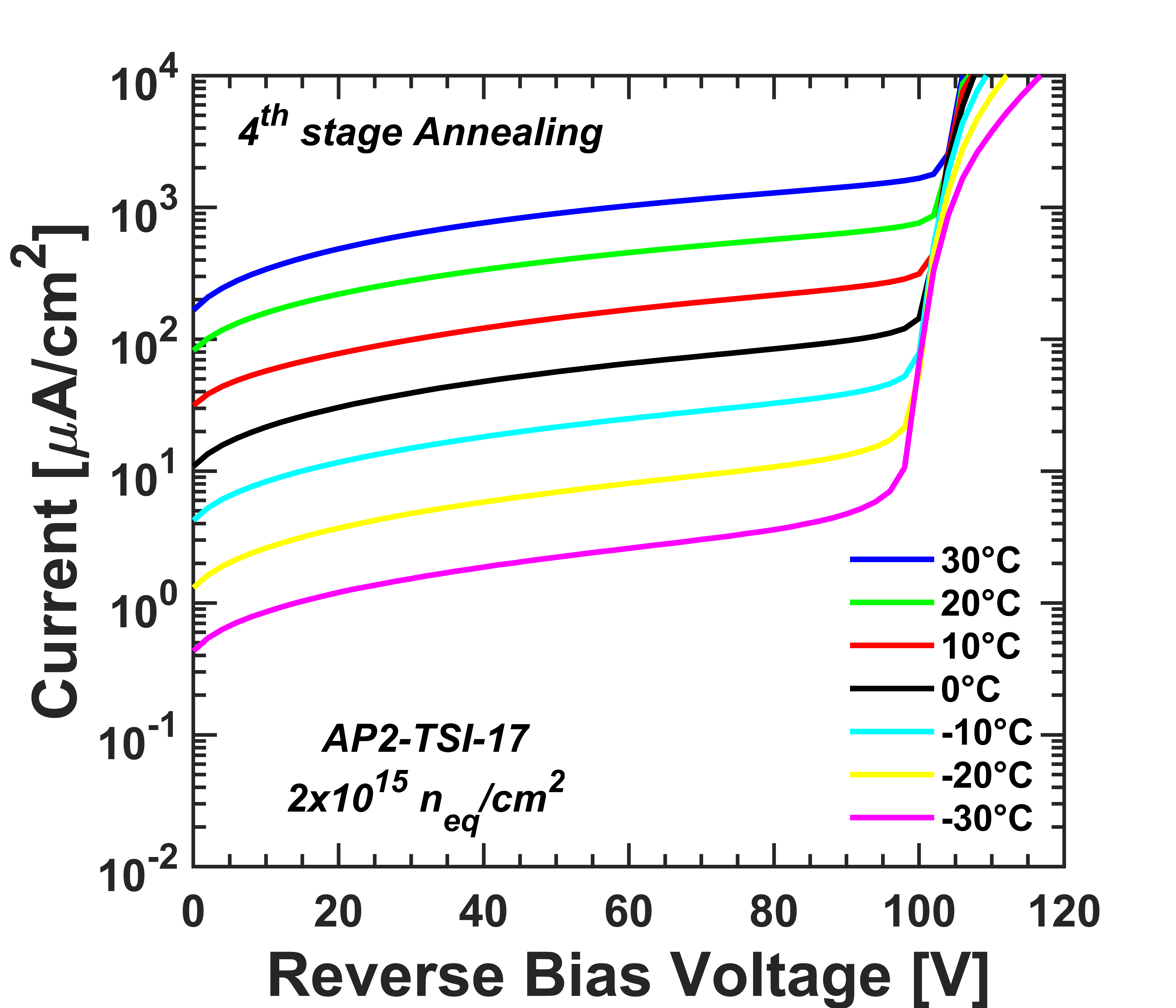}}\\
\subfloat[]{\label{f:fig15d}\includegraphics[width=0.33\textwidth]{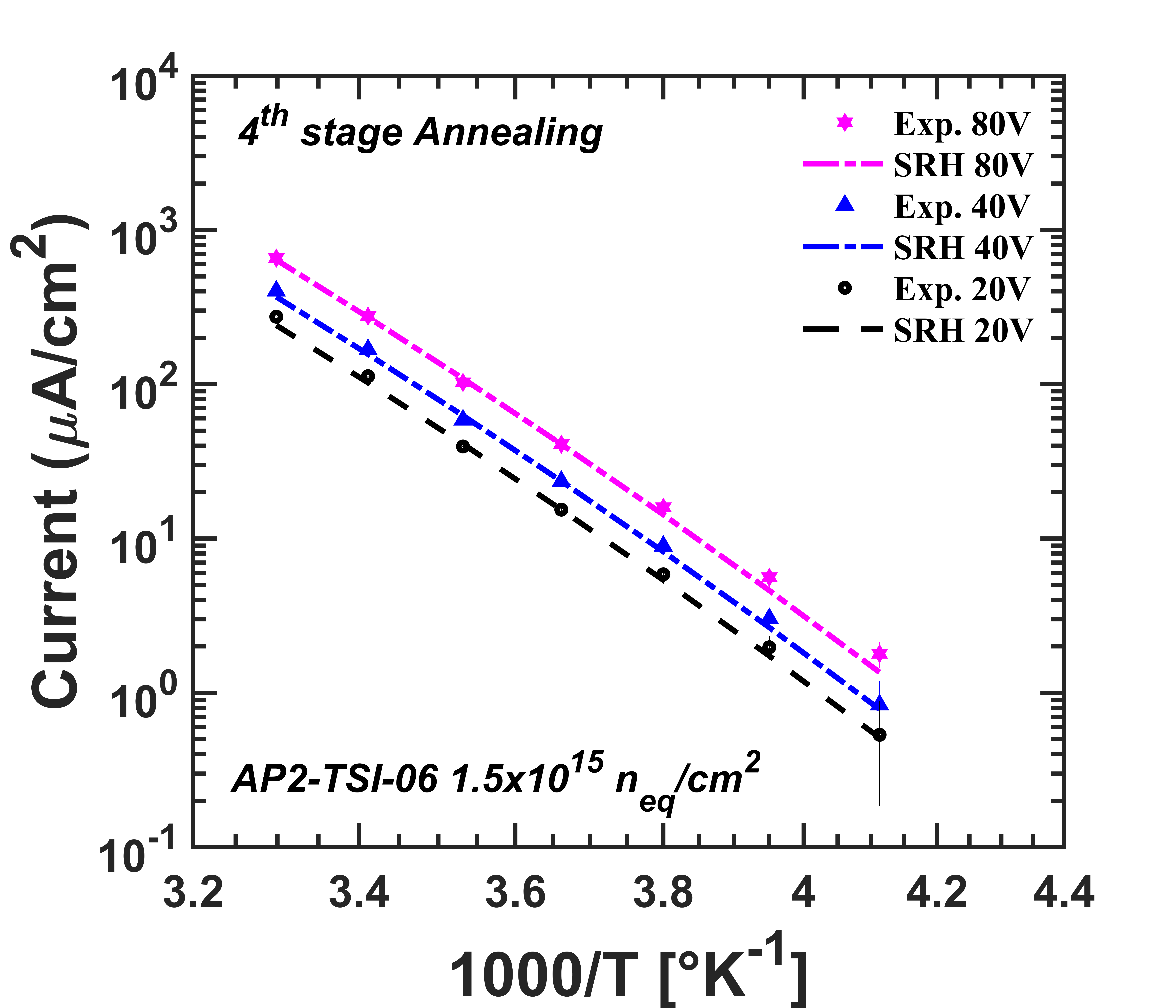}}
\subfloat[]{\label{f:fig15e}\includegraphics[width=0.33\textwidth]{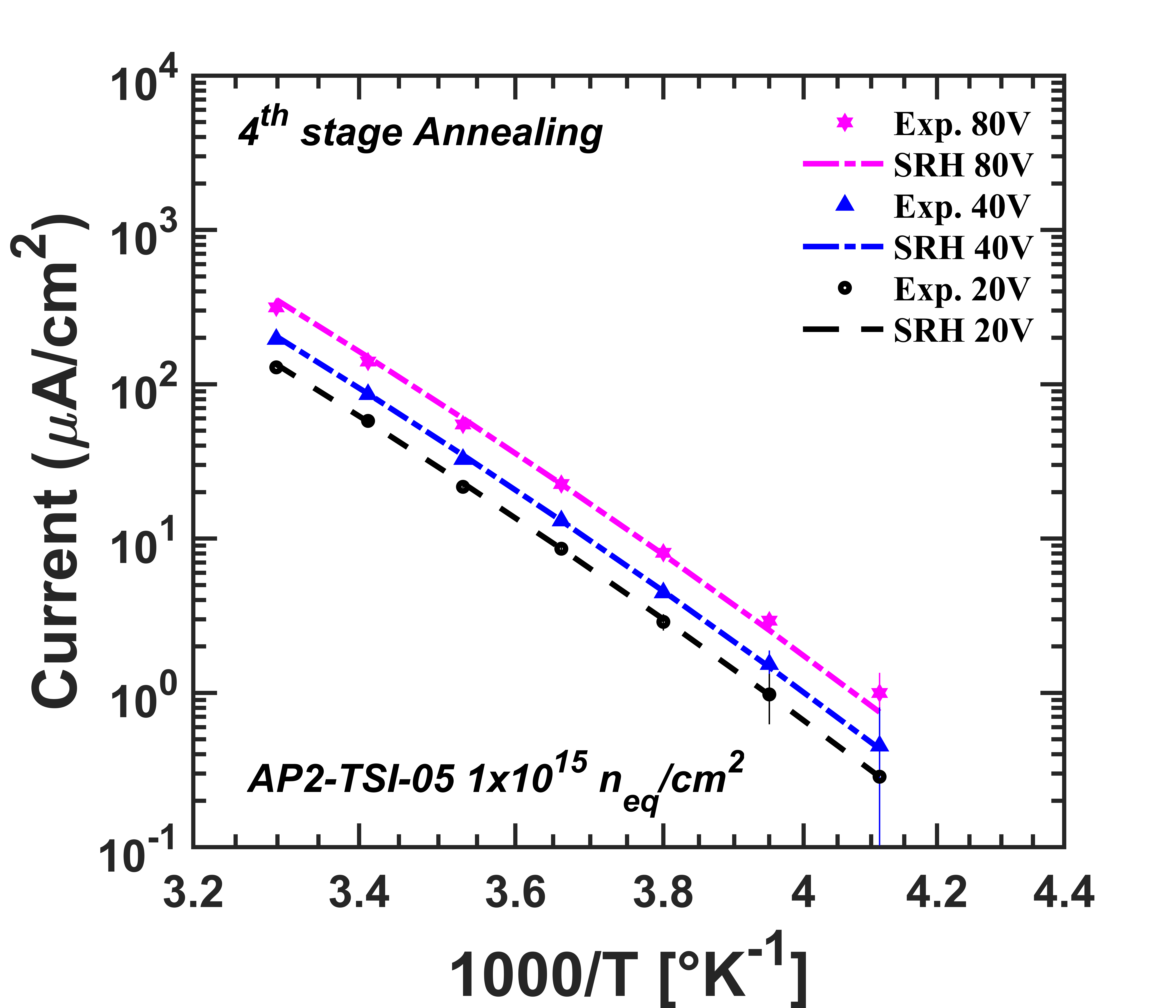}}
\subfloat[]{\label{f:fig15f}\includegraphics[width=0.33\textwidth]{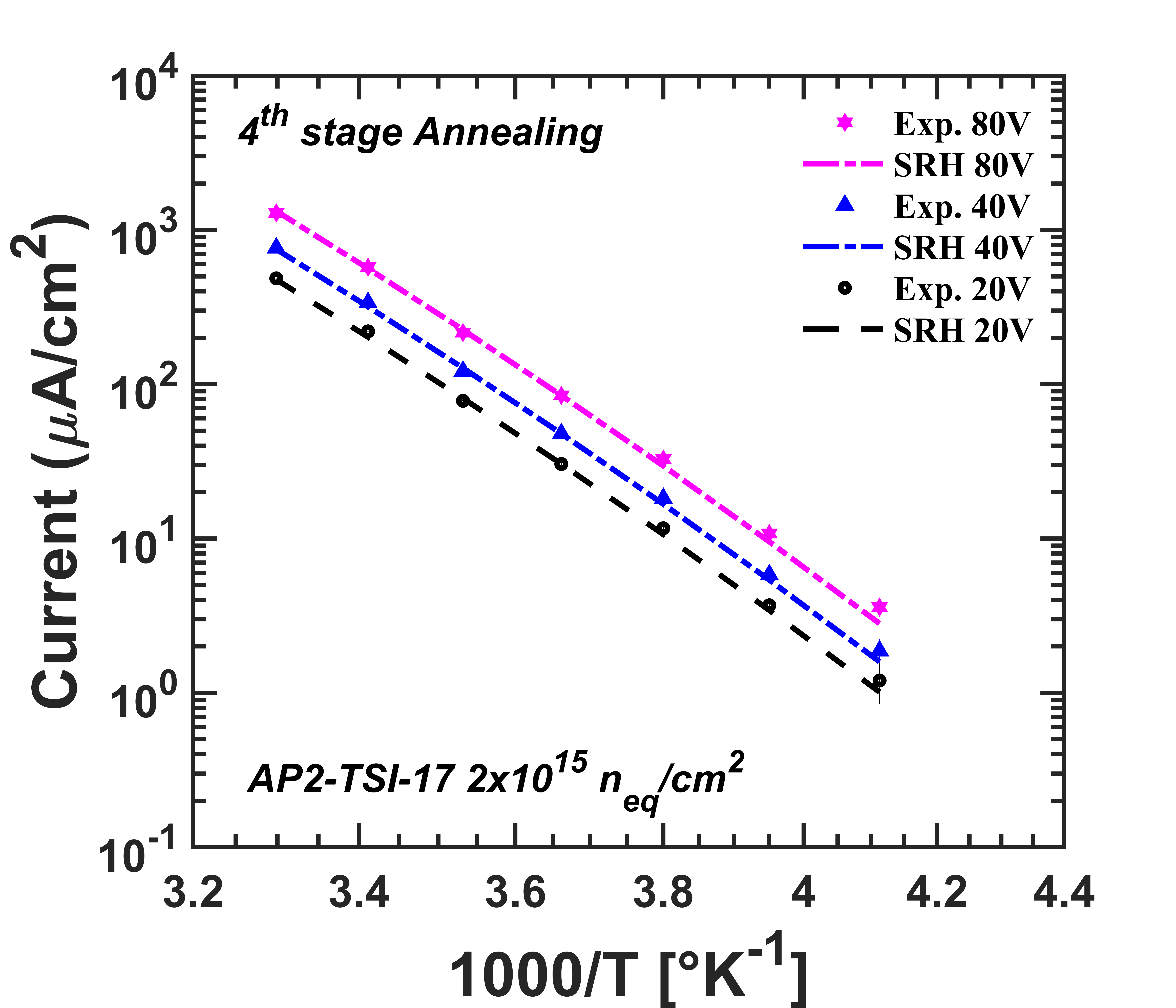}}
\caption{\label{f:fig15}Measured I-V curves (top) and corresponding Arrhenius plots (bottom) of irradiated ATLASPix2 chips: AP2-TSI-05 (irradiated to $1\times 10^{15}~\onemev$, left), AP2-TSI-06 (irradiated to $1.5\times 10^{15}~\onemev$, center) and AP2-TSI-17 (irradiated to $2\times 10^{15}~\onemev$, right).  All chips were further annealed at \SI{80}{\celsius} for 300 min after third stage accelerated annealing. See table~\ref{t:samples} for a description of the samples.}
\end{figure}

Table~\ref{t:results-annealing-4} summarizes all electrical characteristics of the selected samples after the fourth stage annealing. I-V measured at different ambient conditions and the corresponding Arrhenius plots can be found in figure~\ref{f:fig15}. TSI sample’s breakdown after the fourth stage annealing is almost \SI{100}{\volt} (irrespective of the irradiation fluence), as it was seen for TSI non-irradiated prototypes, meaning beneficial annealing played the major role in all annealing stages. A few volts lower breakdown voltage has been observed for the ams sample (table~\ref{t:results-annealing-4}) in comparison to the first stage accelerated annealed data (table~\ref{t:results-annealing-1}). This can be explained from the peripheral current of ams sample. The interface states and traps are recovered by accelerated annealing and the peripheral current contributes to trigger an early impact ionization.    

\begin{table}[htbp]
\small
\centering
\caption{\label{t:results-annealing-4} Electrical characteristics of irradiated ATLASPix2 samples after the fourth annealing stage (\SI{80}{\celsius} for 300 min). The last column indicates the total cumulative equivalent annealing time at \SI{20}{\celsius}.}
\smallskip
\begin{tabular}{|c|l|c|c|c|c|}
\hline
{\bf Fluence} & \multicolumn{1}{c|}{\bf Device ID} & \multicolumn{1}{c|}{\bf $J_\text{lk}$} & \multicolumn{1}{c|}{\bf $\alpha^*$} & {\bf $V_\text{bd}$} & {\bf Equiv. cum.} \\ 
 $[\onemev]$ && \multicolumn{1}{c|}{[\SI{}{\micro\ampere\per\centi\meter\squared}]} & $[\times 10^{-17}\,\rm{A/cm}]$ & [V] & {\bf ann. at \SI{20}{\celsius}} \\ \hline
\multirow{2}{*}{$1\times 10^{15}$} 
 & AP2-AMS-09 & $(9.45\pm 0.08)$ @58 V & $(0.67\pm 0.07)$ @50 V & $60\pm 2$ & \multirow{4}{*}{$\sim$993 days}\\ \cline{2-5}
& AP2-TSI-05 & $(11.48\pm 0.08)$ @94 V & $(0.61\pm 0.11)$ @80 V & $96\pm 2$ &\\ \cline{1-5}
$1.5\times 10^{15}$ & AP2-TSI-06 & $(22.12\pm 0.08)$ @94 V & $(0.78\pm 0.12)$ @80 V & $96\pm 2$ &\\ \cline{1-5}
 $2\times 10^{15}$ & AP2-TSI-17 & $(46.07\pm 0.09)$ @96 V & $(1.18\pm 0.17)$ @80 V & $98\pm 2$ &\\ \hline
\end{tabular}
\end{table}

Figure~\ref{f:fig13b} summarizes the improvement of leakage current related damage rate, $\alpha^*$ in correspondence to the equivalent near room temperature annealing (\SI{20}{\celsius}) (numerically calculated). The very first $\alpha^*$ values for samples of different fluences are representative to the leakage current after irradiattion but before applied annealing steps. 

% The $\alpha^*$ values for samples of all fluences shown at \SI{120}{\minute} annealing period  are representative of annealing treatments, might have experienced during handling and electrical characterizations after irradiation. Later values retrieved from successive accelerated annealing steps showed the decreasing trend with annealing time, even for the samples treated for ~33.1 months long annealing near room temperature. Accelerated annealing effectively helps to freeze out the complex vacancies, the donor-like oxygenated traps (CiOi) and the acceptor-like traps (VP). MCz wafer used for these prototypes production are intrinsically enriched with oxygen impurities, it is thus expected that oxygenated traps remain much larger in quantity than the acceptor-like after neutron irradiation [22]. Elevated temperatures at accelerated annealing treatments reduce these traps in terms of the free carrier recombination time constant ($\tau$) increase.

It is worth noting that $\alpha^*$ has been decreased by a factor of 2 for prototypes (exposed to all fluences). The $\alpha^*$ values are, within uncertainties, in very good agreement with the exponential-logarithmic model of leakage current~\cite{moll}. The ams chip shows a larger damage constant at $\SI{50}{\volt}$ for $1\times 10^{15}\ \onemev$ neutrons irradiation (figure~\ref{f:fig13b}). In fact, the data  for TSI chip has been reported for a higher reverse bias, $\SI{80}{\volt}$. Peripheral current in the ams chip would have contributed to this visible discrepancy.

\section{Conclusions}
\label{sec:conclusions}

Electrical characterizations have been extensively performed on ATLASPix2 prototypes produced by both ams AG and TSI Semiconductors on MCz 20~\ohmcm\ P-type substrate. TSI produced ATLASPix2 prototypes appeared better than ams samples in terms of leakage current and breakdown voltage. Non-irradiated TSI samples reported breakdown voltage at greater than \SI{100}{\volt}, very well in agreement with \SI{180}{\nano\meter} HV-CMOS technology design standards. In contrast, ams ATLASPix2 samples reported a breakdown voltage of \SI{50}{\volt} before irradiation, as seen in their predecessor, the ATLASPix1. The leakage current of TSI samples before irradiation were found to be a fraction of $\rm{nA/cm^2}$ at \SI{-10}{\celsius}, which is a factor of 5 to 6 less than for the ams prototypes. SRH prediction before irradiation shows a good agreement with the measured data (up to \SI{0}{\celsius}) for TSI prototypes and the disagreement seen at the lower temperatures has been driven by carrier ionization rate increase.           

To understand the radiation hardness property of the ATLASPix2 sensor-diodes, samples from both foundries were irradiated up to $2\times 10^{15}~\onemev$ neutrons. After irradiation, measured I-V data for the samples of both foundries were found in good agreement with the SRH model (irrespective of the cumulated fluences), meaning the effect of surface current and ionization rate increase at lower temperature become less significant. The leakage current of TSI samples is a factor of two less than for ams AG prototypes at any respective reverse bias (earlier than avalanche breakdown). For prototypes from both foundries irradiated at a higher fluence, the leakage current increases, as expected. It is also worth noting that the radiation hardness property of pixel-electronics has been beyond the scope of this paper, that would require additional functional-tests.   

Irradiation induces surface interface states and bulk-traps, which help to seize peripheral carriers and delay the impact ionization process, as observed in ams AG samples. Thus, the breakdown voltage has been improved by around $\SI{10}{\volt}$ after irradiation. In the case of irradiated TSI prototypes, an early breakdown voltage driven by impact ionization effect (10 to $\SI{20}{\volt}$ less) is observed in comparison to non-irradiated samples. C-V data measured at irradiated samples shows a few hundreds of \SI{}{\femto\farad}, are comparable to an initial two-dimensional TCAD simulation result. The lateral depletion plateau has also been seen for samples of higher irradiation fluences ($1.5\times 10^{15}$ and $2\times 10^{15}~\onemev$ neutrons) at ~\SI{15}{\volt}, determining the minimal reverse bias required for the sensor operation to obtain a full depletion laterally around the pixel diodes.

In an aim to study the annealing impact on ATLASPix2 prototypes, a systematic annealing study has been performed on prototypes of both foundries. Four stages of accelerated annealing applied for ATLASPix2 prototypes, where the cumulative equivalent time can be estimated to $\sim$33.1 months-long room temperature annealing (\SI{20}{\celsius}). Beneficial annealing seems to play the major role as leakage related damage-rate ($\alpha^*$) decreased even after the fourth stage of accelerated annealing. The breakdown voltage for all samples shows a trend to reach its non-irradiated condition, as expected from annealing driven damage recovery. The lateral depletion observed in C-V data (after annealing treatments) has been seemed unaffected. A dedicated C-V measurement of a thinner sensor (comparable to the actual depletion depth) may give a better picture of what can happen at full depletion, and it is yet to be investigated.      

Regardless of uncertainties, primarily those in the irradiation fluences, depleted volumes and annealing conditions, which would have an impact on the measured values, the leakage current behavior of TSI samples did not show anomalous deviations from the theoretical expectations. The breakdown voltage remained greater than \SI{85}{\volt} for the highest fluence of neutrons considered within this study. The electrical properties variation through annealing are dominated by the  beneficial kind, for a length of time corresponding to a scenario where sensors may remain at room temperature for more than two years in a foreseen long shutdown.

% A similar solution is available for figures via the \texttt{subfigure}
% package (not loaded by default and not shown here).
% All figures and tables should be referenced in the text and should be
% placed on the page where they are first cited or in
% subsequent pages. Positioning them in the source file
% after the paragraph where you first reference them usually yield good
% results. See figure~\ref{fig:i} and table~\ref{tab:i}.

% \begin{figure}[htbp]
% \centering % \begin{center}/\end{center} takes some additional vertical space
% \includegraphics[width=.4\textwidth,trim=30 110 0 0,clip]{example-image-a}
% \qquad
% \includegraphics[width=.4\textwidth,origin=c,angle=180]{example-image-b}
% % "\includegraphics" from the "graphicx" permits to crop (trim+clip)
% % and rotate (angle) and image (and much more)
% \caption{\label{fig:i} Always give a caption.}
% \end{figure}

% \begin{table}[htbp]
% \centering
% \caption{\label{tab:i} We prefer to have borders around the tables.}
% \smallskip
% \begin{tabular}{|lr|c|}
% \hline
% x&y&x and y\\
% \hline
% a & b & a and b\\
% 1 & 2 & 1 and 2\\
% $\alpha$ & $\beta$ & $\alpha$ and $\beta$\\
% \hline
% \end{tabular}
% \end{table}

% We suggest not to abbreviate: ``section'', ``appendix'', ``figure''
% and ``table'', but ``eq.'' and ``ref.'' are welcome. Also, please do
% not use \texttt{\textbackslash emph} or \texttt{\textbackslash it} for
% latin abbreviaitons: i.e., et al., e.g., vs., etc.

\acknowledgments

Authors acknowledge Prof. Vladimir Cindro and Dr. Igor Mandic from JSI, Ljubljana for their support in the neutron irradiation. This project received funding from the SNSF grants 200021\_169015, 200020\_156083, 20FL20\_173601 and 200020\_169000.

\end{document}